\documentclass{article}

\usepackage{amsthm}

\usepackage{amsmath,amssymb,stmaryrd}
\usepackage{xspace}

\usepackage{graphicx}

\usepackage{mathrsfs}
\usepackage{url}

\newcommand{\deriveFor}[1]{\displaystyle\mathop{\Longrightarrow}_{#1}}

\newcommand{\mop}[1]{\ensuremath{\operatorname{#1}}\xspace}
\newcommand{\defn}[1]{\textbf{#1}} \newcommand{\goto}{\mop{goto}}
\newcommand{\action}{\mop{action}} \newcommand{\shift}{\mop{shift}}
\newcommand{\reduce}{\mop{reduce}} \newcommand{\accept}{\mop{accept}}
\newcommand{\error}{\mop{error}} \newcommand{\first}{\mop{FIRST}}

\newcommand{\validDelta}[3]{#1\in\valid{#2}{#3}}
\newcommand{\notValidDelta}[3]{#1\notin\valid{#2}{#3}}
\newcommand{\valid}[2]{\mathcal{V}_{#1}(#2)}
\newcommand{\deltaSet}[2]{\mathcal{#2}_{#1}}
\newcommand{\allDeltas}[1]{\deltaSet{#1}{A}}
\newcommand{\emptyDelta}{\Delta_{\mathrm e}}
\newcommand{\derives}{\displaystyle\mathop{\Longrightarrow}}
\newcommand{\rmDerive}[1]{\derives_{rm}}

\newcommand{\Nu}{\mathrm{N}} 
\newcommand{\TITLE}{Parsing Transformative LR(1) Languages}
\newcommand{\gstring}[1]{\ensuremath{\mathit{#1}}\xspace}
\newcommand{\alt}{\mathrel{|}}

\newcommand{\node}[1]{\ensuremath{\mathsf{#1}}\xspace}

\newcommand{\production}[2]{\ensuremath{\gstring{#1}\to\gstring{#2}}\xspace}

\newcommand{\eofsym}{\ensuremath{\mathord{\dashv}}\xspace}

\newcommand{\conservedBy}[1]{\displaystyle\mathop\simeq_{#1}}
\newcommand{\lang}[1]{L(#1)}
\newcommand{\nontransDerives}[1]{\derives_{#1\text{ nt}}}
\newcommand{\transDerives}[1]{\derives_{#1\text{ t}}}
\newcommand{\semiParse}{\displaystyle\mathrel\rightharpoonup}
\newcommand{\validSemiParse}{\displaystyle\mathrel \rightarrowtail}
\newcommand{\reducesTo}{\displaystyle\mathop{\Longmapsto}}
\newcommand{\transReducesTo}[1]{\displaystyle\mathop{\Longmapsto}_{#1\text{
t}}} 
\newcommand{\nontransReducesTo}[1]{\displaystyle\mathop{\Longmapsto}_{#1\text{
nt}}} 
\newcommand{\code}[1]{\texttt{#1}}
\newcommand{\deltaMachine}{$\Delta$\nobreakdash-\hspace{0pt}machine\xspace}
\newcommand{\sub}[1]{_\mathrm{#1}}
\newcommand{\dash}{\nobreakdash-\hspace{0pt}}
\newcommand{\kparse}[1]{\ensuremath{#1}\dash{}parse}
\newcommand{\PhiLimit}{$\Phi$\dash{}limit\xspace}
\newcommand{\LambdaLimit}{$\Lambda$\dash{}limit\xspace}

\newcommand{\Bfactorization}{B\dash{}factorization}

\newcommand{\properChildString}{\ensuremath{\mathscr S}}
\newcommand{\childString}{\ensuremath{\mathscr C}}
\newcommand{\nodeLabel}{\ensuremath{\mathscr L}}
\newcommand{\treeRoot}{\ensuremath{\mathscr R}}

\newcommand{\connectsBack}[1]{\ensuremath{\displaystyle\mathop\vartriangleright_{#1}}\xspace}

\newcommand{\notConnectsBackEpsilon}{\ensuremath{\not\vartriangleright}\xspace}

\newcommand{\parsePreceeds}[1]{\ensuremath{\displaystyle\mathop\blacktriangleright_{#1}}\xspace}
\newcommand{\parsePreceedsEpsilon}{\ensuremath{\blacktriangleright}\xspace}
\newcommand{\nth}[1]{{#1}^{\text{th}}}
\newcommand{\inv}{^{-1}}

\newtheorem{prop}{Proposition}
\newtheorem{algo}{Algorithm}
\newtheorem{proc}[algo]{Procedure}

\newtheorem{lemma}{Lemma}
\newtheorem{cor}{Corollary}
\newtheorem{theorem}{Theorem}

\newcommand{\dref}[1]{Definition~\ref{#1}}
\newcommand{\sref}[1]{Step~\ref{#1}}
\newcommand{\pref}[1]{Proposition~\ref{#1}}
\newcommand{\lref}[1]{Lemma~\ref{#1}}
\newcommand{\tref}[1]{Theorem~\ref{#1}}
\newcommand{\algoref}[1]{Algorithm~\ref{#1}}
\newcommand{\secref}[1]{Section~\ref{#1}}
\newcommand{\procref}[1]{Procedure~\ref{#1}}
\newcommand{\cref}[1]{Corollary~\ref{#1}}
\newcommand{\condref}[1]{Condition~\ref{#1}}
\newcommand{\ruleref}[1]{Rule~\ref{#1}}
\newcommand{\caseref}[1]{Case~\ref{#1}}
\newcommand{\tableref}[1]{Table~\ref{#1}}
\newcommand{\figref}[1]{Figure~\ref{#1}}
\newcommand{\myabstract}{We consider, as a means of making programming
languages more flexible and powerful, a parsing algorithm in which the
parser may freely modify the grammar while parsing. We are
particularly interested in a modification of the canonical LR(1)
parsing algorithm in which, after the reduction of certain
productions, we examine the source sentence seen so far to determine
the grammar to use to continue parsing. A naive modification of the
canonical LR(1) parsing algorithm along these lines cannot be
guaranteed to halt; as a result, we develop a test which examines the
grammar as it changes, stopping the parse if the grammar changes in a
way that would invalidate earlier assumptions made by the parser. With
this test in hand, we can develop our parsing algorithm and prove that
it is correct. That being done, we turn to earlier, related work; the
idea of programming languages which can be extended to include new
syntactic constructs has existed almost as long as the idea of
high-level programming languages. Early efforts to construct such a
programming language were hampered by an immature theory of formal
languages. More recent efforts to construct transformative languages
relied either on an inefficient chain of source-to-source translators;
or they have a defect, present in our naive parsing algorithm, in that
they cannot be known to halt. The present algorithm does not have
these undesirable properties, and as such, it should prove a useful
foundation for a new kind of programming language.}

\usepackage{txfonts}

\ifx\pdfoutput
\usepackage[pdftex]{hyperref}
\hypersetup{colorlinks=false,citecolor=blue,filecolor=red} 
\fi

\newtheorem{formaldef}{Definition}
\newenvironment{definition}{\begin{formaldef}\rm}{\end{formaldef}}

\newcommand{\algofin}{\hfill\ensuremath{\blacksquare}}

\newenvironment{algodesc}[2]{
\begin{description}
\item[Input] #1
\item[Output] #2
\item[Method]\end{description}}{}

\newenvironment{algorithm}[4]{
\begin{algo}[#1]\rm #2\end{algo}
\begin{algodesc}{#3}{#4}}{\end{algodesc}}

\newenvironment{procedure}[3]{
\begin{proc}\rm #1\end{proc}
\begin{algodesc}{#2}{#3}}{\end{algodesc}}

\newenvironment{conventionTable}
	{\begin{table}[b!]}{\end{table}}
	
\newcommand{\authorcite}[2]{#1 \cite{#2}}
\newcommand{\citeyear}[1]{\cite{#1}}

\newcommand{\preItem}{}
\newcommand{\postItem}{. }

\title{\TITLE}
\author{Blake Hegerle}
	
\begin{document}
\maketitle
%\tableofcontents
\abstract{\myabstract}

\section{Introduction}

Programming is the enterprise of fitting the infinitely subtle
subjects of algorithms and interfaces into the rigid confines of a
formal language defined by a few unyielding rules---is it any wonder
that this process can be so difficult? The first step in this process
it the selection of the language. As we go along in this enterprise,
we might find that our selected language is inadequate for the task at
hand; at which point we can: forge ahead with an imperfect language,
we can attempt to address the problematic section in a different
language, or we can jettison the language for another with its own
limitations, thereby duplicating the effort already put into writing
the program in the first language. With ever larger, more complex
programs, we increasingly find that no single language is especially
well-suited---yet if we try to use multiple languages, we face
significant hurdles in integrating the languages, with rare
exceptions.  A fourth possibility presents itself: we could create a
new programming language that contains all of the features we will
ever need in any section of the program; aside from the fact that
creating a general programming language is a monumental effort in and
of itself, the resulting programming language will likely be a
cumbersome monster. What we seek is a language that is at once general
enough to suffice for very large programs, while also having specific
features for each portion of the program.

There are a great deal of mature programming languages in existence,
each with its own advantages and disadvantages. None of these are the
language we seek.  Ideally, we would like to be able to take an
existing programming language and---without having to duplicate the
tremendous amount of effort which went into its creation and
development, not to mention our own effort in learning it---mold it to
our needs.

We do not have time to survey the major languages, but programs in
these languages do fit a general mold: programs must be syntactically
well-formed, then they must be semantically well-meaning, and finally,
they must specify a program that is free from run-time errors. Moving
from the source code for a program to a run-time executable involves
three phases: syntax analysis, semantic analysis, and code
generation. The first two analysis phases are not separated in
practice, but are performed in concert by a parser which is generated
by a parser generator. The parser generator takes a grammar describing
the syntax of the programming language, in addition to the semantic
value of each production in the grammar, from which it produces a
parser. If we had the source code to the compiler, we could change it
to suit our purposes, producing a \defn{derived language}. However, we
must be careful if we do this, for changes to the code generator could
produce binaries that lack compatibility with existing binaries.

The aforementioned approach is not terribly common: its most glaring
problem is that a program written in a derived language cannot be
compiled by a ``normal'' compiler. An alternative is to make a new
compiler wholesale---one that, rather than outputting a binary,
outputs source code in an existing programming language; such a
compiler is known as a \defn{source-to-source translator}. The
practice of creating source-to-source translators is much more common
that the practice of creating derived languages; two examples are the
cfront compiler for C++ and a WSDL compiler for SOAP. These two
examples illustrate an interesting point: the new language can share
much with the target language, as is the case with cfront; or, the new
language can share nothing with the target language, as is the case
with a WSDL compiler.

Creating a derived language is an attractive concept because we can
directly leverage an existing implementation of a base language, but
modifying any large program---the compiler, in this case---in an
ad-hoc manner is not exactly an easy task. This approach becomes
decidedly less attractive if we seek to radically alter the language:
we will likely find that the code generator is tightly coupled with
the parser, and that the facilities for creating abstract syntax trees
have no more generality than is necessary for the original
language. Creating a source-to-source translator, on the other hand,
is an attractive concept because we can make a language that departs
from the target language as much or as little as we want; however,
perhaps too much information is lost in the conversion to the target
language: data such as debugging information, higher-order typing,
optimization hints, and details necessary for proper error handling
are just a few of the things which might get lost. Another problem is
what I refer to as the ``language tower problem'': say we start with a
language L, then we create a source-to-source translator from L++ to
L, then we create a source-to-source translator for Aspect-L++, then
we create a source-to-source translator for Visual Aspect-L++, ad
nauseam---in short, we end up with far too many parsers.

We will take an approach somewhere between these two. We would like to
develop a \defn{base language} that is general purpose enough to serve
us in its unmodified form, yet can be modified at our pleasure. In
light of our consideration of derived languages, we will create a
general-purpose framework for abstract syntax trees that both the base
language and any derived language can use to capture the full range of
the semantics of a program. Also, we will not require someone wishing
to create a derived language to create an entire grammar: we will
allow modifications of the existing grammar. We have avoided most of
the problems associated with source-to-source translation as well: the
data. As an example consider: run-time metadata (like profiling and/or
debugging data), data type, optimization hints, and error messages;
none of these are likely to be present in the binary if
source-to-source translators are used. Since we are making it easy to
modify the language, we would expect that the language tower problem
would be exacerbated, but this is hardly the case, for there is only
ever a singular parser.

We pause to note that there must be some way of specifying the
semantic actions of a production. We can assume that these actions are
specified in a programming language, probably the base language
itself, and that the parser has an interpreter for that language
included in its implementation.

The code generator only understands so much of what is potentially in
an abstract syntax tree. Everything else---the debugging, type,
optimization, and error data---which gets added to the tree must be,
to a large extent, ignored by the code generator. However, these
data---we will call them \defn{extended semantics data}---are not
valueless; thus, we will allow additional analysis phases to be
performed on the abstract syntax tree between that parsing and the
code generation phases. Here again, an interpreter embedded in the
parser will be invaluable.

How might a language like this be used? We can use it to add gross
language features, for example object-oriented or aspect-oriented
support. Or we could add more behind-the-scenes features, improving
for example, the optimizer. If we know that we are using a particular
library, we can give first-class syntactic support to common
patterns---for example, we could support monitors, as Java does with
the \code{synchronized} keyword. Finally, we could create a modified
grammar to eliminate repetitive code, using the modifiability of the
language as a sort of macro processor.

We will allow the parser to modify itself \emph{during parsing}. From
here on, we will assume that a parser operates strictly
left-to-right. No longer can we treat the syntax analysis and semantic
analysis phases as entirely separate, even conceptually, for some part
of a file may define the syntax and semantics of the remainder of the
file.

The study of formal languages has produced many interesting classes of
languages: regular, context-free, context-sensitive, and
recursive-enumerable being the best known. If $\mathcal X$ is a class
of languages, then the set of \defn{transformative} $\mathcal X$
languages are those languages whose strings $x$ can be decomposed as
$x=y_1 y_2 \dotso y_n$, such that $y_i$ is a substring of an element
of one of the languages in $\mathcal X$, which we term the $\nth{i}$
\defn{instantaneous language}; further, $y_i$ specifies the
instantaneous language $i + 1$.

Our goal in the present work will be to develop a method of parsing a
useful class of transformative languages. Our parser will operate much
like a classical parser, except that, as it moves over the boundary
between $y_i$ and $y_{i+1}$, it will modify itself---more precisely,
it will modify its grammar, and then its parsing tables. Since we are
dealing with a self-modifying parser, we would run into problems if
the parser were to backtrack from $y_{i+1}$ to $y_i$---not
insurmountable problems, to be sure, but we will find a satisfactory
non-backtracking method of parsing that does not have these problems.

\subsection{Applications of Transformative Parsing}

Let us say that we need to write a graphical program in a language
much like Java which can access both a web service and a database;
let us assume that we do not have any visual rapid-development
tools. We must write a lot of GUI code like ``make a window, put a
layout in the window, put the following controls in the layout: \dots,
add a toolbar to the window, add an item to the toolbar with the label
`x', set the callback object to `y''' etc. We must write a lot of
database like ``parse a query, bind the following variables (\dots),
execute the query, create a cursor, advance the cursor, get the first
column, get the second column'' etc. We must write a lot of web
service code like ``create a procedure call, marshal the input, call
the procedure, demarshall the output, handle any exceptions'' etc. The
GUI, database, and web service functionality is most likely handled by
a library.  Would that each library added syntax constructs for the
operations it provides. We could declare the GUI with code like
\begin{verbatim}
window{ layout{...}; toolbar{ item{ label = 'x'; action = y } } }
\end{verbatim}
The interesting thing is that the \verb-y- identifier is bound to the
correct lexical scope. 

We could process our query with code like
\begin{verbatim}
query(select col1, col2 from t1 where col=$z -> (c1, String c2) {
   ...
}
\end{verbatim}
Here, \verb-z- is a variable in the scope containing the \verb-query-
construct, as is \verb-c1-; however, \verb-c2- is local to the block
after the \verb-query-.

 Finally, we could process our web service with
code like: 
\begin{verbatim}
webservice service=ws_connect{ url=http://www/shop, id=shop };
a=shop.lookup(b, c);
\end{verbatim}
There is nothing to prevent the compiler from doing a compile-time
type-safety check: is the column \verb-col- on the table \verb-t1- the
same type as the variable \verb-z-, or is the column \verb-col1- the
same type as \verb-c1-? Something similar can be done for the
webservice. Is the second argument of the \verb-lookup- method of the
webservice the same type as \verb-c-?

The way that we arrive at the functionality requirements for these
examples is to allow the library write to specify new syntactic
constructs to simplify complex, error-prone tasks---along with this
syntax, there must be provided a semantic description of the new
construct. A perfectly viable way to specify semantics is as YACC and
ANTLR do: associating a block of procedural code with each production,
which will be run after that production is used.

We conclude with the observation that the capabilities of the
hypothetical system under discussion in this section are not really
new. Indeed, for the web service example at least, the capabilities
are common. However, continuing with that example, the method used to
achieve web service integration with the host language is by means of
a WSDL compiler; one of our goals is to make obsolete artifacts like
WSDL compilers. What \emph{is} new is a single mechanism which
integrates the language with the program.

\subsection{Conventions}

Will will also observe some (fairly standard) typographical
conventions to denote the type of variables. See
\tableref{table:conventions}. We often deal with grammars in the
sequel, which have either 4 or 6 components; in most cases, we will
label the components of these grammars as $(\Sigma, \Nu, P, S)$ and 
$(\Sigma, \Nu, P,S,T,M)$, as appropriate. 

\begin{conventionTable}
  \begin{tabular}{ll}
    terminals & $a,b,c,d,e,f$\\
    nonterminals & $A,B,C,D,E,F$\\
    terminal strings & $r,s,t,u,v,w,x,y,z$\\
    symbol strings & $\alpha,\beta,\gamma,\delta,\zeta,\eta,\theta,\kappa,\lambda$\\
    symbol & $U,V,W,X,Y,Z$\\
    start symbol & $S,S'$\\
    production & $\pi,\tau,\phi$\\
    node & $\mathsf{A},\mathsf{B},\mathsf{C},\mathsf{D},\mathsf{E},\mathsf{F}$\\
    set of nodes & $\mathsf{U}, \mathsf{V}, \mathsf{W}, \mathsf{X}, \mathsf{Y}, \mathsf{Z}$\\
    transformation & $\Delta$\\
    set of transformations & $\mathcal{V},\mathcal{T},\mathcal{D}$\\
    a  grammar & $G$
  \end{tabular}
  \caption{Typographical conventions.\label{table:conventions}}
\end{conventionTable}

\subsection{The Rest of This Document}

There are 4 main sections in the sequel. In \secref{sec:transformative
  lr(1) languages}, we present the formal definition of a
transformative language---this definition is perhaps surprisingly
complex, but it allows us to establish \tref{thm:valid transformations
  do not extend derivations}, a result which states that parsing
sentences in an appropriate transformative language can be done in a
finite amount of time.

The Algorithm to recognize sentences generated by a given grammar we
present and justify in \secref{sec:tlr algorithms}. We rely on the
determination that a particular object is in a particular set---the
object being a transformation (i.e., a change of grammar) being what
we term ``valid.''  In \secref{sec:checking the validity of a
  transformation}, we present an Algorithm to test a transformation
for membership in the aforementioned set of valid transformations. We
spend a great deal of time proving that this Algorithm is correct.

Finally, in \secref{sec:related work}, we survey related work: there
are the previously mentioned ``extensible languages;'' other
investigations into parsers whose grammar changes while parsing,
frameworks for creating derived (from Java, usually) languages, macro
systems are related in this Section. There are also a few works very near to the
present one.

%%% Local Variables: 
%%% mode: latex
%%% TeX-master: "tlr-plain"
%%% End: 

\section{Transformative LR(1) Languages}
\label{sec:transformative lr(1) languages}

The LR($k$) class of languages, where $k\ge 0$, is the largest known
class of languages which can be parsed without backtracking. The
theory of these languages, along with an algorithm to determine that a
grammar is LR($k$), are due to \authorcite{Knuth}{knuth:lr}, but we
will primarily draw upon the presentation of this theory from
\cite{aho-ullman:top}. We will not concern ourselves with the more
general case of LR($k$) languages for $k \ne 1$. It shall turn out to
be the case that LR(1) languages are convenient as a basis for
creating a practical transformative language. 

We first review LR(1) parsing. We then try to extending the LR(1)
parsing algorithm in the most naive way possible. The first attempt we
make will not be successful, but it will illustrate a subtle problem
in the development of transformative LR(1) languages. Rectifying these
problems will occupy us for much of the rest of this work.

\subsection{LR(1) Languages}
\label{sec:lr(1) languages}

The LR(1) class of languages are a subset of the context-free
languages. Context-free languages are those that are generated by a
context-free grammar, which is a tuple $(\Sigma,\Nu,P,S)$, where
$\Sigma$ is the terminal alphabet, $\Nu$ is the nonterminal alphabet,
$P$ is the set of productions, and $S$ is the start symbol. We
establish the convention that there is a special symbol
$\eofsym\in\Sigma$, that does not appear on the right side of any
production; this symbol is used to terminate strings in the
language. We will specify LR(1) languages by presenting a context-free
grammar for that language; given a context-free grammar, we cannot
tell at first glance whether or not the grammar specifies a LR(1)
grammar, rather, we must utilize the tools of LR(1) theory to make
this determination.

We must clarify some notation we will have occasion to use. Let
$G=(\Sigma,\Nu,P,S)$ be a context-free grammar. Since ``$\to$'' is a
binary relation, it is in fact a subset of the cartesian product
$\Nu\times(\Sigma\cup\Nu)^*$; this subset is $P$. If $A\to\alpha$ is a
production, then there is an ordered pair $(A,\alpha)\in P$. We have
no problem with notation like: $\pi=(A,\alpha)$; we therefore ought to
have no problem with notation like: $\pi=A\to\alpha$. We take the
symbol $\derives$ to mean the replacement of the rightmost nonterminal
in a string, and we take $\derives^*$ to be a rightmost derivation of
zero or more steps.

One way to define LR(1) languages is via \dref{def:lr(k)}, which,
along with \dref{def:first}, is from Chapter~5 of
\cite{aho-ullman:top}.

\begin{definition}
  \label{def:first}
  If $G=(\Sigma,\Nu,P,S)$ is a context-free grammar, then for any
  $\alpha\in(\Sigma\cup\Nu)^*$, we define $\first_k(\alpha)$ to be the
  set of all $y\in\Sigma^*$, where $|y|=k$, such that
  $\alpha\derives_G^* yx$ for some $x\in\Sigma^*$. We understand
  $\first(\alpha)$ to mean $\first_1(\alpha)$. We call
  $\first(\alpha)$ the \defn{first set} of $\alpha$.
\end{definition}

\begin{definition}
  Let $G=(\Sigma, \Nu, P, S)$ be a context-free grammar, and let $S'$
  be a nonterminal not in $\Nu$. Define the context-free grammar
  $(\Sigma, \Nu \cup \{S'\}, P \cup \{S' \to S \}, S')$ as the
  \defn{augmented grammar} associated with $G$.
\end{definition}
We are interested in augmented grammars because it is an easy way of
ensuring that the start symbol does not appear on the right side of
any production: this is a necessary condition for the construction of
LR($k$) parser tables.

\begin{definition} 
  \label{def:lr(k)}
  Let $G=(\Nu,\Sigma,P,S)$ be a CFG and let $G'=(\Nu',\Sigma,P',S')$ be
  its augmented grammar. We say that $G$ is LR($k$), $k \ge 0$ if the
  three conditions
  \begin{enumerate}
  \item $S' \deriveFor{G'}^* \alpha A w \deriveFor{G'} \alpha\beta w,$

  \item $S' \deriveFor{G'}^* \gamma B x \deriveFor{G'}
    \alpha\beta y,$ and
    
  \item $\first_k(w)=\first_k(y)$
  \end{enumerate}
  imply that $\alpha A y=\gamma B x$. (That is, $\alpha=\gamma$, $A=B$,
  and $x=y$.)
\end{definition}

\subsubsection{Shift-Reduce Parsing}

\label{sec:shift-reduce parsing}

A shift-reduce parser is a deterministic pushdown automaton with a
stack of binary tuples, controlled by a \defn{parsing table},
calculated from the language's context-free grammar before parsing
begins. Each tuple on the stack is in the set $\mathbb Z_k \times (\Sigma \cup
\Nu \cup \{\epsilon\})$, where $K$ is a finite set of integers. The
parser can do one of three things:
\begin{enumerate}
\item it can remove the first symbol from the input string, and put it
  and a state onto the stack, a move which we will call a
  \defn{shift};

\item it can remove zero or more tuples from the stack, and replace
  them with a single new tuple, a move which will will call a
  \defn{reduction}; or,

\item it can halt.
\end{enumerate}
Should the automaton be in an accepting state when it halts, then we
know that $x \in \lang G$, and we say that the parser \defn{accepts} the
string $x$. Should the parser halt in any other state, then we know
that $x \notin \lang G$, and we say that the parser \defn{rejects} the
string $x$. The parser is in an accepting state if and only if it just
reduced by the production $S' \to S$.

The automaton examines the current input symbol, which will will call
$a$, and takes an action based upon the value of $a$ and the value of
the integer in the tuple on top of the stack, which we will call $k$;
if $\action[k,a]=\shift m$, then we set $a$ to be the next input
symbol and we push $(m, a)$ onto the stack; if $\action[k,a]=\reduce
\text{``} A\to\alpha \text{''}$, then we pop $|\alpha|$ states off of
the stack and, letting $(k', X)$ be the tuple on the top of the stack
after popping those items off, we push $(\goto[k',A], A)$ onto the
stack; if $\action[k,a]=\error$, then we halt in a non-accepting
state; finally, if $\action[k,a]=\accept$, then we halt in an
accepting state.

\subsubsection{The Canonical Shift-Reduce Parser for an LR(1) Grammar}

As we just saw in \secref{sec:shift-reduce parsing}, all of the
decisions on how to parse a string are deferred to the construction of
the parsing tables. It is this construction we turn to now.

Given a context-free grammar, there are many ways of constructing
parsing tables for a shift-reduct parser.  Not every method will
succeed for a given context-free grammar, but if a grammar is LR(1),
there is one method which is guaranteed to work: this is the original
method of Knuth \cite{knuth:lr}, and we refer to the parser (the
tables used by the shift-reduce parser, specifically) produced by this
method as the \defn{canonical LR(1) parser} for that grammar.

The previously cited source does present the algorithm for the
construction of LR(1) parsing tables \cite{knuth:lr}, but in an
indirect form; a more direct presentation is \cite{aho-ullman:top}.
Perhaps the most friendly presentation of this algorithm is in
\cite[chap.~4]{dragon}. We need only summarize the algorithm here,
following the presentation from \cite{wiki:canonical-lr}. Let
$G=(\Sigma,\Nu,P,S)$ be a context-free grammar. We begin by augmenting
the grammar. We then construct sets of \defn{LR(1) items}; in general,
these items are of the form $[A\to\alpha\cdot\beta,a]$, where
$\production{A}{\alpha\beta}$ is a production, and
$a\in\Sigma$. Intuitively, we think of an item as a memo to ourselves
that we are trying to match the production
$\production{A}{\alpha\beta}$, we have so far matched the $\alpha$,
and we expect to match $\beta$ later; the meaning of the ``$a$'' is
this: when we have matched $\alpha \beta$, we reduce by
$\production{A}{\alpha \beta}$ if and only if the lookahead is $a$.
We begin our construction of the item sets with the \defn{initial
  item} $[S'\to\cdot S, \eofsym]$; we let
\label{initial state is zero}
$I_0$ be the closed item set containing the initial item, where we
define an item set to be \defn{closed} if for every item of the form
$[A\to\alpha\cdot B\beta,a]$ in that set, such that $B\to\gamma$ is a
production, we have that $[B\to\cdot\gamma,b]$ is in that item set,
for all $b$ in $\first(\beta a)$ (see \dref{def:first}). For item sets
$I_k$ and $I_m$, and a grammar symbol $X\in(\Sigma\cup\Nu)$, we define
the \defn{goto function} on $I_k$ and $X$ to be the closed item set
$I_m$, which we write as $m=\goto[k,X]$, if $[A\to\alpha\cdot
X\beta,a]\in I_k$ and $[A\to\alpha X\cdot\beta,a]\in I_m$. Finally, we
define the \defn{action function} for an item set $I_k$ and a terminal
$a$ in one of two ways: if there is an item $[A\to\alpha\cdot a
\beta,b]\in I_k$, then we let $\action[k,a]=\shift m$, where
$m=\goto[k,a]$; otherwise, if there is an item $[A\to\alpha\cdot,a]\in
I_k$, then we let $\action[k,a]=\reduce \text{``}A\to\alpha\text{''}$.
The only exception to this last rule is if the item is the initial
item: if we reduce by the production $S' \to S$, then we \defn{accept}
the string, or recognize that the string is a member of the language. 

We can now encode the functions $\goto$ and $\action$ into two tables,
as suggested by the bracketed notation. For any entry on the action
table corresponding to an undefined value of the $\action$ function,
we give that entry the value of ``$\error$''. These are the
\defn{canonical LR(1) parsing tables}.

\subsection{A Note on Parse Trees}

If we have a context-free grammar $G=(\Sigma,\Nu,P,S)$ and we have
some $x\in\Sigma^*$, then we can prove that $x\in\lang G$ by supplying
a derivation of the form
\begin{equation}
  \label{eq:derivation proving x in lang G}
  S \derives \alpha_1 \derives \alpha_2 \derives \dotsb \derives
  \alpha_n=x.
\end{equation}

\begin{definition}
\label{def:simple parse tree}
  If \node T is a tree such that:
  \begin{enumerate}
  \item the root of \node T is labeled $S$; 

  \item every interior node, with label $X$, the children of which are
    labeled $Y_1,Y_2,\dotsc,Y_n$, such that $X\to Y_1Y_2\dotso Y_n$ is
    a production;

  \item every leaf is labeled with a terminal or $\epsilon$; and

  \item the yield of \node T (removing $\epsilon$'s) is $x$,
  \end{enumerate}
  then we say that \node T is a \defn{simple parse tree} for $x$.
\end{definition}

The existence of a simple parse tree for $x$ is a necessary and
sufficient condition for $x\in\lang G$.

\begin{definition}
  Let \node T be a tree whose nodes are labeled with terminals,
  productions or $\epsilon$. Let $\treeRoot(\node T)$ be the root of
  \node T. For any node \node N, define
  \begin{align*}
    \nodeLabel (\node N) &= \begin{cases}
      a & \text{if \node N is labeled with terminal $a$}\\
      \pi & \text{if \node N is labeled with production $\pi$}\\
      \epsilon& \text{if \node N is labeled with  $\epsilon$}\\
    \end{cases},
    \intertext{and define}
    \nodeLabel \sub H (\node N) & =\begin{cases}
      a & \text{if \node N is labeled with terminal $a$}\\
      A & \text{if \node N is labeled with production $A\to\alpha$}\\
      \epsilon& \text{if \node N is labeled with  $\epsilon$}\\
    \end{cases}.
    \end{align*}
  If \node N is an interior node labeled with the production $B \to
  \alpha$, whose children are $\node M_1, \node M_2, \dotsc, \node
  M_m$, then we define two more functions---$\childString$ and
  $\properChildString$---as follows: define
  \[
  \childString (\node N)=\nodeLabel \sub H (\node M_1) \nodeLabel \sub
  H (\node M_2) \dotso \nodeLabel \sub H (\node M_m),
  \]
  and define 
  \[
  \properChildString(\node N)=\alpha.
  \]
\end{definition}

We can reformulate \dref{def:simple parse tree} as follows.
\begin{definition}
  \label{def:parse tree}
  If \node T is a tree such that:
  \begin{enumerate}
  \item every interior node is labeled with a production;

  \item if \node N is an interior node, then $\childString(\node
    N)=\properChildString (\node N)$;

  \item every leaf is labeled with a terminal or $\epsilon$; and

  \item the yield of \node T (removing $\epsilon$'s) is $x$,
  \end{enumerate}
  then we say that \node T is a \defn{parse tree} for $x$.
\end{definition}

We developed the nonstandard definition of a parse tree in
\dref{def:parse tree} because we will have occasion, particularly in
\secref{sec:checking the validity of a transformation}, to do
nontrivial work on parse trees that would be impossible with simple
parse trees, as removing child nodes from a node destroys information
about that node.

\begin{definition}
  Let \node T be a tree and let \node N and \node M be nodes in \node
  T, such that \node N is an ancestor of \node M. One of the children,
  which we will call \node A, is a child of \node N such that either
  \node A is \node M itself, or \node A is an ancestor of \node M. In
  either case, we say that \node A is \defn{autoancestral} to \node M.
\end{definition}

We finish with a note on ordering parse trees. A parse tree is
inherently an ordered tree. If nodes $\node N_1$ and $\node N_2$ share
then same parent, then $\node N_1$ and $\node N_2$ a comparable; call
this partial order $\le \sub T$.  We will find it convenient to give
parse trees the following total order. 

\begin{definition}
  Let an ordered tree be given, with \node A and \node B nodes in that
  tree. Define $\node A\le \node B$ if any of the following are true:
  \begin{enumerate}
  \item $\node A=\node B$;

  \item \node B is a descendant of \node A; or

  \item there exist distinct nodes \node C and \node D with a shared
    parent and $C \le \sub T D$, such that \node C is autoancestral to
    \node A, and \node D is autoancestral to \node B.
  \end{enumerate}
\end{definition}

Our total ordering of the nodes in a tree suggests a method of
diagramming trees. We can illustrate this, along with some of the
other ideas of this section, with an example. Let $G=(\{a, b\}, \{S\},
P, S)$ be a context-free grammar, with $P=\{\production{S}{a S b \alt
\epsilon}\}$. Consider the derivation of the string $aabb$:
\[
S \derives a S b \derives aa S bb \derives aa bb.
\]
We can represent this using the diagram in \figref{fig:first parse
  tree}; 
\begin{figure}[t!]
  \begin{center}
    \includegraphics[scale=.75]{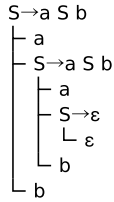}
  \end{center}
  \caption{\label{fig:first parse tree} The parse tree for the string
    $aabb$.}
\end{figure}
in that diagram, nodes appear least to greatest from top to
bottom.

\subsection{Grammars and Transformations}

When the grammar is changed during a parse, we will often want to
change only a part of the grammar, rather than the entire
grammar. Therefore, we will treat a change of grammar as the act of
adding and removing productions from the grammar; we call this act a
\defn{transformation of grammar}. Only after reduction by certain
productions will we add or remove productions from the grammar; we
call these productions \defn{transformative productions}.

We can begin to see how the parsing algorithm will have to be modified
to parse a transformative language: when the parser performs a
reduction, it checks to see if the production was a transformative
production; assuming that it was, we perform a transformation of the
grammar and calculate the parsing tables for the new grammar. The
parsing stack is completely dependent upon the particulars of the
parsing tables; hence, when we stop to perform a transformation of
grammar, we will endeavor to construct a new stack which will allow
parsing to continue from that point. \lref{lem:valid deltas maintain
  viable prefixes} provides a sufficent condition for a new stack to
be constructed; we look at the exact conditions for constructing a new
stack in \secref{sec:correctness}.

A method of determining which productions to add to or remove from a
grammar must be supplied with the grammar. These productions are
encoded in some manner in the portion of the sentence that the parser
has already scanned, and must be translated into a form the parser
understands. Conceivably, the parser could dictate the manner of this
encoding, but we will not allow the parser to do so. Rather, the
grammar will include a mechanism for decoding the change to the
grammar. This mechanism will take the form of a Turing machine that
will take as input the portion of the input already scanned, and will,
upon halting, contain upon one of its tapes an encoding of the
productions to add to or remove from the grammar in a form that the
parser understands; the contents of this tape encode the \defn{grammar
  transformation} to apply to the grammar. We will let $G \sub T$ be a
context-free grammar with terminal set $\Sigma \sub T$; grammars and
grammar transformations are encoded as strings in $\lang{G \sub T}$.

We will generally use the symbol ``$\Delta$'', or some variation on
it, to represent a grammar transformation. Owing to this terminology,
we will refer to the Turing machine which produces the transformation
as a \deltaMachine. An LR(1) grammar, a source sentence, together with
a \deltaMachine will form the input for our modified parsing
algorithm. We will define these objects precisely.

\begin{definition}
  Let $\Sigma$ be a terminal set. Let $M$ be a Turing machine with
  three semi-infinite tapes that have tape alphabets $\Gamma_1$,
  $\Gamma_2$, and $\Gamma_3$, respectively, where $\Gamma_1, \Gamma_2
  \supset \Sigma$ and $\Gamma_3 \supset \Sigma \sub T$. We will
  require that the machine does not output blanks on the second and
  third tapes; this allows us to define $w\in\Gamma_2^*$ as the
  contents of the second tape up to the first blank cell after the
  machine halts; similarly, we define $z\in\Gamma_2^*$ for the
  contents of the third tape up to the first blank cell after the
  machine halts. If we can guarantee that $M$ halts and that
  $w\in\lang{G \sub T}$, then we say that $M$ is a
  \defn{$\mathbf\Delta$-machine over $\mathbf \Sigma^*$}.  We define
  the output of a \deltaMachine as $(w, \Delta)$, where $\Delta$ is
  the transformation encoded in $z$.
\end{definition}

\begin{definition}
  \label{def:trans}
  A \defn{transformative context-free grammar} (TCF grammar) is a
  tuple $(\Sigma,\Nu,P,S,T,M)$, where:
  \begin{itemize}
  \item $\Sigma$ is the terminal alphabet,

  \item $\Nu$ is the nonterminal alphabet,

  \item $P$ is the production set,

  \item $S$ is the start symbol,

  \item $T$ is a subset of $P$ whose elements are the
    transformative productions of this grammar, and

  \item $M$ is a \deltaMachine over $\Sigma^*$.
  \end{itemize}
\end{definition}

\begin{definition}
  Let $G=(\Sigma,\Nu,P,S,T,M)$ be a TCF grammar be given, and let
  $G_0=(\Sigma, \Nu, P, S)$ be the associated context-free grammar.
  Let $\alpha, \beta \in (\Sigma \cup \Nu)^*$ be such that
  \begin{equation}
    \label{eq:[non]trans derivation}
    \alpha \derives_{G_0}^* \beta.
  \end{equation}
  If no transformative productions are used in \eqref{eq:[non]trans
    derivation}, then we say that \eqref{eq:[non]trans derivation} is
  a \defn{nontransformative derivation}, and we write $\alpha
  \nontransDerives{G}^* \beta$.  If only transformative productions
  are used in \eqref{eq:[non]trans derivation}, then we say that
  \eqref{eq:[non]trans derivation} is a \defn{transformative
    derivation}, and we write $\alpha \transDerives{G}^* \beta$. We
  take $\derives_G$ to mean $\nontransDerives{G}$.
\end{definition}

\begin{definition}
  If $G=(\Sigma, \Nu, P, S, T, M)$ is a TCF grammar, and $(\Sigma,
  \Nu, P, S)$ is LR(1), then we say that $G$ is a \defn{transformative
    LR(1) grammar} (TLR grammar).
\end{definition}

\begin{definition}
  A \defn{grammar transformation} for a TLR grammar
  $G=(\Sigma,\Nu,P,S,T,M)$ is a tuple $(N,P_+,P_-)$, where:
  \begin{itemize}
  \item $N$ is a set of nonterminals to add to $\Nu$;

  \item $P_+$ is a set of productions to add to $P$,
    where $P\cap P_+=\emptyset$; and

  \item $P_-$ is a set of productions to remove from
    $P$, where $P_-\subset P$.
  \end{itemize}
\end{definition}

This immediately implies that $P_-$ and $P_+$ are disjoint. 

Given a grammar and a transformation, we define the symbol 
\[
  \Delta G \equiv (\Sigma,\Nu\cup N,(P \cup P_+) \setminus P_-,S,T,M).
\]
For a given TLR grammar $G$, let $\deltaSet{G} A$ be the set of all
grammar transformations $\Delta$ for $G$.

A couple of observations.  Note that $T$ is constant; because $T
\subset P$, we must have that $P_- \cap T = \emptyset$. We note also that
$\emptyDelta=(\emptyset, \emptyset, \emptyset)$ is an identity.

\subsection{TLR Languages}

\label{sec:tlr languages}

At what point should we allow the parser to stop and change the
grammar? There are two options: after a shift, or after a
reduction. We do not include the option of stopping before a shift
because it is not substantively different from stopping after a shift;
likewise, we do not include the option of stopping before a
reduction. We will adopt the later option: the parser will stop and
change the grammar after a reduction. 

As a basis for a method of parsing transformative languages, LR(1)
parsing would seem to be ideal: LR(1) languages parse from left to
right, as is required for transformative languages; LR(1) parsing
requires no backtracking; and LR(1) requires single-character
lookahead.

Creating an exact definition of transformative LR(1) languages
requires the creation of a fair amount of machinery. This will occupy
us for the rest of this section. To see why this is work is required,
we consider an example in the next section.

\subsubsection{A Naive Approach to Transformative Languages}
\label{sec:naive approach}
Let us attempt to define the transformative language generated by a
TLR grammar in the most obvious way, and see what goes wrong.

\begin{definition}
  If $\alpha\in(\Sigma\cup\Nu)^*$ is a viable prefix for $G$, and
  there is some sentential form $\alpha a x$, where $a\in\Sigma$ and
  $x\in\Sigma^*$, then we say that $\alpha$ is a \defn{viable prefix
    followed by $a$}.
\end{definition}

\begin{definition}
  \label{def:semiparse}
  Let $G=(\Sigma, \Nu, P, S, T, M)$ and $G'=(\Sigma, \Nu', P', S, T,
  M)$ be two TLR grammars. Let the alphabets for the three tapes of
  $M$ be $\Gamma_1$, $\Gamma_2$, and $\Gamma_3$, respectively. Let $g
  \in \Sigma \sub T ^*$ be an encoding of $G$ in $\lang{G \sub T}$.
  Let $\alpha, \beta \in (\Sigma \cup \Nu)^*$ and $x, z \in \Sigma^*$
  be given, and let $u, w \in \Gamma_2^*$ also be given. We say that
  $(\beta z, w, G')$ is a \defn{semiparse} for $(\alpha x, u, G)$, a
  relationship we denote with the symbol $(\alpha x, u, G) \semiParse
  (\beta z, w, G')$, in either of two cases. The first case is that we
  have all of the following:
  \begin{enumerate}
  \item $\beta z=S$,  
    
  \item $\beta z \nontransDerives{G}^* \alpha x$,
    
  \item $G' = G$.
  \end{enumerate}
  The second case is that we have all of the following:
  \begin{enumerate}
  \item $\beta z = \beta' B z \transDerives G \beta' \gamma z
    \nontransDerives G^* \alpha yz = \alpha x,$ where $y
    \in \Sigma^*$ and $\gamma \in (\Sigma \cup \Nu)^*$,
    
  \item $\beta$ is a viable prefix followed by $b$ for $G'$, where
    $b=\first(z)$,

  \item the output of $M$ with input $(y, u, g)$ is $(w, \Delta)$ such
    that $G' = \Delta G$.
  \end{enumerate}
\end{definition}

We could now try to define the language generated by a TLR
grammar. Surely, the language generated by a TLR grammar $G$ consists
of those strings $x$ such that we have:
\[
(x,\epsilon,G)
\semiParse(\alpha_1,u_1,G_1)
\semiParse(\alpha_2,u_2,G_2)
\semiParse\dotsb
\semiParse(S,u_n,G_n).
\]

As an example of the kind of problem this naive definition of
transformative languages poses can be illustrated by example. We let
$G$ be the TLR grammar with terminal alphabet $\{c,d\}$, nonterminal
alphabet $\{S,A,B,C,D\}$, production set
\[
\production{S}{A\alt B}
,\ \ 
\production{A}{C}
,\ \ 
\production{B}{D}
,\ \ 
\production{C}{c}
,\ \ 
\production{D}{d},
\]
start symbol $S$, transformative productions \production A C and
\production B D and \deltaMachine $M$. Consider the production sets
\begin{equation}
  \label{gram:second grammar, infinite parse example}
  \production{S}{A}
  ,\ \ 
  \production{A}{C}
  ,\ \ 
  \production{B}{D}
  ,\ \ 
  \production{C}{B}
  ,\ \ 
  \production{D}{d},
\end{equation}
and
\begin{equation}
  \label{gram:third grammar, infinite parse example}
  \production{S}{B}
  ,\ \ 
  \production{A}{C}
  ,\ \ 
  \production{B}{D}
  ,\ \ 
  \production{C}{c}
  ,\ \ 
  \production{D}{A}.
\end{equation}
Let $G_1$ and $G_2$ be the grammars that are identical to $G$, only
with the production sets in \eqref{gram:second grammar, infinite parse
  example} and \eqref{gram:third grammar, infinite parse example},
respectively. Let $\Delta_0$, $\Delta_1$, and $\Delta_2$ be grammar
transformations such that
\begin{align*}
  G_1 &= \Delta_0 G\\
  G_2 &= \Delta_1 G_1\\
  G_1 &= \Delta_2 G_2
\end{align*}
We now let the \deltaMachine $M$ be such that: when the instantaneous
grammar is $G$, the return value of $M$ is $\Delta_0$; when the
instantaneous grammar is $G_1$, the return value of $M$ is $\Delta_1$;
and finally, when the instantaneous grammar is $G_2$, the return value
of $M$ is $\Delta_2$.

We consider now how to parse the string ``$c$'' using our naive
method:
\[
(d,\epsilon,G)
\semiParse(B,u_1,G_1)
\semiParse(A,u_2,G_2)
\semiParse(B,u_3,G_1)
\semiParse(A,u_4,G_2)
\dotsb.
\]
Since the value of $u_i$ does not affect the \deltaMachine, for any
$i$, we can see that this sequence of semiparses has no end. This is
not a theoretical difficulty, for by \dref{def:trans}, we only require
that the sequence of semiparses terminates; thus, the string ``$c$''
is not generated by the grammar $G$.

Immediately before every transformation of grammar, we have a
sentential form which can be derived from the start symbol in the
nontransformative grammar associated with the instantaneous grammar in
a finite number of steps, say $n$ steps. The problem that arises in
this example is that, after the transformation of grammar, the
sentential form is now derivable from the start symbol in greater than
$n$ steps---in this case, we will say that the parse has been
\defn{extended}. Analyzing a given \deltaMachine to answer the
question of whether or not it will always produce a grammar
transformation that will extend the parse is not a task that we can
expect the parser to do; indeed, this question is undecidable.

What we can and will do is look for some property of $\Delta$ such
that, should we require that any transformation emitted by the
\deltaMachine must have this property, then as a result, the parse
will not be extended, hence it can be completed in a finite number of
steps.

\subsubsection{Allowable Transformations}
\label{sec:allowable transformations}

In this section, we will construct the test the parser can perform to
determine if it will accept or reject a transformation emitted by a
\deltaMachine. In the next section, we shall see if this test does
indeed perform as advertised.

The basic idea of the test is to examine the grammar, examine the
symbol stack at the time the transformation is to be applied, and
identify a certain set of productions called \defn{conserved
  productions}. The transformation will be considered acceptable if,
for each conserved production, there is a corresponding production in
the transformed grammar, such that these two productions share a head
and a certain prefix, which we refer to as the \defn{conserved
  portion} of that production. The function we define now determines
if a production is in this set, and if so, how long this prefix is.

For these definitions, we will take $G=(\Sigma,\Nu,P,S,T,M)$ to be
given.

\begin{definition}
  \label{def:conservation function for sentential form}
  Let the sentential form $\beta\in(\Sigma\cup\Nu)^*$ be given, such
  that $\beta\notin\Sigma^*$; that is, there is at least one
  nonterminal in $\beta$, which we call $B$. We thus write
  $\beta=\alpha B a x$, where $\alpha \in(\Sigma\cup\Nu)^*$,
  $a\in\Sigma$ and $x\in\Sigma^*$; we know that $a$ exists, for at the
  very least the end-of-file marker appears after the last appearance
  of $B$.  Let $y \in \Sigma^*$ be such that $\alpha B a
  x\derives^*y$; we consider the parse tree \node T for
  $y\in\Sigma^*$.  Let \node B and \node A be the nodes in \node T
  that  correspond to the symbols $B$ and $a$ in $\alpha B a
  x$. Let \node P be an interior node with $n$ children; we define
  $N_{\beta, \node T}(\node P)$ in one of \ref{item:last conservation    case} ways:
  \begin{enumerate}
  \item If \node P is an ancestor of \node B, but not an ancestor of
    \node A, then let $N_{\beta, \node T}(\node P) = n+1$.

  \item If \node P is an ancestor of \node A, then one of the children
    of \node P is autoancestral to \node A. Assuming the children of
    \node P are $\node{X}_1, \node{X}_2,\dotsc,\node{X}_n$, let
    $\node{X}_i$ be the node autoancestral to \node A; let
    $N_{\beta, \node T}(\node P)=i$.

  \item 
    \label{item:conservation:P shares an ancestor with both B and A,
      but is an ancestor of neither}
    If \node P shares an ancestor with both \node B and \node A, but
    is an ancestor of neither, and $\node B<\node P<\node A$, then let
    $N_{\beta, \node T}(\node P)=n+1$.

  \item 
    \label{item:last conservation case}
    If none of the above conditions holds, then let $N_{\beta, \node T}(\node P)
    = -1$.
  \end{enumerate}
  For every $\pi\in P$, we define
  \[
  V_\beta(\pi)=\begin{cases}
    \max\{N_{\beta, \node T}(\node P)\colon \nodeLabel(\node P)=\pi\} &
    \text{there is some \node P such
      that $\nodeLabel(\node P)=\pi$}\\
    -1 & \text{there is no \node P such that $\nodeLabel(\node
      P)=\pi$}
  \end{cases}.
  \]
  We call $V_\beta$ the \defn{conservation function for $\beta$}.
\end{definition}

\begin{prop}
  \label{prop:conservation function independent of sentential form} 
  Let the TLR grammar $G=(\Sigma, \Nu, P, S, T, M)$, the production
  $\pi \in P$, and the viable prefix $\beta$, followed by $a$, all be
  given. If $y, z \in \Sigma^*$ are such that both $\beta a y$ and
  $\beta a z$ are sentential forms for $G$, then $V_{\beta a
    y}(\pi)=V_{\beta a z}$.
\end{prop}

As a result of \pref{prop:conservation function independent of
  sentential form}, we can amend the definition of the conservation
function: if $\beta$ is a viable prefix followed by $a$, then the
value of $V_{\beta a}(\pi)$ is the value of $V_{\beta a x}(\pi)$ for
any terminal string $x$ such that $\beta a x$ is a sentential form.

\begin{definition}
  \label{def:conserved portion of a production}
  Let the sentential form $\beta\in(\Sigma\cup\Nu)^*$ be given.  If
  $V_\beta(\pi)=-1$, then we say that $\pi$ is a \defn{free production
    for $\beta$}.  If $V_\beta(\pi)=i$, for $\pi$ equal to $A\to \beta
  $, and $0 < i \le |\beta| $, then we define the first $i$ grammar
  symbols on the right side of $\pi$ to be the \defn{conserved
    production for $\beta$} of $\pi$. If $V_\beta(\pi)=i$, for $\pi$
  equal to $A\to \beta$, and $i = |\beta| + 1 $, then we define $\pi$
  to be an \defn{entirely conserved production for $\beta$}. Entirely
  conserved productions are also conserved.
\end{definition}

\begin{definition}
  \label{def:production is conserved by production}
  Let the transformative grammar $G=(\Sigma,\Nu,P,S,T,M)$, and the
  sentential form $\beta$ be given such that
  $\beta\in(\Sigma\cup\Nu)^*$, but $\beta\notin\Sigma^*$. Let $\phi$
  be a production, not necessarily in $P$, denoted $B\to Y_1 Y_2
  \dotso Y_m$. Let $\pi$ be a conserved production for $\beta$ in
  $P$. Let $\pi$ be a production in $P$, denoted $A\to X_1 X_2 \dotso
  X_n$. If $V_\beta(\pi)\le n$, for $1\le i\le V_\beta(\pi)$ we have
  that $X_i=Y_i$, and $A=B$, then we write $\pi\conservedBy{\beta}
  \phi$. If $V_\beta(\pi)=n+1$ and $\pi=\phi$, then we again write
  $\pi\conservedBy{\beta}\phi$. If $P'$ is a set of productions with
  terminal set $\Sigma$ and nonterminal set $\Nu' \supset \Nu$, such
  that there exists some $\phi \in P'$ for every conserved production
  $\pi \in P$ satisfying $\pi \conservedBy{\beta} \phi$, then we say
  that $P'$ \defn{conserves $\beta$ for $P$}.
\end{definition}

\begin{definition}
  \label{def:valid transformation}
  Let the transformative grammar $G=(\Sigma,\Nu,P,S,T,M)$ and the
  grammar transformation $\Delta\in\allDeltas G$ be given. Let $\Delta
  G=(\Sigma,\Nu_{\Delta G}, P_{\Delta G}, S, T,M)$. Let $\alpha A$, a
  viable prefix followed by $a$, be given where
  $\alpha\in(\Sigma\cup\Nu)^*$, $B\in\Nu$, and $a\in\Sigma$. If, for
  all $x\in\Sigma^*$ such that $\alpha A a x$ is a sentential form of
  $G$, we have that $P_{\Delta G}$ conserves $\alpha A a x$ for $P$,
  then we say that $\Delta$ is \defn{valid for $\alpha B a$ in
    $G$}. The set of all transformations which are valid for $\alpha B
  a$ we denote as $\valid{G}{\alpha B a}$. We include
  $\emptyDelta=(\emptyset,\emptyset,\emptyset)$ in $\valid{G}{\alpha B
    a}$. 
\end{definition}

To get a feel for how this test accomplishes our goal, we consider the
manner in which an LR(1) parser operates. The value on the top of the
state stack corresponds to an item set. The item set contains those
productions that the parser might be able to reduce. The parser's
initial item set is the one containing the item $[S'\to\cdot
S,\eofsym]$ and zero of more other items of the form
$[A\to\cdot\alpha,a]$; the parser is in the initial item set only when
it has scanned nothing. The parser can either shift or reduce. After
the parser shifts a terminal---for example, the terminal ``$a$''---the
parser goes into a new state, related to the old, as follows: for
every item of the form $[A\to\alpha\cdot a\beta,b]$ in the first
state, there is an item of the form $[A\to\alpha a\cdot\beta,b]$ in
the second state.  The parser does not stop considering an item just
because that item does not have $a$ to the right of the dot; for each
item $[A \to \gamma \cdot B \delta, c]$ that the parser is
considering, it looks for an item of the form
\begin{equation}
  \label{eq:item found by closure}
  [D \to \cdot \zeta, g];
\end{equation}
should the parser be able to find a sequence of items like
\eqref{eq:item found by closure}, the last of which has the dot to the
left of the terminal ``a'', then the whole sequence of productions
remains under consideration.

After the parser reduces a production, say $C \to \delta$, it pops the
appropriate number of states off of the stack, after which, ``$C$''
will be immediately to the right of the dot in one of the items in the
new current state; the parser will move to a new state, in which $C$
is to the left of the dot.

So we can summarize the operation of the parser as follows: the parser
considers several production in parallel; for each production, as it
shifts terminals, it either:
\begin{itemize}
\item advances along that production, 

\item it records its position in that production, shifting its
  attention to those productions with the appropriate nonterminal at
  their head, or

\item it drops that production.
\end{itemize}
This process continues until a reduction, at which time, one of the
productions currently under consideration is selected. The parser then
recalls its previous state. We want to view the parser stack as the
parser's method of keeping track of those productions it is
considering. In this light, requiring the transformation to be valid
for the current viable prefix means that each production under
consideration by the pre-transformation parser has a counterpart that
is under consideration by the post-transformation parser---and the
portion of the production that the parser has already matched remains
unchanged.

The current input symbol is what causes the parser to select the
production to use in a reduction; since a grammar transformation takes
place immediately after a reduction, the parser will have already
``seen'' the current input symbol. Say the transformative production
just reduced has the nonterminal ``$B$'' at its head: we know that the
current input symbol---say it's ``$a$''---is in the follow set of $B$
because one of the productions the parser had been considering before
it matched the transformative production---the production $\pi$, for
instance---derived some string such that the current input symbol $a$
follows the nonterminal $B$. Requiring the transformation to be valid
for the current viable prefix followed by the current input symbol
means that $\pi$ has a counterpart in the post-transformation grammar
that also derives some string such that the current input symbol $a$
follows the nonterminal $B$, and in this latter derivation is
different only inasmuch as it affects parts of the string strictly
after the ``$a$''.

In other words: any part of the grammar that the parser was using
right after the reduction by the transformative production must exist
in the new grammar unchanged. Productions not in use, and unused
suffixes of productions that were in use, can be modified freely,
provided the grammar remains LR(1).

\subsubsection{The Language Generated by a TLR Grammar}

Since we are interested in those languages for which a parser can be
constructed, we found it necessary to restrict those transformations
that will be acceptable for a \deltaMachine to emit. Now that we have
given a precise description of which transformations will be allowed,
we can define the concept of a transformative LR(1) language. We begin
with \dref{def:semiparse}: the definition of ``semiparse''. The
following definition is \dref{def:semiparse}, with the requirement
that the transformation be valid.

\begin{definition}
  \label{def:valid semiparse}
  Let $G=(\Sigma, \Nu, P, S, T, M)$ and $G'=(\Sigma, \Nu', P', S, T,
  M)$ be two TLR grammars. Let the alphabets for the three tapes of
  $M$ be $\Gamma_1$, $\Gamma_2$, and $\Gamma_3$, respectively.  Let $g
  \in \Sigma \sub T ^*$ be an encoding of $G$ in $\lang{G \sub T}$.
  Let $\alpha, \beta \in (\Sigma \cup \Nu)^*$ and $x, z \in \Sigma^*$
  be given, and let $u, w \in \Gamma_2^*$ also be given. We say that
  $(\beta z, w, G')$ is a \defn{valid semiparse} for $(\alpha x, u,
  G)$, a relationship we denote with the symbol $(\alpha x, u, G)
  \validSemiParse (\beta z, w, G')$, in either of two cases. The first
  case is that we have all of the following:
  \begin{enumerate}
  \item $\beta z=S$,  
    
  \item $\beta z \nontransDerives{G}^* \alpha x$,
    
  \item $G' = G$.
  \end{enumerate}
  The second case is that we have all of the following:
  \begin{enumerate}
  \item $\beta z = \beta' B z \transDerives G \beta' \gamma z
    \nontransDerives G^* \alpha yz = \alpha x,$ where $y
    \in \Sigma^*$ and $\gamma \in (\Sigma \cup \Nu)^*$;
    
  \item $\beta$ is a viable prefix followed by $b$ for $G'$, where
    $b=\first(z)$;

  \item the output of $M$ with input $(y, u, g)$ is $(w, \Delta)$ such
    that:
    \begin{enumerate}
    \item $G' = \Delta G$, and

    \item $\validDelta \Delta \beta G$.
    \end{enumerate}
  \end{enumerate}
\end{definition}

Earlier, we tried to define the language generated by a TLR grammar in
the obvious way using the ``semiparse'' relationship
(\secref{sec:naive approach}), and we ran into a computational
problem. When we define the language generated by a TLR grammar, we do
so in much the same way we did before, except that we require the
semiparses to be valid.

\begin{definition}
  \label{def:language generated by G}
  Let $G_0=(\Sigma,\Nu,P,S,T,M)$, a TLR grammar, and $x\in\Sigma^*$ be
  given. Let $w_0=\epsilon$ and $\alpha_0=x$. If there is some $k>0$
  such that
  \[
  (\alpha_0,w_0,G_0)\validSemiParse
  (\alpha_1,w_1,G_1)\validSemiParse
  (\alpha_2,w_2,G_2)\validSemiParse
  \dotsb\validSemiParse
  (\alpha_k,w_k,G_k)=(S, w_k, G_k),
  \]
  then $x\in\lang G$. The set of all strings $x$ is the \defn{language
    generated by $G$}.
\end{definition}

\begin{theorem}
  If $L \subset \Sigma^*$ is a recursively enumerable language, then
  there is a TLR grammar that generates $L$.
\end{theorem}
\begin{proof}
  We know that there is a Turing machine $T$ that recognizes $L$. We
  will construct the TLR grammar $G=(\Sigma,\Nu,P,S,T,M)$ that
  generates $L$. Let $\Sigma=\{a_1, a_2, \dotsc, a_n\}$, let $\Nu=\{S,
  A, B\}$, let
  \[
  P=\{   \production{S }{A}
  ,\ \ 
  \production{A }{B \alt AB}
  ,\ \ 
  \production{B }{a_1 \alt a_2 \alt \dotsb \alt a_n}  \},
  \]
  and let $T=\{S \to A\}$. We now describe the operation of $M$. The
  input on the first tape of $M$ will be some $x \in \Sigma^*$. If $x
  \in L$, then output $\emptyDelta$; if $x \notin L$, then output
  $\Delta=(\emptyset, \emptyset, P)$---that is, the transformation
  that removes all productions.

  Clearly, if $x \in L$, then $x \in \lang G$, but if $x \notin L$,
  then $x \notin \lang G$ because the transformation $\Delta$ is not
  valid.
\end{proof}

\subsection{A Fundamental Theorem of TLR Parsing}

We saw before that the basic problem with the naive approach to
defining the language generated by a TLR grammar is that the
derivation of the string on the symbol stack can be extended. In the
last section, we defined the language generated by a TLR grammar using
the valid semiparse relation. In this section, we will prove that
requiring valid transformations prevents the extension of the string
on the symbol stack. This is essential to proving that the TLR parsing
algorithm, the subject of the next section, is correct.

We begin with some technical Lemmas, leading up to the main Lemma of
this section: \lref{lem:valid deltas maintain viable prefixes}. The
title theorem is \tref{thm:valid transformations do not extend
  derivations}.

\begin{lemma}
  \label{lem:used to show that AB => Aax survives the transformation}
  Let $G$ be a TLR grammar and let $\alpha A$ be a viable prefix
  followed by $c$. If $\validDelta{\Delta}{G}{\alpha A c}$, and
  \[
  \beta B c\derives_G^* \alpha A c,
  \]
  then $\validDelta{\Delta}{G}{\beta B c}$.
\end{lemma}
\begin{proof}
  By contradiction. Let $\Delta G=(\Sigma,\Nu_{\Delta G}, P_{\Delta
    G}, S, T, M)$. Assume that $\notValidDelta{\Delta}{G}{\beta B
    c}$. Thus, there is some $x\in\Sigma^*$ such that $\beta B c x$ is
  a sentential form, and yet $\Delta$ is not valid for $\beta B c x$
  in $G$; there is some production $\pi$ that is conserved for $\beta
  Bcx$, where $\pi$ appears in the derivation
  \begin{equation}
    \label{eq:derivation of beta B c x}
    S\derives_G^* \beta B cx,
  \end{equation}
  such that for no $\phi\in P_{\Delta G}$ do we have $\pi
  \conservedBy{\beta B c x} \phi$.

  Isn't $\pi$ conserved for $\alpha A c x$? Let $y\in\Sigma^*$ such that
  \[
  \label{eq:derivation of alpha A c x, from beta B c x}
  S\derives_G^* \alpha A c x\derives_G^* y;
  \]
  use the parse tree $\node T_y$ for $y$. Let the nodes corresponding to $B$ and $c$
  in \eqref{eq:derivation of beta B c x} be \node B and \node C,
  respectively. Let \node U be the set of all nodes \node P be a node
  such that either:
  \begin{enumerate}
  \item \label{item:anc B, not C} \node P is an ancestor of \node B, but
    is not an ancestor of \node C;
    
  \item \node P is an ancestor of \node C; or
    
  \item \label{item:shared anc} \node P shares an ancestor with
    \node B and \node C, but is an ancestor of neither, such that
    $\node B < \node P < \node C$.
  \end{enumerate}
  These three possilities correspond to the first three possibilities in
  \dref{def:conservation function for sentential form}. Note that
  \begin{equation}
    \label{eq:derivation of beta gamma A c x}
    S\derives_G^* \beta B cx \derives_G^* \alpha A c x\derives_G^* y.
  \end{equation}
  Let \node A be the node corresponding to $A$ in the derivation
  \eqref{eq:derivation of beta gamma A c x}. Let \node W be the set of
  all nodes \node Q such that either:
  \begin{enumerate}
  \item \node Q is an ancestor of \node A, but is not an ancestor of
    \node C;
    
  \item \node Q is an ancestor of \node C; or
    
  \item \node Q shares an ancestor with \node A and \node C, but
    is an ancestor of neither, such that $\node B < \node Q < \node C$.
  \end{enumerate}

  Let $\node P \in \node U$. Either \node B is an ancestor of \node A,
  or it shares an ancestor, which we will call \node T, with both
  \node A and \node C. We go through the three possibilities for \node
  P.
  \begin{enumerate}

  \item \node P is an ancestor of \node B but not \node C. There are
    two ways that this can arise.
    \begin{enumerate}
    \item \node B is an ancestor of \node A. This means that \node P
      is an ancestor of \node A but not \node C, so $\node P \in \node
      W$. Thus, $\nodeLabel(\node P)$ is entirely conserved for
      $\alpha Acx$.
      
    \item \node B shares \node T as an ancestor with \node A and \node
      C, but is an ancestor of neither; thus, \node P shares \node T
      as an ancestor with \node A and \node C, but is an ancestor of
      neither, so $\node P\in\node W$, which means that $\nodeLabel
      (\node P)$ is an entirely conserved production for $\alpha Acx$.
    \end{enumerate}
    Either way, $N_{\alpha Acx, \node T_y}(\node P)=N_{\beta Bcx}(\node P)$.
    
  \item \node P is an ancestor of \node C, in which case $\node P \in
    \node W$. Thus, $N_{\alpha Acx, \node T_y}(\node P)=N_{\beta Bcx}(\node P)$.
    
  \item \node P shares an ancestor \node R with \node B and \node C,
    but is an ancestor of neither. There are two ways that this can
    arise.
    \begin{enumerate}
    \item \node B is an ancestor of \node A. This means that \node R
      is an ancestor of \node B and \node C, so $\node P\in\node
      W$. Therefore, $\nodeLabel(\node P)$ is an entirely conserved
      production.
      
    \item \node B shares \node T as an ancestor with \node A
      and \node C. This means that \node T is an ancestor of
      \node A, \node P, and \node C. Therefore, we have that
      $\nodeLabel(\node P)$ is an entirely conserved production
    \end{enumerate}
    Either way, $N_{\alpha Acx, \node T_y}(\node P)=N_{\beta B cx, \node T_y}(\node P)$.
  \end{enumerate}
  We have here established that $\node U\subset\node W$, and that, for
  all $\node P\in\node U$ such that $N_{\beta Bcx, \node T_y}(\node P)>0$, then
  we have that $N_{\alpha Acx, \node T_y}(\node P)=N_{\beta Bcx, \node T_y}(\node P)$. It is
  still possible that $N_{\alpha Acx, \node T_y} > 0$. Thus we have the following
  inequality:
  \begin{equation}
    \label{eq:N inequality}
    N_{\beta Bcx, \node T_y}(\node P) \le N_{\alpha A c x, \node T_y}(\node P).
  \end{equation}

  The fact that $\node P\in\node U$, together with \eqref{eq:N
    inequality}, lets us conclude that, for any production $\pi$ that is
  conserved for $\beta Bcx$, we have that
  \begin{equation}
    \label{eq:V inequality}
    V_{\beta Bcx}(\pi)\le V_{\alpha Acx}(\pi).
  \end{equation}
  Since we know that $\validDelta{\Delta}{G}{\alpha A c}$, there is
  some $\phi\in P_{\Delta G}$ such that $\pi\conservedBy{\alpha A c
    x}\phi$. By \eqref{eq:V inequality}, we conclude that
  $\pi\conservedBy{\beta B c x}\phi$.
\end{proof}

\begin{lemma}
  \label{lemma:alpha C derives alpha beta A x is conserved}
  If
  \[
  \alpha C \derives^*_G \alpha \beta A z \derives^*_G \alpha
  \beta\gamma B y z
  \]
  and $\validDelta{\Delta}{G}{\alpha \beta \gamma B y}$, then
  \[
  \alpha C \derives^*_{\Delta G} \alpha \beta A x .
  \]
  furthermore, $\first(y)=\first(w)$.
\end{lemma}
\begin{proof}
  Assume that $\alpha C \derives^*_G \alpha \beta A z \derives^*_G
  \alpha \beta\gamma B y z$.

  We proceed by induction on the number of steps in the derivation
  $\alpha C \derives^*_G \alpha \beta A z$.

  If $\alpha C \derives_G \alpha \beta A z$, then there is a production
  $\production{C}{\beta A z}$. Therefore, there is a production
  $\production{C}{\beta A \delta}$ in $\Delta G$; for any $x\in\Sigma^*$
  such that $\delta \derives^*_{\Delta G} x$, we find that $\alpha C
  \derives^*_{\Delta G} \alpha \beta A x $.

  Assume this result for all derivations less than $n$ steps long, for
  $n>1$. Assume that $\alpha C \derives_G^* \alpha \beta A z$ in $n$
  steps. We write
  \[
  \alpha C\derives_G \alpha \zeta D \eta \derives_G^* \alpha \zeta D v
  \derives_G^* \alpha \zeta \theta A u v = \alpha \beta A z.
  \]
  The derivation $D\derives_G^* \theta A u$ is fewer than $n$ steps
  long, hence the induction hypothesis implies that
  \[
  \alpha' D \derives_{\Delta G}^* \alpha' \theta A s,
  \]
  where $\alpha'=\alpha\zeta$.  Since $\validDelta{\Delta}{G}{\alpha
    \beta \gamma B y}$, there must be a production $\production{C}{\zeta
    D\kappa}$ in $\Delta G$; therefore, for any $t\in\Sigma^*$ such that
  $\kappa\derives_{\Delta G}^* t$, we have
  \[
  \alpha C
  \derives_{\Delta G}   \alpha  \zeta D \kappa 
  \derives_{\Delta G}^*  \alpha \zeta D t
  \derives_{\Delta G}^*  \alpha \zeta \theta A s t = \alpha \beta A x.
  \qedhere
  \]
\end{proof}

\begin{lemma}
  \label{lemma:gamma derives epsilon is maintained}
  If 
  \begin{equation}
    \label{eq:gamma derives epsilon in G}
    \alpha A \gamma a \derives_G^* \alpha A a,
  \end{equation}
  and $\validDelta{\Delta}{G}{\alpha Aa}$, then
  \[
  \alpha A \gamma a \derives_{\Delta G}^* \alpha A a.
  \]
\end{lemma}
\begin{proof}
  By induction on the number of steps in the first
  derivation\eqref{eq:gamma derives epsilon in G}. If $\alpha A \gamma
  x \derives_G \alpha A x$, then there must be a production
  $\production{B}{\epsilon}$ in $G$, where $\gamma=B$. Evidently, this
  production is conserved.

  Assume the result for derivations $n \ge 1$ steps long. If there are
  $n+1$ steps, then let $\gamma = \delta C$. We know that $\gamma$ is
  composed entirely of nonterminals, for any terminals ion $\gamma$
  would remain between $A$ and $x$ in the final string of the
  derivation. We have a production (possibly with $\epsilon$ on the
  right-hand side) $\production{C}{\zeta}$ in $G$, which is entirely
  conserved. Thus
  \[
  \alpha A \delta C x \derives_G \alpha A \delta \zeta x \derives_G^*
  \alpha A x.
  \]
  We can use the induction hypothesis to establish that 
  \[
  \alpha A \delta \zeta x \derives_{\Delta G}^* \alpha A x;
  \]
  since $\production{C}{\zeta}$ is also a production in $\Delta G$, we
  have our result.
\end{proof}

\begin{lemma}
  \label{lemma:alpha A derives beta B is maintained}
  If $G$ is an TLR grammar, and
  \begin{equation}
    \label{eq:lemma alpha A derives beta B}
    \alpha A a \derives_G^* \beta B a,
  \end{equation}
  and $\validDelta{\Delta}{G}{\beta B a}$, then
  \[
  \alpha A a \derives_{\Delta G}^* \beta B a.
  \]
\end{lemma}
\begin{proof}
  By induction. If $\alpha A a \derives_G \beta B a$, then there must be
  some production $\production{A}{\gamma B}$ in $G$ such that
  $\alpha\gamma=\beta$. In any case, this is a conserved production,
  so this production is also in $\Delta G$.

  Let us assume this Lemma for derivations of length $n \ge 1$ and
  that there are $n+1$ steps in \eqref{eq:lemma alpha A derives beta
    B}. We can rewrite derivation \eqref{eq:lemma alpha A derives beta
    B} as
  \[
  \alpha A a \derives_G
  \alpha \delta C \zeta a \derives_G^*
  \alpha \delta C a \derives_G^*
  \alpha \delta \theta B a = \beta B a.
  \]
  We can use \lref{lemma:gamma derives epsilon is maintained} to
  establish that
  \[
  \alpha \delta C \zeta a \derives_{\Delta G}^* \alpha \delta C a.
  \]
  By \caseref{item:conservation:P shares an ancestor with both B and
    A, but is an ancestor of neither} of \dref{def:conservation
    function for sentential form}, we see that the production
  $\production{A}{\delta C \zeta}$ is entirely conserved; thus,
  $\alpha A a \derives_{\Delta G} \alpha \delta C \zeta a$. By the
  induction hypothesis, we have that
  \[
  \alpha \delta C a \derives_{\Delta G}^* \beta B a;
  \]
  therefore, 
  \[
  \alpha A a \derives_{\Delta G}^* \beta B a. \qedhere
  \]
\end{proof}

\begin{lemma}
  \label{lemma:alpha AB derives alpha A ax is maintained}
  Let $G$ be a TLR grammar.  If
  \begin{equation}
    \label{eq:AB derives Aax in G}
    \alpha AB\derives_G^*\alpha Aax
  \end{equation}
  then
  \[
  \alpha AB\derives_{\Delta G}^*\alpha Aay,
  \]
  for some $y \in \Sigma^*$, provided that
  $\validDelta{\Delta}{G}{\alpha A a}$.
\end{lemma}
\begin{proof}
  By induction on the number of steps in \eqref{eq:AB derives Aax in
    G}. If $\alpha AB\derives_G \alpha Aax$, then there is a
  production $\production{B}{ax}$ in $G$; the conserved portion of
  this production includes at least the $a$, therefore, there is a
  production $\production{B}{a\beta}$ in $\Delta G$. For any $z \in
  \Sigma^*$ such that $\beta \derives_{\Delta G}^* z$, we thus have
  \[
  \alpha AB\derives_{\Delta G} \alpha A a \beta \derives_{\Delta
    G}^*\alpha Aaz.
  \]

  Assume the result for all derivations of length not greater than
  $n$, for some $n \ge 1$. If there are $n+1$ steps, then let the
  first production used be $\production{B}{\beta}$.

  If the $a$ appears on the right side of this production---that is
  $\beta=\gamma a \delta$---then we have
  \[
  \alpha AB\derives_G \alpha A \gamma a \delta \derives_G^* \alpha
  A\gamma a x \derives_G^* \alpha A a x;
  \]
  we can use the preceding Lemma to establish that 
  \begin{equation}
    \label{eq:gamma disappears because of the previous lemma}
    \alpha A\gamma a x
    \derives_{\Delta G}^* \alpha A a x.
  \end{equation}
  Since the conserved portion of $\production{B}{\beta\gamma a
    \delta}$ is at least $\gamma a$, we must have a production
  $\production{B}{\gamma a \zeta}$ in ${\Delta G}$. Thus, for any $w$
  such that $\zeta\derives_{\Delta G}^* w$, we have
  \[
  \alpha AB\derives_{\Delta G} \alpha A \gamma a \zeta \derives_{\Delta
    G}^* \alpha A \gamma a w \derives_{\Delta G}^* \alpha A a w,
  \]
  in light of \eqref{eq:gamma disappears because of the previous
    lemma}.

  If, however, $a$ does not appear on the right side of the production
  $\production{B}{\beta}$, then there must be some nonterminal $C$ in
  the conserved portion of the right-hand side of this production
  deriving the $a$. That is, there is a production
  $\production{B}{\eta C \theta}$ in $G$, such that $\eta C \derives
  _G^* at$ for some $t \in \Sigma^*$.  If $\eta=\epsilon$, then note
  that
  \[
  \alpha AB\derives_G \alpha AC \theta \derives_G^* \alpha AC s
  \derives_G^* \alpha Aats=\alpha Aax;
  \]
  we can apply the induction hypothesis to the derivation $\alpha
  AC\derives_G^* \alpha Aat$; in this case we have
  \[
  \alpha AB\derives_{\Delta G} \alpha AC \theta \derives_{\Delta G}^* \alpha AC s' \derives_{\Delta G}^* \alpha Aat's',
  \]
  for appropriate $t', s' \in \Sigma^*$. Now assume that
  $\eta\ne\epsilon$.  Note that
  \[
  \alpha AB \derives_G \alpha A \eta C \theta \derives_G^* \alpha A
  \eta C u \derives_G^* \alpha A \eta \lambda a v u \derives_G^*
  \alpha A \eta a v u \derives_G^* \alpha A a v u=\alpha A a x,
  \]
  for appropriate $u, v \in \Sigma^*$.  As $\eta $ is a nonterminal
  string, let us write $\eta=\eta' E$. Note that
  \[
  \alpha A \eta' E C u \derives_G^* \alpha A \eta' E a v u;
  \]
  in other words,
  \begin{equation}
    \label{eq:E derivation}
    \alpha ' EC \derives_G^* \alpha' Eav,
  \end{equation}
  where $\alpha' = \alpha A \eta'$.  By \lref{lem:used to show that AB
    => Aax survives the transformation},
  $\validDelta{\Delta}{G}{\alpha' E a}$, thus, for every production
  $\pi$ in derivation \eqref{eq:E derivation}, there is some
  production $\phi$ such that $\pi\conservedBy{\alpha' E
    a}\phi$. Also, derivation \eqref{eq:E derivation} is not more than
  $n$ steps long. Therefore, we may use the induction hypothesis to
  yield the derivation
  \[
  \alpha ' EC\derives_{\Delta G}^* \alpha ' Eav'.
  \]
  From the previous Lemma, we get
  \[
  AB\derives_{\Delta G}^*A\eta'ECu'\derives_{\Delta G}^*A\eta'Eav'u'\derives_{\Delta G}^* A av'u'. \qedhere
  \]
\end{proof}

\begin{definition}
  If $G=(\Sigma, \Nu, P, S)$ is an LR(1) grammar and $\beta$ is a
  viable prefix followed by $a$, then we say that the parser for $G$
  will \defn{shift $a$ after $n$ reductions of $\beta$} if there is a
  sequence of $\beta_1,\beta_2,\dotsc,\beta_n\in(\Sigma\cup\Nu)^*$
  such that
  \[
  \beta_n\derives \dotso \beta_2 \derives \beta_1 \derives \beta_0 =
  \beta,
  \]
  and $\beta_n a$ is a viable prefix, but for no $0 \le k < n$ is it
  the case that $\beta_k a$ is a viable prefix.
\end{definition}

Sometimes it will be useful to use the ``converse'' of the $\derives$
symbol.

\begin{definition}
  If $G=(\Sigma,\Nu,P,S)$ is a context-free grammar, and for some
  $\alpha,\beta\in(\Sigma\cup\Nu)^*$, we have that
  $\alpha\derives\beta$, we say that \defn{$\beta$ reduces to
    $\alpha$}, and we write $\beta\reducesTo\alpha$. Similarly, if
  $\alpha\derives^*\beta$, then we write $\beta \reducesTo^* \alpha$.
\end{definition}

If $G$ is a transformative grammar, then we give the symbols
$\transReducesTo G$ and $\nontransReducesTo G$ the obvious meanings as
the converses of the symbols $\transDerives G$ and $\nontransDerives
G$, respectively.

\begin{lemma}
  \label{lem:valid deltas maintain viable prefixes}
  Let $G$ be a transformative grammar and let $\alpha A$ be a viable
  prefix followed by $a$. If $\validDelta{\Delta}{G}{\alpha A a}$, then
  $\alpha A$ is a viable prefix followed by $a$ in $\Delta G$, assuming
  $\Delta G$ is TLR.
\end{lemma}
\begin{proof}
  There exists some $x\in\Sigma^*$ such that $\alpha Aax$ is a
  sentential form in $G$. We will show that there is some
  $w\in\Sigma^*$ such that $\alpha Aaw$ is a sentential form for
  $\Delta G$. Let $t \in \Sigma^*$ be such that $\alpha A a y
  \derives_G^* t$, and consider the parse tree for $t$; there must be
  one node \node P that is an ancestor of both the node representing
  $A$ and the node representing $a$, such that the child of \node P
  that is autoancestral to the node representing $A$ is different from
  the child of \node P that is autoancestral to the node representing
  $a$; let these two children of \node P be labeled \node C and \node
  X, respectively. Let the production corresponding to \node P be
  $P\to\beta C \gamma X \delta$. Note that
  \begin{align}
    X & \derives^*_G ay \text{;} \label{eq:derivation of ay}\\
    \gamma &\derives^*_G \epsilon \text{; and} \label{eq:gamma to
      epsilon}\\
    C &\derives^*_G \eta A\text . \label{eq:derivation of eta A}
  \end{align}
  We use \node X to remind us of the possibility that $X=a$; that is, we
  could have $X$ be either a terminal or a nonterminal.

  The derivations in \eqref{eq:derivation of ay}, \eqref{eq:gamma to
    epsilon}, and \eqref{eq:derivation of eta A} appear in the
  derivation for $\alpha Aax$ in sequence. Immediately before we begin
  deriving according to \eqref{eq:derivation of ay}, there is a
  sentential form $\theta P z$. We therefore have that
  $\theta\beta\eta=\alpha$ and $yz=x$.

  In light of \eqref{eq:derivation of eta A}, we have that $C
  \derives_{\Delta G}^* \eta A$ by \lref{lemma:alpha A derives beta B
    is maintained}.

  In light of \eqref{eq:gamma to epsilon}, we have that $\gamma
  \derives_{\Delta G}^*$ by \lref{lemma:gamma derives epsilon is
    maintained}.

  From the assumption that $\validDelta{\Delta}{G}{\alpha Aa}$, we can
  conclude that there is a production $P\to\beta C \gamma X \zeta$ in
  $\Delta G$.

  We have two more things to establish: that $X\derives^*_{\Delta G}au$
  for some $u\in\Sigma^*$; and that there is a sentential form $\theta P
  v$ for some $v\in\Sigma^*$.

  Since
  \[
  S \derives_G^* \theta P z \derives_G^* \theta \beta \eta A ax,
  \]
  we have, by \lref{lemma:alpha C derives alpha beta A x is
    conserved}, that
  \[
  S \derives_{\Delta G}^* \theta P v.
  \]

  Therefore, we turn to \eqref{eq:derivation of ay}. If $X=a$, then we
  are done. So assume that $X$ is a nonterminal; let $X=D$. We know
  that $C\gamma$ ends is a nonterminal string, so let $C\gamma=\lambda
  E$. By \lref{lem:used to show that AB => Aax survives the
    transformation}, and \lref{lemma:alpha AB derives alpha A ax is
    maintained}, we see that
  $\validDelta{\Delta}{G}{\theta\beta\lambda Ea}$. As
  \[
  \theta\beta\lambda ED \derives_G^* \theta\beta\lambda E a y,
  \]
  we therefore have
  \[
  \theta\beta\lambda ED \derives_{\Delta G}^* \theta\beta\lambda E aw.
  \qedhere
  \]
\end{proof}

\begin{theorem}
  \label{thm:valid transformations do not extend derivations}
  Let $G$ be TLR, let $B \to \alpha$ be a production in $G$, and let
  $\validDelta{\Delta}{G}{\gamma Ba}$. If $\gamma \alpha$ is a viable
  prefix followed by $a$ in $G$, such that $a$ will be shifted after
  $n$ reductions of $\gamma\alpha$ in $G$, then $a$ will be shifted
  after $n-1$ reductions of $\gamma B$ in $\Delta G$.
\end{theorem}
\begin{proof}
  By induction on $n$. If $n=1$, then the only reduction possible is
  $\gamma\alpha \reducesTo_{G} \gamma B$. Since $a$ can be shifted, we
  know that $\gamma B a$ is a viable prefix for $G$; by \lref{lem:valid
    deltas maintain viable prefixes}, we see that $\gamma B a$ is a
  viable prefix for $\Delta G$.

  Assume the conclusion for some $k \ge 1$, and assume that $k=n +
  1$. Now,
  \[
  \gamma\alpha \reducesTo_{G} \gamma B \reducesTo_{G}^* \delta.
  \]
  There are $n$ steps in the reduction $\gamma B \reducesTo_{G}^*
  \delta$; we can thus write
  \[
  \gamma B =\zeta \eta B\reducesTo_{G} \zeta C \reducesTo_{G}^*
  \delta.
  \]
  Note that
  \[
  \zeta C a \derives_{G}^* \gamma B a,
  \]
  therefore, by \lref{lem:used to show that AB => Aax survives the
    transformation}, we have $\validDelta{\Delta}{G}{\zeta C a}$. By the
  induction hypothesis, we see that the $a$ can be shifted after $n-1$
  reductions of $\zeta\eta B$ in $\Delta G$.
\end{proof}

%%% Local Variables: 
%%% mode: latex
%%% TeX-master: "tlr-plain"
%%% End: 

\section{TLR Algorithms}
\label{sec:tlr algorithms}

\begin{algorithm}
  {TLR Parsing Algorithm\label{algo:TLR}}
  {Parse the string $x$ using the LR parsing algorithm with the parsing
    table for the CFG grammar associated with $G$ until such a time as a
    transformative production from $G$ is reduced; at this time, apply a
    transformation to the grammar, recalculate the parsing tables, and
    then continue parsing with the new grammar.}
  {$G$: a TLR grammar, where $G=(\Sigma,\Nu,P,S,T,M)$; $x$: a string
    over $\Nu^*$}
  {$e$: a boolean which is true only if $x\in\lang G$}
  \begin{enumerate}
  \item Let $w_\Delta=\epsilon$ and $z_\Delta=\epsilon$.

  \item Calculate the parse table for $G$.
    
  \item 
    \label{step:TLR:initialize the stack}
    Push $(0, \epsilon)$ onto the stack.
    
  \item Set $a$ to the first terminal of $x$.
    
  \item \label{item:get state} Set $s$ to be the state on the
    top of the state stack.
    
  \item If $\action[s,a]=\shift$, and $a=\eofsym$, then return
    true.
    
  \item Otherwise, if $\action[s,a]=\shift$, then do the
    following:
    \begin{enumerate}
    \item Push $(\goto[s,a],a)$ onto the symbol stack.
      
    \item Set $w_\Delta=w_\Delta a$.

    \item Set $a$ to the next input symbol.

    \item Goto~\ref{item:get state}.
    \end{enumerate}
    
  \item Otherwise, if $\action[s,a]=\reduce \pi$, then do the
    following (letting $\pi=A\to\beta$):

    \begin{enumerate}
    \item Pop $|\beta|$ items off the stack.
      
    \item \label{item:call gta} If $\pi\in T$, then execute the
      Grammar Transformation Algorithm (\algoref{algo:gta}); set $G$
      and the stack to the returned values.

    \item Set $w_\Delta=\epsilon$.
      
    \item Set $s'$ to be state on the top of the state stack.
      
    \item Push $(\goto[s', A], A)$ onto the symbol stack.

    \item Goto~\ref{item:get state}.
    \end{enumerate}
    
  \item \label{item:syntax error} Otherwise, if $\action[s,a]=\error$,
    then return false.  \algofin
  \end{enumerate}
\end{algorithm}
This algorithm closely follows the presentation of the LR parsing
algorithm found Section~4.7 of \cite{dragon}. Indeed, the only
essential difference is in \sref{item:call gta}. We turn now to the
previously referenced Grammar Transformation Algorithm.

\begin{algorithm}
  {Grammar Transformation Algorithm\label{algo:gta}}
  {Compute the new grammar and its parsing tables. Assuming the
    transformation valid, put the parser into the correct state to
    continue parsing.}
  {$G$: a TLR grammar, where $G=(\Sigma,\Nu,P,S,T,M)$; $\sigma$: a
    parsing stack; $w_\Delta$: a string in $\Sigma^*$; $z_\Delta$: a
    string in $\Gamma$}
  {$G$: a TLR Grammar; $\sigma$: a Parser Stack; $z_\Delta$: a string
    in $\Gamma$}

  \begin{enumerate}
  \item \label{item:gta compute delta} Execute the \deltaMachine with
    $(w_\Delta, z_\Delta, G)$ as input, and $(w_\Delta, z_\Delta,
    \Delta)$ as output. 
    
  \item \label{step:assert valid Delta} Assert that $\validDelta\Delta
    G \alpha$.
    
  \item \label{step:apply g} Set $G=\Delta G$.
    
  \item \label{item:prepop} Calculate the parsing tables
    for $G$.
    
  \item \label{item:gta popall} Pop $|\sigma|$ states off of the
    stack.

  \item Do the following until the stack is empty:
    \begin{enumerate}
    \item Let the top item on the stack be $(s, X)$.

    \item Set $\alpha=X\alpha$.

    \item Pop the top item off of the stack.
    \end{enumerate}
        
  \item\label{item:gta init stack} Push $(0, \epsilon)$ onto the
    stack.

  \item \label{item:gta loop over symbols} Do the
    following, for $i$ from $1$ to $|\alpha|$:
    \begin{enumerate}
    \item Let $X$ be the $\nth{i}$ symbol of $\alpha$.
      
    \item Let $s$ be the state on the top of the
      stack.
      
    \item \label{item:push goto} Push $(\goto[s,a], a)$
      onto the  stack.
    \end{enumerate}
    
  \item\label{item:gta return}
    Return the new values for $G$, $\sigma$ and $z_\Delta$.
    \algofin
  \end{enumerate}
\end{algorithm}

If $\goto[s,a]$ in \sref{item:push goto} were ever undefined, then the
Grammar Transformation Algorithm fails, which will cause the TLR
Parsing Algorithm to fail as well. However, in light of
\lref{lem:valid deltas maintain viable prefixes}, we can be sure that
the Grammar Transformation Algorithm will fail at \sref{step:assert
  valid Delta} first. It is straightforward to give a useful (to a
human) error message in this case.

\subsection{Efficiency of The TLR Parsing Algorithm}

The efficiency of the Algorithm in the absence of grammar
transformations is essentially that of LR parsing. The computation of
LR parsing tables is expensive, but since the tables being generated
are not wholly independent of the tables that were used up to the
point of transformation, an incremental approach is available to
us. That is, we need calculate only the portion of the table that has
changed.

This idea---incrementally generating parsing tables---was first
introduced in the context of an interactive parser generator: the
language designer would enter in productions, or modifications to
productions, one at a time. After each production was entered, the
parser generator would recalculate the affected portion of the parsing
tables. As such a system was meant to be interactive, a high premium
was placed on response time---hence the development of more efficient
algorithms for computing parsing tables.

The two options---a full generation or an incremental generation of
parsing tables during a grammar switch---are identical for the
consideration of the worst-case performance of a grammar switch
operation, because the addition of a single production can cause an
exponential increase in the number of parsing states
\cite{horspool:incremental-lr-parsers}.

The TLR parsing algorithm generates canonical LR(1) parsing tables,
which are more general, but also far larger, than the more common
LALR(1) parsing tables. It is widely quoted (see \cite{dragon}) that
LR(1) parsing tables are much larger that LALR(1) parsing
tables. However, LR(1) parsing tables are easy to analyze compared
with LALR(1) parsing tables---hence the trade-off. It would be
interesting to see how the ideas, algorithms and analysis presented in
this work could apply to LALR(1) parsing. 

\subsection{Correctness}

\label{sec:correctness}

\begin{definition}
  Let $G=(\Sigma, \Nu, P, S)$ be an LR(1) grammar, and let $X_i
  \in (\Sigma \cup \Nu)$ for $1 \le i \le n$. We define $g \colon
  (\Sigma \cup \Nu)^* \to \mathbb Z_k$ as follows:
  \[
  g(X_1 X_2 \dotso X_n)=
  \begin{cases}
    0 & n=0 \\
    \goto[g(X_1 X_2 \dotso X_{n-1}), X_n] & n > 0
  \end{cases}.
  \]  
\end{definition}

\begin{theorem}
  \label{thm:gta works on viable prefixes}
  If $\alpha$ is a viable prefix for $G$, then $g(\alpha)$ is
  defined. Furthermore, the items within $I_{g(\alpha)}$ are valid for
  $\alpha$.
\end{theorem}
\begin{proof}
  Let the viable prefix $\alpha$ be given.

  If $|\alpha|=0$, then $g(\alpha)=0$. Since the item set containing
  $[S'\to \cdot S, \eofsym]$ is always $I_0$ by our convention
  (established on page~\pageref{initial state is zero}) and since
  $I_0$ is closed, the second conclusion is true in this case.

  Assume now that $g$ is defined for all viable prefixes of length not
  more than $n$, for some $n\ge 0$, and assume that $|\alpha| = n +
  1$; thus, we write $\alpha=\alpha' X$. Let $x$ be a terminal string such
  that $\alpha' X x$ is a sentential form. Consider the derivation of
  $\alpha' X x$:
  \begin{equation}
    \label{eq:derivation of alpha X x}
    S' \derives^* \beta A_1 z \derives \beta \gamma X \delta_0 z 
    \derives^* \beta \gamma X y z=\alpha' X x.
  \end{equation}

  If $\gamma\ne\epsilon$, then it must be true that
  \[
  [ B\to \gamma \cdot X \delta_0, u ] \in I_{g(\alpha')}
  \]
  by the assumption that $I_{g(\alpha')}$ contains all valid items for
  the viable prefix $\alpha'$.

  What if $\gamma=\epsilon$? Clearly, $\beta = \alpha'$. There is a
  sequence of $M$ steps in \eqref{eq:derivation of alpha X x} that are
  of the form
  \begin{equation}
    \label{eq:derivation within the same item set}
    \alpha' A_{m+1} w_{m+1} v \derives \alpha' A_m \delta_m w_{m + 1} v
  \end{equation}
  when going from $S'$ to $\alpha' X y z$, where we have $A_0 = X$ and
  $w_0 = y$. Between each step of the form \eqref{eq:derivation within
    the same item set}, there is a derivation
  \[
  \alpha' A_m \delta_m w_{m + 1} v \derives^* \alpha' A_m w_m v,
  \]
  for an appropriate $w_m \in \Sigma^*$.  If we consider the steps
  prior to the appearance of $\alpha' A_M w_M z$, we see that
  \[
  S' \derives^* \zeta C v \derives \zeta \eta A_M \delta_M v
  \]
  where $\zeta \eta = \alpha'$.
  Since $M$ is maximal, we see that $\eta\ne\epsilon$. There is thus an
  item
  \[
  [C\to\eta \cdot A_M \delta_M, u]\in I_{g(\alpha')}.
  \]

  Going through our sequence of productions $A_{m+1}\to A_m \delta_m$ in
  descending order, we see that
  \[
  [A_M \to \cdot A_{M-1} \delta_{M-1}, u_{M-1}] \in I_{g(\alpha')}\text{;}
  \]
  in general
  \[
  [ A_{m+1} \to \cdot A_m \delta_m, u_m] \in I_{g(\alpha')}
  \]
  for all $0 \le m < M$ because $I_{g(\alpha')}$ is closed.

  Consider what we have established: There is an item of the form
  \[
  [Y\to \gamma' \cdot X \delta', u']\in I_{g(\alpha')}.
  \]
  Thus, since 
  \[
  g(\alpha)=g(\alpha' X)\equiv \goto[g(\alpha'),X] \text{,}
  \]
  the first conclusion of this Theorem is established.

  For the second conclusion of this Theorem, we can use Theorem 5.10 of
  \cite{aho-ullman:top}, which justifies the construction of the item
  sets, and in particular, the item set $I_{g(\alpha' X)}$.
\end{proof}

It is clear that, in \sref{item:gta loop over symbols} of
\algoref{algo:gta} calculates $g$ in a bottom-up fashion; it will
succeed when that function is defined. Therefore, \tref{thm:gta works
  on viable prefixes} gives sufficient condition for the success of
that Algorithm.

\begin{definition}
  Let $G=(\Sigma, \Nu, P, S)$ be an LR(1) grammar. Let $X_i$ be a
  grammar symbol for $1 \le i \le n$ such that $X_1 X_2 \dotso X_n$ is
  a viable prefix for $G$. Define $\mathscr P \colon (\Sigma \cup
  \Nu)^* \to ( \mathbb Z_k^* \times ( \Sigma \cup \Nu )) ^*$ as
  \[
  X_1 X_2 \dotso X_n \mapsto
  ( (0, \epsilon),
    (g(X_1), X_1), 
    (g(X_1 X_2), X_2), 
    \dotsc,
    (g(X_1 X_2 \dotso X_n), X_n).
  \]
\end{definition}

\begin{definition}
  Let $G=(\Sigma, \Nu, P, S)$ be an LR(1) grammar. If $\alpha = X_1
  X_2 \dotso X_n$ is a sentential form for $G$, then let $1 \le i \le
  n$ such that $X_i \in \Nu$, and $X_{i + 1} X_{i + 2} \dotso X_n \in
  \Sigma^*$. Let $\beta \equiv X_1 X_2 \dotso X_i$ and let $x = X_{i +
    1} X_{i + 2} \dotso X_n$. Define the \defn{\Bfactorization} of
  $\alpha$ as $\beta$ and $x$.
\end{definition}

\begin{prop}
  If $G$ is an LR(1) grammar, and $\alpha$ is a sentential form for
  $G$, then the \Bfactorization of $\alpha$ is unique.
\end{prop}

\begin{algo}\rm
  \label{algo:modified TLR}
  We modify \algoref{algo:TLR} to include a viable prefix as an input
  parameter; this viable prefix will be used to initialize the
  stack. We do this by letting $\alpha$ be the new viable prefix
  paramater, and we replace \sref{step:TLR:initialize the stack} with
   \begin{enumerate}
   \item[\preItem{}3$'$\postItem] Set the stack to $\mathscr P(\alpha)$.
   \end{enumerate}
\end{algo}

\begin{lemma}
  \label{lem:modified TLR accepts}
  Let $G_0=(\Sigma, \Nu, P, S, T, M)$. Let $x \in \lang G$. Let the
  semiparse sequence for $x$ be
  \[
  (x, u_0, G) \validSemiParse 
  (\alpha_1, u_1, G_1) \validSemiParse 
  \dotsb
  (\alpha_n, u_n, G_n).
  \]
  For $1 \le i \le n$, if $\alpha_n$ is $B$-factored into $\beta$ and
  $x$, then \algoref{algo:modified TLR} will accept, given $\beta$,
  $x$, and $G_i$ as input.
\end{lemma}
\begin{proof}
  If $n - i = 0$, then $\beta=S$ and $x=\eofsym$. By \lref{thm:gta
    works on viable prefixes}, the prefix parse stack contains all
  valid items for the form $S \eofsym$, which is to say that the item
  set contains $[S' \to S \cdot, \eofsym]$. Reducing this is an
  accepting action.

  If $n - i > 0$, then we proceed by induction on $n - i$. If we have
  that $n - i = 1$, then we let $y \in \Sigma^*$ such that $\beta x
  \derives_{G_i}^* y$. By Theorem~5.12 of \cite{aho-ullman:top}, an
  LR(1) parser will accept $y$, and at some point during the parsing,
  the parser will have $\beta$ on its stack, and $x$ will be its
  unshifted input. Since 
  \[
  (\alpha_i, u_i, G_i) \validSemiParse (S, u_n, G_n), 
  \]
  \algoref{algo:modified TLR} will not apply any grammar
  transformations; instead, it will execute the same series of actions
  that an LR(1) parser would once it reaches the aforementioned
  configuration. Hence, \algoref{algo:modified TLR} will accept.
  
  Assume the result when the input appears as the
  $\nth{j}$\dash{}to\dash{}last form in the parse sequence, for some
  $j > 0$. Assume that $n - i = j + 1$. By \dref{def:valid semiparse},
  we know that there is some $\alpha'$ such that 
  \[
  \alpha_i \reducesTo_{G_i}^* \alpha' \transReducesTo{G_i} \alpha_{i + 1}.
  \]
  By Theorem~5.12 of \cite{aho-ullman:top}, the parser will correctly
  trace $\alpha_i \reducesTo_{G_i}^* \alpha'$, at which point the
  parser will reduce by a transformative production. This will leave
  the stack string as the viable prefix $\gamma$ followed by $a$;
  since the transformation which brings $G_i$ to $G_{i + 1}$ is valid
  for $\gamma$, we have by \lref{lem:valid deltas maintain viable prefixes}
  that $\gamma$ is a viable prefix for $G_{i + 1}$. Thus, we can use
  the induction hypothesis to claim that the parser will accept
  $\gamma$.
\end{proof}

\begin{lemma}
  \label{lemma:lookahead shifted after same number of reductions in LR
    and TLR parsers}
  Let $G=(\Sigma, \Nu, P, S, T, M)$ be a TLR grammar. If $\alpha$ is a
  viable prefix followed by $a$, and an LR(1) parser for the LR(1)
  grammar $(\Sigma, \Nu, P, S)$ would shift $a$ after $n$ reductions
  if $\alpha$ is on the parsing stack as $a$ is the lookahead, then an
  TLR parser will shift $a$ after $n$ reductions if $\alpha$ is on the
  parsing stack as $a$ is the lookahead
\end{lemma}
\begin{proof}
  By induction on $n$. If $n=0$, then both parsers will immediately
  shift $a$.

  Assume the result for some $k \ge 0$, and assume that $n=k+1$.
  Since both of the parsers will initially have identical stacks, they
  will reduce by the same production. If it happens that this
  production is not transformative, then the parsers will have
  identical stacks after the first reduction, after which we can apply
  the induction hypothesis. If it happens that this production is
  transformative, then, letting the new grammar by $\Delta G$, we can
  apply the induction hypothesis by virtue of \tref{thm:valid
    transformations do not extend derivations}. 
\end{proof}

\begin{theorem}
  \label{thm:correctness of the TLR parsing algo}
  Let $G=(\Sigma, \Nu, P, S, T, M)$ be a TLR grammar. The TLR Parsing
  Algorithm (\ref{algo:TLR}) recognizes $\lang G$.
\end{theorem}
\begin{proof}
  Let $x \in \Sigma^*$. 

  If $x \in \lang G$, then, because \algoref{algo:modified TLR}
  operates as \algoref{algo:TLR} does when $\alpha=\epsilon$, we know
  that the parser will accept $x$, given \lref{lem:modified TLR
    accepts}. 

  Assume, then, that $x \notin \lang G$. There are several ways in
  which this could happen. First, the \deltaMachine could emit a
  transformation that is not valid; if this happens, then the parser
  will clearly reject $x$. Second, after a shift or a reduction by a
  production that is not transformative, it could be that the stack
  string is not a viable prefix, or the lookahead might not follow the
  stack string; in either case, by Theorem~5.12 of
  \cite{aho-ullman:top}, the parsing tables will call for an error
  action, and so the parser will reject the string.

  The only other possibility is that there is a sequence of tuples
  \[
  (x, u_0, G) = 
  (\alpha_0, u_0, G_0) \validSemiParse
  (\alpha_1, u_1, G_1) \validSemiParse
  (\alpha_2, u_2, G_2) \validSemiParse
  \dotsb
  \]
  with no upper bound on the length of this sequence.  We shall
  dispose of this possibility presently. Assume that such a sequence
  exists.  Choose some $i \ge 0$, and let $\alpha_i = \beta B y$,
  where $\beta \in (\Sigma \cup \Nu)^*$, $B \in \Nu$, and $y \in
  \Sigma^*$. Note that $\beta B \equiv \gamma$ is a viable prefix. We
  proceed by induction on $|y|$.  If $|y|=1$, then we note that an
  LR(1) parser would shift the first symbol of $y$ (specifically:
  $\eofsym$) after $m$ reductions. By \lref{lemma:lookahead shifted
    after same number of reductions in LR and TLR parsers}, the TLR
  parser will shift the first symbol of $y$ after $m$
  reductions. Assume that the parse always terminates for strings of
  length $k \ge 1$, and assume that $|y| = k + 1$. The parser will, in
  light of \lref{lemma:lookahead shifted after same number of
    reductions in LR and TLR parsers}, shift the first symbol of $y$
  after a finite number of reductions. At this point, we have a viable
  prefix, followed by a $k$\dash{}character string, allowing us to
  apply the induction hypothesis. Therefore, the parse always
  completes.

  Since we have exhausted the possible reasons why $x \notin \lang G$,
  we conclude that the parser recognizes $\lang G$.
\end{proof}

%%% Local Variables: 
%%% mode: latex
%%% TeX-master: "tlr-plain"
%%% End: 
 
\section{Checking the Validity of a Transformation}
\label{sec:checking the validity of a transformation}

In the previous section, we considered the set of valid
transformations for a given viable prefix and lookahead symbol. In
this section, we develop an algorithm to determine if a particular
transformation is valid, and we provide a correctness proof of the
same.

\subsection{Computing the Conservation Function}

We have discussed the criteria for membership of a transformation in
the set of valid transformations; these criteria must be met by
transformations emitted b the \deltaMachine. It is not immediately
clear how we are to determine whether or not a transformation is in
this set. We consider a method of making this determination presently.

\begin{algorithm}
  {An Algorithm to Compute the Conservation Function
    \label{algo:compute conservation func}}
  {This algorithm computes a conservation function much like
    $V_\beta(\pi)$. It is straightforward to test the transformation for
    validity, given this function. The construction of the set is
    accomplished by tracing all of the different ways we might decide
    that the lookahead gets parsed from the start symbol. As we trace
    through the different productions in the grammar, we record our
    progress in sets of ordered pairs. The inclusion of an ordered pair
    like $(\pi,k)$ in one of these sets, labeled $V_{\text{something}}$,
    means that one of the procedures invoked during the execution of
    this Algorithm visited the first $k$ symbols of $\pi$; fortuitously,
    this turns out to be exactly what we need to generate the
    conservation function.}
  {$G=(\Sigma,\Nu,P,S,T,M)$: a TLR grammar; $\sigma=((s_0, \epsilon),
    (s_1, X_1), \dotsc, (s_m, X_m))$: a parse stack; $a$: a terminal
    called the ``lookahead''}
  {$V \sub P$: a set of ordered pairs $(C \to \delta, i)$, where $C
    \to \delta \in P$, and $i\le |\delta|$}
  \begin{enumerate}
  \item Calculate the item sets for $G$; let them be $I_0, I_1,
    \dotsc, I_p$.

  \item Let $V \sub T$ be an empty set of ordered pairs of the same
    type as $V_P$.

  \item Let $(s, B)$ be the item on the top of the stack.

  \item For every item of the form $j=[A\to\alpha B\cdot
    \gamma,b]$ in $I_s$, do the following:

    \begin{enumerate}
    \item Call \procref{proc:find following in item sets} with $G$,
      the stack, $j$, and $a$ as input; let $V \sub F$ and $f$ be its
      output.

    \item Set $V \sub T = V \sub T \cup V \sub F$.

    \item If $f$ is true, then call \procref{proc:trace ancestors in
        item sets} with $G$, the stack, $j$, and $a$ as input; let $V
      \sub A$ be its output, and set $V \sub T = V\sub T\cup V\sub A$.

    \end{enumerate}

  \item For every production, $\pi \in P$, define $V \sub x$ as
    follows:
    \[ 
    V \sub x(\pi)=
    \begin{cases}
      \max\{ i\in\mathbb{Z}\colon (\pi,i)\in V \sub T\} &
      \text{there exists some such $i$}\\
      -1 & \text{otherwise}
    \end{cases}.  \]

  \item Let $V \sub P=\{(\pi,V \sub x(\pi) \colon \pi \in P\}$. Return
    $V \sub P$.\algofin
  \end{enumerate}
\end{algorithm}

\begin{procedure}
  {\label{proc:find following in item sets}
    This procedure starts from an item in the current item set and
    searches for all of the ways that the lookahead could be included
    by that item, if we assume that the production in the given item
    must eventually be reduced. It does this by considering $\gamma$,
    the ``tail'' of the item in question. Each symbol of $\gamma$ is
    considered, continuing as long as $\epsilon$ can be derived from
    the current symbol, until $\epsilon$ cannot be derived. If it
    turns out that $\gamma\derives^*\epsilon$, then we back up in the
    symbol stack to where the parser first started to consider the
    current item, and we recursively retry this Procedure from that
    point. Upon halting, we return $V \sub F$ and $f$. The set of
    ordered pairs $V \sub F$ records which productions we have visited
    during the execution of this Procedure, or one of the procedures
    invoked during its execution. The flag $f$ indicates whether we
    found any way of deriving the lookahead.}
  {$G=(\Sigma,\Nu,P,S,T,M)$: a TLR grammar; $\sigma=((s_0, \epsilon),
    (s_1, X_1), \dotsc, (s_m, X_m))$: a parse stack; $a$, a terminal;
    $j=[A\to\alpha B\cdot \gamma,b]$: an item}
  {$V \sub F$: a set of ordered pairs $(C \to \delta, i)$, where $C
    \to \delta \in P$, and $i\le |\delta|$; $f$: a boolean flag}
  \begin{enumerate}
  \item Calculate the item sets for $G$; let them be $I_0, I_1,
    \dotsc, I_p$.

  \item Set $f$ to false.

  \item Let $V \sub Z$ be an empty set of ordered pairs, of the same
    type as $V \sub F$.

  \item Let $\gamma=Y_1 Y_2 \dotso Y_n$. 

  \item Let $\pi$ be the production $A\to\alpha B \gamma$.

  \item Let $i$ range from $1$ to $n$, and do the following:
    \label{item:find in item sets:i from 1 to n}
    \begin{enumerate}
    \item If $Y_i=a$, then add $(\pi,|\alpha|+1+i)$ and the contents
      of $V \sub Z$ to $V \sub F$.

    \item If $Y_i$ is a terminal, then return $V \sub F$.

    \item \label{item:find in item sets:call first for sym}
      If $Y_i$ is a nonterminal, then call
      \procref{proc:transformative first for symbol}
      with $G$, $Y_i$, $a$, and $\emptyset$ as the input;
      let the output be $V \sub E$, $f_\Nu$, and $e$
      (we ignore $\Pi$). 

    \item If $e$ is true or $f_\Nu$ is true, then set $V \sub Z=V \sub
      Z\cup V \sub E$.

    \item If $f_\Nu$ is true, then add $(\pi, |\alpha|+1+i)$ and the
      contents of $V \sub Z$ to $V \sub F$, set $V \sub Z=\emptyset$
      and set $f$ to true.

    \item If $e$ is false, then return $V \sub F$ and $f$. 
    \end{enumerate}

  \item Add $(\pi,|\alpha B\gamma|+1)$ to $V \sub Z$.

  \item \label{item:find following:pop |alpha|+1} Pop $|\alpha|+1$
    items off of the stack. Let $s$ be the state in the top item of
    the stack.

  \item Set $J=\{j_0\}$, where $j_0$ is the item
    $[A\to\cdot\alpha B\gamma,b]$ that is in $I_s$.

  \item Repeat the following until no more items can be added to
    $J$.
    \begin{enumerate}
    \item If there is an item of the form
      $[\production{C}{\cdot\delta},c]$ in $J$ and an item
      of the form $[\production{D}{\zeta\cdot C\eta},d]$
      in $I_s$, then add the latter item to $J$, if it is
      not already in $J$.
    \end{enumerate}

  \item For each item $[\production{E}{\theta \cdot \kappa},e]$ in
    $J$, do the following:
    \begin{enumerate}
    \item Let $\phi$ be the production \production{E}{\theta \kappa}.

    \item Add the ordered pair $(\phi,|\theta|+1)$ to $V \sub Z$.

    \item Call this Algorithm recursively with $G$, $a$, and the
      current value of the stack as input, along with
      $[\production{E}{\theta\cdot\kappa},e] $ in place of $j$; let $V
      \sub P$ and $f \sub P$ be the output.

    \item If $f \sub P$ is true, then set $f$ to true, set $V \sub Z=V
      \sub Z\cup V \sub P$, and set $V \sub Z=\emptyset$.
    \end{enumerate}

  \item Return $V \sub F$ and $f$. \algofin
  \end{enumerate}
\end{procedure}

\begin{procedure}
  {\label{proc:trace ancestors in item sets}
    This procedure takes an item $j$ in the current item set, and
    finds all of the items in one of the preceding item sets that
    might be reduced, if we assume that $j$ must be reduced. These
    ``ancestor'' items of our item $j$ do not need to be totally
    conserved: they only need those symbols to the left of and
    immediately to the right of the dot to be conserved. Once this
    procedure has popped some item sets off of the stack, then it
    calls itself recursively.}
  {$G=(\Sigma,\Nu,P,S,T,M)$: a TLR grammar; $\sigma=((s_0, \epsilon),
    (s_1, X_1), \dotsc, (s_m, X_m))$: a parse stack; $a$, a terminal
    called the ``lookahead''; $j=[A\to\alpha B\cdot \gamma,b]$: an
    item}
  {$V \sub A$: a set of ordered pairs $(C \to \delta,i)$, where $C \to
    \delta \in P$, and $i\le |\delta|$}
  \begin{enumerate}
  \item Calculate the item sets for $G$; let them be $I_0, I_1,
    \dotsc, I_p$.

  \item Pop $|\alpha|+1$ items off of the stack. Let $s$ be the state
    in the element on the top of the symbol stack.

  \item Let $J=\{j\}$.

  \item Repeat the following until no more items can be added to $J$.
    \begin{enumerate}
    \item If there is an item of the form
      $[\production{C}{\cdot\delta},c]$ in $J$ and an item
      of the form $[\production{D}{\cdot C\zeta},d]$ in
      $I_s$, then add the latter item to $J$, if it is not
      already in $J$.
    \end{enumerate}

  \item For every item $k=[\production{E}{\eta\cdot\theta},d]$
    in $J$, do the following:
    \begin{enumerate}
    \item Let $\pi$ be the production
      \production{E}{\eta\theta}.

    \item Add $(\pi,|\eta|+1)$ to $V \sub A$.

    \item If $\eta\ne\epsilon$, then call this Algorithm
      recursively with $G$ and $a$, the current values of
      the stacks as input, along with $k$ in place of $j$;
      let the output be $V \sub A'$.

    \item Set $V \sub A=V \sub A\cup V \sub A'$. \algofin
    \end{enumerate}

  \item Return $V \sub A$.
  \end{enumerate}
\end{procedure}

\begin{procedure}
  {\label{proc:transformative first for symbol}
    Determine all of the ways that a given grammar symbol can derive
    the lookahead. If the grammar symbol is a terminal, then do
    nothing. If the grammar symbol is a nonterminal, then return $V
    \sub S$, a set representing the portions of the productions that
    we visited during the execution of this and the following
    Procedure. This procedure calls itself recursively, so care must
    be taken if we are dealing to avoid an infinite loop; we use
    $\Pi$, a set of the productions that this Procedure has already
    visited, that is both input to and output from this Procedure. We
    also return two flags: the flag $e$ is true if and only if
    $X\derives^*\epsilon$; the flag $f$ is true if and only if
    $X\derives^*ax$, for some $x\in\Sigma^*$.  }
  {$G=(\Sigma,\Nu,P,S,T,M)$: a TLR grammar; $X$: a symbol in
    $\Sigma\cup\Nu$; $\Pi$: a set of productions; $a$: a terminal }
  {$V \sub S$: a set of ordered pairs $(\pi,n)$; $\Pi$: a set of
    productions; $e$: a boolean flag; $f$: a boolean flag}
  \begin{enumerate}
  \item Set both $e$ and $f$ to false.

  \item If $X$ is a terminal, then do the following:
    \begin{enumerate}
    \item Set $V \sub S=\emptyset$.

    \item If $X=a$, then set $f$ to true.

    \item Return $V \sub S$, $\Pi$, $f$ and $e$.
    \end{enumerate}

  \item For every production $X \to \alpha$ , such that $X \to \alpha
    \notin \Pi$, do the following:
    \begin{enumerate}
    \item Add $X \to \alpha$ to $\Pi$.

    \item Execute \procref{proc:transformative first} with $G$,
      $\alpha$, $\Pi$, and $a$ as input, and let $V_*$, $\Pi_*$, $f_*$
      and $k$ be the output.

    \item If $k=|\alpha|+1$, then set $e$ to true.

    \item Set $\Pi=\Pi\cup\Pi_*$.

    \item If $f_*$ is true, then set $f$ to true.

    \item If $f_*$ is true or $e$ is true, then 
      set $V \sub S=V \sub S\cup V_*$, and put $(\pi,k)$ in $V \sub S$.
    \end{enumerate}

  \item Return $V \sub S$, $\Pi$, $f$ and $e$. \algofin
  \end{enumerate}
\end{procedure}

\begin{procedure}
  {\label{proc:transformative first}
    Determine all of the ways that a given string of grammar symbols
    can derives a string beginning with the lookahead. Return this
    information in $V_*$, a set representing the portions of the
    productions that we visited during the execution of this and the
    preceding Procedure. The set $\Pi$ is used for the same purpose as
    in \procref{proc:transformative first for symbol}. The integer $k$
    encodes the result of this Procedure's execution as follows: if
    $\gamma\derives^*\epsilon$, then we return $k=|\gamma|+1$;
    otherwise, if $\gamma\derives^*ax$, for some $x\in\Sigma^*$, then
    we let $1\le k\le|\gamma|$; otherwise, we let $k=-1$. If we have
    both that $\gamma\derives^*\epsilon$ and that
    $\gamma\derives^*ay$, for some $y\in\Sigma^*$, then we let
    $k=|\gamma|+1$ and we let $f$ be true.  }
  {$G=(\Sigma,\Nu,P,S,T,M)$: a TLR grammar; $\gamma$: a string over
    $(\Sigma\cup\Nu)^*$; $\Pi$: a set of productions; $a$: a terminal}
  {$V_*$: a set of ordered pairs $(\pi,n)$; $\Pi$: a set of productions;
    $f$: a boolean flag; $k$: an integer with $-1\le k\le|\gamma|+1$ }
  \begin{enumerate}
  \item Set $k=-1$. 

  \item Set $f$ to false.

  \item Let $\gamma=X_1 X_2 \dotso X_n$.

  \item For each $i$ from $1$ to $n$, do the following:
    \begin{enumerate}
    \item Execute \procref{proc:transformative
        first for symbol} with $G$, $X_i$, and $\Pi$
      as input and $V \sub S$, $\Pi \sub S$, $e$, and $f \sub S$
      as output.

    \item Set $\Pi=\Pi\cup\Pi \sub S$.

    \item If $f \sub S$ is true or $e$ is true, then set $k=i$.

    \item If $f \sub S$ is true, then set $f$ to true.

    \item Set $V_*=V_*\cup V \sub S$.

    \item If $e$ is false, then go to
      \sref{item:transformative first:return when
        doesn't derive epsilon}.
    \end{enumerate}

  \item Return $V_*$, $\Pi$, $f$ and $n + 1$.

  \item \label{item:transformative first:return when
      doesn't derive epsilon} If $f$ is false, then set
    $V_*=\emptyset$ and set $k=-1$.

  \item Return $V_*$, $\Pi$, $f$ and $k$.  \algofin
  \end{enumerate}
\end{procedure}

In \procref{proc:find following in item sets}, we call
\procref{proc:transformative first for symbol} in \sref{item:find in
  item sets:call first for sym}. Since calling
\procref{proc:transformative first for symbol} twice with the same
nonterminal does not yield any new information, as an optimization, we
could keep track of the nonterminals that have already been passed to
\procref{proc:transformative first for symbol}, calling that Procedure
only if we have not called it with that nonterminal before. As an
additional optimization, we could retain $\Pi$ in the same step, and
not pass $\emptyset$ to \procref{proc:transformative first for
  symbol}. We leave the algorithm as it is because it makes it a
little easier to analyze, a task which we turn to now.

\subsubsection{A Model of the Operation of the Algorithm}

We will endeavor to prove that the Algorithm is correct. This will be
be exceedingly dull and difficult if we attempt to do so
directly. Rather, we will model the operation of the algorithm in
simple, formal terms in this section. With this model in hand, it will
be possible to produce the desired proof. Note that we do not formally
assert the equivalence of the Algorithm presented in this section and
the model presented here: we will, however, take the ``model'' to be
authoritative.

\begin{definition}
  Let the transformative context-free grammar $G=(\Sigma, \Nu, P, S,
  T, M)$ be given such that the context-free grammar $(\Sigma, \Nu, P,
  S)$ is LR($k$). Let the item set $I$ be given. If
  $i=[\production{A}{\cdot \beta}, y]$ and $j=[\production{C}{\gamma
    \cdot A \delta}, z]$ are two items in $I$, then we write $j
  \connectsBack{k} i$.
\end{definition}

\begin{definition}
  \label{def:parse precedes} 
  Let the transformative context-free grammar $G=(\Sigma, \Nu, P, S,
  T, M)$ be given, such that the context-free grammar $(\Sigma, \Nu,
  P, S)$ is LR($k$). Let the collection of item sets for $G$ be $I_0,
  I_1, \dotsc, I_n$. Finally, let the parser stack
  \[
  ((s_0, \epsilon), (s_1, X_1), \dotsc, (s_m, X_m))
  \]
  be given. Let $i$ and $j$ be two items, and let $1 \le p \le m$ such
  that $i \in I_{s_p}$. If either:
  \begin{enumerate}
  \item $j \connectsBack{k} i$; or
    
  \item
    \label{item:parse precedes:different item sets}
    $j$ is in $I_{s_p - 1}$ and it is of the form $[A \to \alpha \cdot
    X\sub p \beta, z]$, such that $i$ is of the form $[X \sub p \to \cdot
    \gamma, y]$; 
  \end{enumerate}
  then we write $j \parsePreceeds{k} i$.
\end{definition}

\begin{definition}
  \label{def:k-parse precession}
  Let the transformative context-free grammar $G=(\Sigma, \Nu, P, S,
  T, M)$ be given, such that the context-free grammar $(\Sigma, \Nu,
  P, S)$ is LR($k$). Let the collection of item sets for $G$ be $I_0,
  I_1, \dotsc, I_n$. Finally, let the parser stack
  \[
  ((s_0, \epsilon), (s_1, X_1), \dotsc, (s_m, X_m))
  \]
  be given. Let $j_1, j_2, \dotsc, j_q$ be a
  sequence of items, and let $\rho_1, \rho_2, \dotsc, \rho_q \in \{0,
  1, \dotsc, n\}$, such that, for $1 \le i < q$, we have that $
  \rho_{i + 1} - \rho_i$ is $0$ or $1$.  If we have all of the
  following, then we say that $(j_1, j_2, \dotsc, j_q;\rho_1, \rho_2,
  \dotsc, \rho_q)$ is a \defn{\kparse{k} precession}:
  \begin{enumerate}
  \item $j_q \in I_s$, where $s=s_{\rho_q}$;

  \item for $1\le r < q$,
    \[
    j_r \parsePreceeds{k} j_{r + 1};
    \]

  \item 
    \label{item:k-parse precession:first item has dot at beginning}
    $j_1$ is of the form $[C \to \cdot \gamma, x]$, where $|x|=k$; and

  \item 
    \label{item:k-parse precession:no repeats}
    there do not exist indices $p$ and $q$ such that $j_p = j_q$ and
    $\rho_p = \rho_q$, and for all $p \le h < q$ we have
    $j_h \connectsBack{k} j_{h + 1}$ and for some $p < r < q$, we
    have that either $j_p = j_r$ or that $j_q = j_r$.
  \end{enumerate}
\end{definition}

We often omit the second component of a \kparse{k} precession---namely,
the sequence of indices $\rho_1, \rho_2, \dotsc, \rho_q$---unless they
are explicitly called for.

\begin{definition}
  \label{def:upward link, ancestral link and sidelink}
  Let $G=(\Sigma, \Nu, P, S, T, M)$ be a TLR grammar, let $\beta B=X_1
  X_2 \dotso X_p$ be a viable prefix followed by $a$, and let $((s_0,
  \epsilon), (s_1, X_1),\dotsc,(s_p,X_p))$ be a parse stack.  Let
  $J=(j_1, j_2, \dotsc, j_n; \rho_1, \rho_2, \dotsc, \rho_n)$ be a
  \kparse{0} precession, let $U=(u_1, u_2, \dotsc, u_m; \tau_1,
  \tau_2, \dotsc, \tau_m)$ be a \kparse{1} precession, and let
  $s=[\production{A}{\delta \cdot \gamma}]$ be an LR(0) item.  Let
  $u_i=[\production {C_i} {\zeta_i \cdot \eta_i}, a]$, for $1 \le i
  \le m$.  Consider the following criteria:
  \begin{enumerate}
  \item $j_1=[\production{S'}{\cdot S}]$;

  \item either of the following:
    \begin{enumerate}
    \item 
      \label{item:upward link, ancestral link and sidelink:the upward
        link is trivial}
      $m=0$, in which case, $s$ is of the form
      $[\production{A}{\delta' B \cdot \gamma}]$ where
      $\delta'B=\delta$, and we have that $j_n \parsePreceedsEpsilon
      s$ and that $\rho_n=p-1$, or

    \item 
      \label{item:upward link, ancestral link and sidelink:the upward
        link is nontrivial}
      $m>0$, and all of the following:
      \begin{itemize}
      \item $s$ is of the form $[\production{A}{\delta' C \cdot
          \gamma}]$ for $\delta' C=\delta$,

      \item $u_m$ is of the form $[\production{D}{\theta B \cdot
          \kappa}, a]$,

        % \item
        %   \label{item:upward link, ancestral link and sidelink:the
        %     upward link is nontrivial:j parse precedes u, but u does
        %     not connect back to j}
        %   $j_n\parsePreceedsEpsilon u_1'$ but $j_n
        %   \notConnectsBackEpsilon u_1'$ where $u_1'=[\production {C}
        %   {\zeta \cdot \eta}]$, and

      \item $\tau_m=p$;
      \end{itemize}
    \end{enumerate}

  \item $\gamma\derives^* a x$, for some $x\in\Sigma^*$; 

  \item all $\eta_h\derives^*\epsilon$, for $h \ge 1$.
  \end{enumerate}
  If all of these criteria hold, then we say that $U$, $J$, and $s$
  constitute an \defn{upward link} to $a$, an \defn{ancestral link},
  and a \defn{sidelink} to $a$, respectively for $\beta B$, and we say
  that $U$ and $J$ \defn{join} $s$.
\end{definition}

\begin{definition}
  Let $\beta B$ be a viable prefix.  Let $U=(u_1, u_2, \dotsc, u_n)$
  be a \kparse{1} precession. Let $P=\{p_1, p_2, \dotsc, p_m\}$ be the
  set of indices such that $p_1 < p_2 < \dotsb < p_m$ and for $1 \le h
  \le m$, we have that $j_{p_h}$ is of the form
  $[\production{A}{\alpha \cdot \gamma}]$ where $\alpha \ne \epsilon$,
  and we also have that, for any index $h_0\notin P$, we have that
  $j_{h_0}$ is of the form $[\production{A}{\cdot \gamma}]$. We let
  $j_{p_h}=[\production{A_h}{\alpha_h X_h \cdot \gamma_h}]$ for $1\le
  h \le n_p$, where $X_h$ is a grammar symbol. Let $\delta\equiv X_1
  X_2 \dotso X_m$; we call $\delta$ the \defn{trace} of $U$.
\end{definition}

\begin{definition}
  \label{def:trace}
  Let $\beta B$ be a viable prefix followed by $a$. Let $U$, $J$ be an
  upward and an ancestral link, joining the sidelink $s$. Let
  $\delta_J$ be the trace of $J$. There are two cases that we will
  consider:
  \begin{enumerate}
  \item 
    \label{item:trace:|U|=0}
    if $|U|=0$, then we know that $s$ is of the form $[\production {A}
    {\alpha B\cdot \gamma}]$. Let $\delta\equiv \delta_J B$;
    otherwise,

  \item if $|U|>0$, then we let $\delta \equiv \delta_J \delta_U$,
    where $\delta_U$ is the trace of $U$.
  \end{enumerate}
  We call $\delta$ the \defn{trace} of the tuple $(U, J, s)$.
\end{definition}

\begin{theorem}
  \label{thm:the trace of (U, J, s) is beta B}
  Let $\beta B$ be a viable prefix followed by $a$. Let $U$, $J$ be an
  upward and an ancestral link, joining the sidelink $s$. The trace of
  $(U, J, s)$ is $\beta B$.
\end{theorem}
\begin{proof}
  Let $\sigma=(s_0, s_1, s_2, \dotsc, s_n)$ be the state stack, and
  let $I_0, I_1, I_2, \dotsc, I_m$ be the item sets for $G$. Let
  $U=(u_1, u_2, \dotsc, u_p; \tau_1, \tau_2, \dotsc, \tau_p)$ be a
  \kparse{1} precession, and let $J=(j_1, j_2, \dotsc, j_q; \rho_1,
  \rho_2, \dotsc, \rho_q)$ be a \kparse{0} precession. Let $\beta=X_1
  X_2 \dotso X_n$.

  We will first define the LR(0) item $r$:
  \begin{enumerate}
  \item if $|U|=0$, then let $r=s$;

  \item if $|U|>0$, then let $u_1=[\production {A \sub U} {\alpha \sub
      U \cdot \gamma \sub U}, a]$ and let $r=[\production {A \sub U}
    {\alpha \sub U \cdot \gamma \sub U}]$.
  \end{enumerate}

  We begin by proving that the trace of $J$ is a prefix of $\beta B$
  that is of length $\rho_q$.  We shall proceed by induction on
  $\rho_q$. Assume that $\rho_q=0$. For all $j \in J$, we know that
  $j$ is of the form $[\production {A \sub j} {\cdot \alpha \sub
    j}]$. Thus, the trace of $J$ is $\epsilon$.

  Assume that we know that the trace of $J$ is a prefix of $\beta B$
  that is $\rho_q$ symbols long when $|J|=n \sub j$, where $n \sub j
  \ge 0$. Let us assume that $\rho_q=n \sub j + 1$ symbols long. Let
  $q \sub g$ be the greatest index such that $\rho_{q_g}=n \sub j$. We
  know that $\rho_{q \sub g}=n \sub j + 1$. Thus,
  \[
  j_{q \sub g} \parsePreceedsEpsilon j_{q \sub g + 1} \quad \text{but}
  \quad j_{q \sub g} \notConnectsBackEpsilon j_{q \sub g + 1}.
  \]
  Since $j_{q_g + 1}$ is of the form $[\production {A \sub g} {\alpha
    \sub g X \sub g \cdot \gamma \sub g}]$, we must have, by
  \condref{item:parse precedes:different item sets} of
  \dref{def:parse precedes}, that
  \[
  X \sub g=X_{n \sub j + 1}.
  \]
  Since we know, by the induction hypothesis, that the trace of
  $J'=(j_1, j_2, \dotsc, j_{q \sub g})$ is
  \[
  X_1 X_2 \dotso X_{n \sub j},
  \]
  we conclude that the trace of $J$ is 
  \[
  X_1 X_2 \dotso X_{n \sub j + 1}.
  \]

  We now consider the possibility that $|U|=0$. In this case, we
  have---by \condref{item:upward link, ancestral link and sidelink:the
    upward link is trivial} of \dref{def:upward link, ancestral link
    and sidelink}---that $\rho_q = n - 1$. Also, we know that $s$ is
  of the form
  \[
  [\production {A \sub s} {\delta \sub s B \cdot \gamma \sub s}].
  \]
  Thus, the trace of $J$ is $\beta$, so by \condref{item:trace:|U|=0}
  of \dref{def:trace}, we see that the trace of $(U, J, s)$ is $\beta
  B$.

  If $|U| > 0$, then note that
  \[
  \tau_1 = \rho_q + 1.
  \]
  Let $t=\tau_p - \rho_q$. We shall proceed by induction on $t$.

  Assume that $t=1$. Now, as we know that $\tau_h = \rho_q + 1$ for
  all $1 \le h \le p$, we conclude that $u_1$ is of the form
  \[
  [\production {C \sub u} {\zeta \sub u X \sub u \cdot \eta \sub u},
  a]
  \]
  and that when $1 < h \le p$, we have that $u_h$ is of the form
  \[
  [\production {C_h} {\cdot \theta_h}, a].
  \]
  However, we know that $u_p$ is of the form
  \[
  [\production {C_*} {\zeta_* B \cdot \eta_*}, a].
  \]
  Therefore, we note that $|U|=1$, and by \condref{item:upward link,
    ancestral link and sidelink:the upward link is nontrivial} of
  \dref{def:upward link, ancestral link and sidelink}, we note that
  $s(j_q)=n - 1$. Thus, the trace of $U$ is $B$, and since the trace
  of $J$ is $\beta$, we conclude that the trace of $(U, J, s)$ is
  $\beta B$.

  Assume now that the trace of $|U|$ is known to be 
  \[
  X_{n-t+2} X_{n-t+3} \dotso X_n B
  \]
  when $t=k$, for $k \ge 1$. Assume that $t=k+1$. Consider $u_1$: let
  $u_1=[\production {A_{u_1}} {\alpha_{u_1} \cdot \gamma_{u_1}}, a]$;
  now let $u \sub f=[\production {A_{u_1}} {\alpha_{u_1} \cdot
    \gamma_{u_1}}]$.  We know that
  \[
  j_p \parsePreceedsEpsilon u \sub f \quad \text{but} \quad j_p
  \notConnectsBackEpsilon u \sub f;
  \]
  thus, $s(j_p)=s(u_1) - 1$. Since $j_p \notConnectsBackEpsilon u \sub
  f$, we conclude that $u_1$ is of the form $[\production {A_{u_1}}
  {\alpha_{u_1}' X_{u_1} \cdot \gamma_{u_1}}, a]$, where
  $\alpha_{u_1}' X_{u_1}=\alpha_{u_1}$. As $u_1$ is a live item in
  $I_{\tau_1}$, we conclude that $X_{u_1}=X_{n-t+2}$. Therefore, by
  induction, the trace of $U$ is
  \[
  X_{n-t+2} X_{n-t+3} \dotso X_n B.
  \]

  We now know that the trace of $J$ is $\beta_J=X_1 X_2 \dotso
  X_{\rho_q}$. Also, when we assume that $|U|>0$, we also know that
  the trace of $U$ is $\beta_U=X_{\tau_1} X_{\tau_2} \dotso
  X_{\tau_{p-1}} B$. Since $\tau_1=\rho_q + 1$, and since $\tau_p=n$,
  we have therefore established that $\beta_J \beta_U=\beta B$.
\end{proof}

\begin{definition}
  Let $G=(\Sigma, \Nu, P, S, T, M)$ be a TLR grammar.  Let $a$ and
  $\gamma$ be given, such that $\gamma\derives^*a x$, where
  $x\in\Sigma^*$. Let $\gamma=X_1X_2\dotso X_m$. If $k$ is an index
  such that
  \begin{equation}
    \label{eq:def of downlink: preceding derives epsilon}
    X_1 X_2 \dotso X_{k-1} \derives^* \epsilon
  \end{equation}
  and
  \[
  \label{eq:def of downlink: first derivation of a}
  X_k\derives^* ay,
  \]
  where $y\in\Sigma^*$, then $(k, \gamma)$ comprises a \defn{partial
    downward link to} $a$ \defn{for} $\gamma$. Of course, if $k=1$,
  then we take the symbol $X_1 X_2 \dotso X_{k-1}$ to be a synonym for
  $\epsilon$, in which case \eqref{eq:def of downlink: preceding
    derives epsilon} is trivial.
\end{definition}

\begin{definition}
  Let $G=(\Sigma, \Nu, P, S, T, M)$ be a TLR grammar.  Let $a$ and
  $\gamma$ be given, such that $\gamma\derives^*a x$, where
  $x\in\Sigma^*$. Let $\gamma=X_1X_2\dotso X_m$. If $k$ is an index
  such that
  \[
  X_1 X_2 \dotso X_{k-1} \derives^* \epsilon
  \]
  and $X_k=a$, then we say that $(k, \gamma)$ is a \defn{terminal link
    to} $a$ \defn{for} $\gamma$.
\end{definition}

\begin{definition}
  Let $L$ be a partial downward link to $a$ for $\gamma$, with value
  $k$. Let $\gamma=X_1X_2\dotso X_n$. If $L_d$ is a partial downward
  link to $a$ for $\gamma_d$, then $L_d$ is \defn{chained to $L$} if
  $\production {X_k} {\gamma_d}$ is a production. If $L_t$ is a
  terminal link to $a$ for $\gamma_t$, then $L_t$ is \defn{chained to
    $L$} if $\production {X_k} {\gamma_t}$ is a production. If $L_1$
  and $L_2$ are two links, then we define the \defn{chain production}
  to be either $\production {X_k} {\gamma_d}$ or $\production {X_k}
  {\gamma_t}$, as appropriate, and we represent this production with
  the symbol $P(L_1,L_2)$.
\end{definition}

\begin{definition}
  \label{def:complete downward link to a}
  Let $L_1, L_2, \dotsc, L_n$ be a sequence of chained links, where
  for each $L_i=(\gamma_i,a)$ such that $(\gamma_i,k_i)$ is a partial
  downward link for $\gamma_i$ to $a$ when $i<n$, while
  $(\gamma_i,k_i)$ is a terminal link for $\gamma_i$ to $a$ when
  $i=n$.  If, for $1\le j<n$, there is most one other index $1\le k<n$
  such that $j\ne k$ but $P(L_j,L_{j+1})=P(L_k,L_{k+1})$, then we say
  that this sequence of strings, partial downward links and this
  terminal link comprises a \defn{complete downward link from
    $\gamma_1$ to $a$}.
\end{definition}

\begin{definition}
  Let $G=(\Sigma, \Nu, P, S, T, M)$ be a TLR grammar, and let $\beta$
  be a viable prefix followed by $a$.  Let $U$, $J$ be an upward and
  an ancestral link, joining the sidelink $s$. Letting
  $s=[\production{A}{\delta \cdot \gamma}]$, let the  complete
  downward link $D$ from $\gamma$ to $a$ be given.
  Let $U=(u_1, u_2, \dotsc, u_n ; \tau_1, \tau_2, \dotsc, \tau_n)$
  such that, for $1 \le i \le n$, we have $u_i$ of the form 
  \[
  u_i=[C_i \to \alpha_i \cdot \delta_i, a].
  \]
  If we have that such that $\delta_i \derives^* \epsilon$ for $1 \le
  i \le n$, then we call the ordered quadruple $(U,J,s,D)$ a
  \defn{parse path for the viable prefix $\beta$ followed by $a$}.
\end{definition}

We will argue that \algoref{algo:compute conservation func} operates
by enumerating all parse paths for the viable prefix $\beta$ followed
by $a$.

Let us say that we have a TLR grammar $G=(\Sigma,\Nu,P,S,T,M)$, and
that we are parsing a sentence $x$. The parser has just reduced by a
transformative production, leaving the stack as $\beta B$, with
lookahead $a$. \algoref{algo:compute conservation func} begins with
all of the items in the item set on the top of the stack that are of
the form $[\production{A}{\beta B\cdot \gamma},b]$, which is
incidentally the form of the sidelink. The Algorithm's next step is to
invoke \procref{proc:find following in item sets} to find the
downlinks.

\procref{proc:find following in item sets} scans the ``remainder'' of
the current item---that is, the portion to the right of the dot---to
determine if this remainder can be used to derive $a$ or
$\epsilon$. The way that it makes this determination is with
Procedures~\ref{proc:transformative first for symbol}
and~\ref{proc:transformative first}. If these Procedures successfully
find such a derivation, then they return the portions of each
production that they used in $V \sub S$ and $V_*$; if they are not
successful, then those two sets are empty. If these Procedures
determine that the remainder derives $\epsilon$, then
\procref{proc:find following in item sets} will find all items that
preceeds the current item in the parse precession. The way that these
items are found is by first ``rewinding'' the parse stack until such a
time as the parse first started to consider the current item. At this
point, we create $J$, which is like the closure of an item set, taken
in reverse. We consider the items in $J$ one at a time. For a parse
precession, we do not allow an item to be present more than twice,
unless we go to the previous item set; in \procref{proc:find following
  in item sets}, we create $J$ first, then we call the Procedure
recursively for each item in the set. Since, in \sref{item:find
  following:pop |alpha|+1} of the Procedure, we pop at least one item
off of the stack, and since we call the Procedure recursively exactly
once for each item in $J$, we can be sure that the sequence of items
we trace does not violate condition~\ref{item:k-parse precession:no
  repeats} of \dref{def:k-parse precession}.

We do not allow a production to appear more than twice in the complete
downward link to $a$, so we use the production set $\Pi$, which we
initialize to $\emptyset$ when we invoke \procref{proc:transformative
  first for symbol} in \procref{proc:find following in item sets}. 

During this process, each of the upward links to $a$ are enumerated,
as are the complete downward links to $a$, with the appropriate
sidelink for one of the parse precessions.

As for the ancestral links, we have \procref{proc:trace ancestors in
  item sets}, which is invoked only if \procref{proc:find following in
  item sets} succeeds in finding a way to derive $a$ from an item.

What of $V \sub P$, the set returned by \algoref{algo:compute
  conservation func}? We have gone to some effort to construct a model
of the operation of \algoref{algo:compute conservation func}, but we
have no counterpart for the set $V \sub P$. We now construct a
function which, given a parse path and a production $\pi$, returns the
value of $V \sub P(\pi)$ that the Algorithm would produce as it traces
out that parse path.

\begin{definition}
  \label{def:parse path conservation function}
  Let $\mathbf P=(U,J,s,D)$ be a parse path for the viable prefix $\beta$
  followed by $a$.  Letting $U=(u_1, u_2, \dotsc, u_n)$, we let
  \[
  u_k=[\production{A_k}{\alpha_k\cdot\gamma_k},a]
  \text{ and }
  \pi_k=\production{A_k}{\alpha_k\gamma_k}
  \]
  for $1\le k\le n$.  Letting $J=(j_1,j_2,\dotsc,j_m)$, we let:
  \[
  j_k= [\production{C_h}{\zeta_h\cdot D_h\eta_h}]
  \text{ and }
  \phi_h= \production{C_h}{\zeta_h D_h\eta_h},
  \] 
  for $1 \le h \le m$.  Letting $D=(L_1, L_2, \dotsc, L_p)$, we let:
  \[
  L_i=(\theta_i,k_i),
  \]
  for $1 \le i \le p$.  We first define six functions.
  \begin{enumerate}
  \item 
    \label{item:parse path conservation function:V_Pi}
    Define $V_\Pi\colon P\to\mathbb Z$ as
    \[
    V_\Pi(\pi)=
    \begin{cases}
      |\alpha_k|+|\gamma_k|+1 & \text{$\pi=\pi_k$ for some $k$}\\
      -1 & \text{otherwise}
    \end{cases}.
    \]

  \item Define $V_{\Pi,\epsilon} \colon P\to \mathbb Z$ as
    \[
    V_{\Pi,\epsilon}(\production{A}{\delta})=
    \begin{cases}
      |\delta|+1 & \text{\production{A}{\delta} is used in the
        derivation $\gamma_k\derives^*\epsilon$ for some $k$}\\
      -1 & \text{otherwise}
    \end{cases}.
    \]

  \item Define $V_\Phi\colon P\to\mathbb Z$ as 
    \[
    V_\Phi(\phi)=
    \begin{cases}
      |\zeta_k|+1 & \text{$\phi=\phi_k$ for some $k$}\\
      -1 & \text{otherwise}
    \end{cases}.
    \]

  \item Define $V_\Psi\colon P \to \mathbb Z$ as
    \[
    V_\Psi(\psi)=
    \begin{cases}
      k_{h+1} & \text{$\psi=P(L_h,L_{h+1})$ for some $h$}\\
      -1 & \text{otherwise}
    \end{cases}.
    \]

  \item Letting $\theta_h=X_{h,1} X_{h,2} \dotso X_{h,q_h}$, define
    $V_{\Psi,\epsilon} \colon P \to \mathbb Z$ as
    \[
    V_{\Psi,\epsilon}(\production{E}{\kappa})=
    \begin{cases}
      |\kappa|+1 & \text{\production{E}{\kappa} is used in the
        derivation $X_{h,i}\derives^*\epsilon$ for some $h$}\\
      -1 & \text{otherwise}
    \end{cases}.
    \]

  \item
    \label{item:parse path conservation function:V_Omega}
    Letting $s=[\production {D \sub s} {\gamma \sub s \cdot \delta
      \sub s}]$, define $V_\Omega \colon P \to \mathbb Z$ as
    \[
    V_\Omega(\omega)=
    \begin{cases}
      |\gamma \sub s| + |\theta_1| + 1 & \omega=\production {D \sub s}
      {\gamma \sub s \delta \sub s}\\
      -1 & \text{otherwise}
    \end{cases}.
    \]

  \end{enumerate}
  We have now come to our goal: define $L_{\mathrm V, \mathbf P}\colon P\to\mathbb Z$ as 
  \[
  L_{\mathrm V, \mathbf P}(\chi)= \max\{V_\Pi(\chi), V_{\Pi,\epsilon}(\chi),
  V_\Phi(\chi), V_\Psi(\chi), V_{\Psi,\epsilon}(\chi),
  V_\Omega(\chi)\}.
  \]
  We call $L_{\mathrm V, \mathbf P}$ the \defn{parse path conservation
    function}.
\end{definition}

We intend $V \sub P$ to correspond exactly to the function $L \sub V$,
and vice versa. We can justifiably use the output of the Algorithm to
determine the validity of a transformation if we can justifiably use
the function $L \sub V$ for that task. We will first recast the
validity test for transformations in the next section, after which we
will provide the promised justification.

\subsection{An Alternative Test for Allowable Transformations}

\label{sec:simplified tree test}

The conservation function of \secref{sec:allowable transformations}
may consider an infinite number of parse trees; thus, it is not self
evident that any analysis of parse paths will be able to reproduce the
conservation function, unless an infinite number are considered. In
this section, we consider a subset of the parse trees considered by
the conservation function which is finite in number and which does
reproduce the conservation function. Moreover, this subset of parse
trees will be ``isomorphic,'' in a sense, to the set of parse
paths. After constructing this set of parse trees, and establishing
the claimed properties, we will have shown that the parse path
conservation function, and by extension, \algoref{algo:compute
  conservation func}, correctly calculate the conservation function in
a finite amount of time and guarantee the successful execution of
\algoref{algo:TLR}.

Consider \dref{def:conservation function for sentential form}, wherein
we define the function $N_{\alpha, \node T}$. In that section, we were given a
TLR grammar and a sentential form, and we chose a sentence derivable
from that sentential form. \pref{prop:conservation function
  independent of sentential form} justified our choice of an arbitrary
sentence derivable from the given sentential form: the portion of the
tree consisting of nodes that were descendants of nodes representing
symbols in the original sentential form make no contribution to the
value of the function $N_{\alpha, \node T}$. By inspection of the definition of
the function $N_{\alpha, \node T}$, we can also conclude that nodes that are
ordered greater than the node representing the symbol $a$ make no such
contribution either. Let us investigate what would happen to
$N_{\alpha, \node T}$ were we to remove those nodes from a parse tree.

\begin{definition}
  \label{def:simplified tree}
  Let $G=(\Sigma,\Nu,P,S,T,M)$ be a TLR grammar and let $\alpha=\beta
  B a x$ be a sentential form for $G$ such that
  $\beta\in(\Sigma\cup\Nu)^*$ and $x\in\Sigma^*$, while $B$ and $a$
  are a nonterminal and a terminal, repectively. Let $y$ be a sentence
  in $G$ such that $\alpha\derives^*y$, and let \node T be the parse
  tree for $y$. Let $\node A_1,\node A_2,\dotsc,\node A_n$ be the
  nodes representing the symbols $\beta B$, and let \node B and \node A
  represent the $B$ and $a$, as they appear in $\alpha$,
  respectively. We define two operators $\mu$ and $\mu'$ which acts on
  trees.  The action of $\mu'$ is to remove all nodes \node X if
  either:
  \begin{enumerate}
  \item \node X is a descendent of some node \node Y, where $\node
    Y=\node A_i$ for some $1\le i\le n$; otherwise

  \item \node X is not a descendent of any node \node Y, where $\node
    Y=\node A_i$ for some $1\le i\le n$, and in addition $\node X >
    \node A$.
  \end{enumerate}
  Let $\mu \node T$ be the tree formed from $\mu' \node T$ by
  replacing every leaf node labeled by a production $C \to \gamma$
  with a node labeled $C$.  The tree $\mu \node T$ we shall refer to
  as the \defn{simplified tree for $\beta B a$.} The operator $\mu$ is
  the \defn{simple-tree projection operator.}
\end{definition}

\begin{definition}
  Let $G=(\Sigma, \Nu, P, S)$ be a context-free grammar.  Let $\node
  T$ be a tree labeled with productions and grammar symbols from $G$,
  along with $\epsilon$. Let $\node N$ be an interior node in $\node
  T$ that is labeled with the production $\production {A}
  {\alpha}$. If the child-string of $\node N$ is a prefix of $\alpha$,
  then we say that $\node N$ is \defn{parse-proper}.
\end{definition}

\begin{prop}
  Every simplified tree is parse-proper.
\end{prop}

\begin{definition}
  \label{def:proper above B, and prefix-ancestral}
  Let \node T be a simplified tree for $\beta B a x$. Let \node B and
  \node A be the nodes corresponding to $B$ and $a$, respectively. Let
  \node U be the set of nodes that are ancestral to \node B. Let
  \begin{align*}
    \node Z= & \{ \text{\node C in \node T} \colon \text{\node C
      corresponds to one of the symbols in $\beta B$} \} \\
    & \cup \{ \text{\node C in \node T} \colon \text{\node C is the least not
      autoancestral to \node A but not \node B}\}
  \end{align*}
  For every $\node C \in \node Z$, the parent of $\node C$ is in
  $\node U$. Let $\node W \subset \node U$ be such that, for every
  node $\node D\in\node W$, there exists some $\node C \in \node Z$
  such that \node D is the parent of $\node C$.  Let us put the
  elements of \node into an ascending sequence we call the
  \defn{prefix-ancestral sequence}: this sequence is $(\node M_0,
  \node M_1, \dotsc, \node M_m)$.  For $0\le i\le m$, define
  \[
  \node V_i=
  \begin{cases}
    \{\node F\in\node U\colon\node F<\node M_1\} & i=0 \\
    \{\node F\in\node U\colon\node M_i<\node F<\node
    M_{i+1}\} & 1 \le i < m 
  \end{cases}.
  \]
  Call $(\node V_i)_{i=0}^m$ the \defn{prefix-ancestral interstitial
    sequence} for \node B and \node A. If, for all $0\le i\le m$,
  there are no more than two distinct nodes $\node A_1$ and $\node
  A_2$ in $\node V_i$ such that $\nodeLabel(\node A_1)=\nodeLabel
  (\node A_2)$, then we say that \node T is \defn{proper above \node
    B.}
\end{definition}

\begin{definition}
  \label{def:proper above a}
  Let \node T be a simplified tree for $\beta B a$. Let \node B and
  \node A be the nodes corresponding to $B$ and $a$, respectively. Let
  \node X be the set of nodes that are ancestral to \node A but not
  \node B.  If there exist no more than two nodes $\node D_1,\node
  D_2\in\node X$ such that $\nodeLabel(\node D_1)=\nodeLabel(\node
  D_2)$, then we say that \node T is \defn{proper above} \node A.
\end{definition}

\begin{definition}
  Let \node T be a simplified tree for $\beta B a $. Let \node B and
  \node A be the nodes corresponding to $B$ and $a$, respectively. If
  \node T is proper above both \node A and \node B, then we say that
  \node T is a \defn{proper simplified tree for $\beta B ax$}.
\end{definition}

\begin{definition}
  Let $G=(\Sigma,\Nu,P,S,T,M)$ be a TLR grammar and let $\beta$ be a viable prefix followed by the terminal $a$, for the
  nonterminal $B$ is a nonterminal. Define a set of trees which we
  will refer to as the \defn{simplified tree set for $\beta a$} as
  follows:
  \[
  F_{\beta a}=\{\node T'\colon\text{$\node T'$ is a simplified tree
    for $\beta a$}\}.
  \]
  We also define the following set of trees, which we will refer to as
  the \defn{proper simplified tree set for $\beta a$}:
  \[
  T_{\beta a}=\{\node T'\colon\text{$\node T'$ is a proper simplified
    tree for $\beta a $}\}.
  \]
\end{definition}

\begin{definition}
  \label{def:[proper] simplified conservation function}
  Let $G=(\Sigma,\Nu,P,S)$ be an LR(1) grammar. Let $\beta$ be a viable prefix
  followed by $a$. Let $\node T\in T_{\beta a}$.  Let $\node
  P\in\node T$, and let the children of \node P be $\node A_1,\node
  A_2,\dotsc,\node A_n$. Define
  \[
  H_{\node T}(\node P)=
  \begin{cases}
    n+1 & \text{\node P is an ancestor of \node B but not
      \node A}\\
    i & \text{\node P is an ancestor of \node A, and $\node A_i$ is
      autoancestral to \node A}\\
    n+1 & \text{\node P shares an ancestor with both \node
      A and \node B, and $\node B<\node P<\node A$}
  \end{cases}.
  \]
  For  $\pi\in P$ and $\node T\in F_{\beta a}$ define 
  \[
  M( \node T, \pi)=\max\{H_{\node T}(\node P)\colon \nodeLabel(\node
  P)=\pi\}.
  \]
  We define the \defn{simplified conservation function for $\beta a$};
  \[
  Y_{\beta a}(\pi)=
  \begin{cases}
    M(\node T,\pi) & \text{there exists some
      $\node T\in F_{\beta a}$ containing \node P such
      that $\nodeLabel(\node P)=\pi$}\\
    -1 & \text{otherwise}
  \end{cases}.
  \]
  Finally, we define the \defn{proper simplified conservation function
    for $\beta a$};
  \[
  Z_{\beta a}(\pi)=
  \begin{cases}
    M( \node T,\pi) & \text{there exists some $\node T\in T_{\beta a}$
      containing \node P such
      that $\nodeLabel(\node P)=\pi$}\\
    -1 & \text{otherwise}
  \end{cases}.
  \]
\end{definition}

\begin{definition}
  Let \node T be a tree, and let \node A, \node B, and \node C be
  three nodes such that \node A is ancestral to \node B, which is
  ancestral to \node C. A function which takes \node T, along with
  \node A, \node B, and \node C as argument, whose range is $\{0, 1\}$
  we refer to as a \defn{tree-projection decision function.}
\end{definition}

\begin{definition}
  Let \node T be a tree. We call $f$ as a \defn{triplet location
    function} if
  \[
  f(\node T)=
  \begin{cases} 
    (\node A, \node B, \node C) \\ 
    \emptyset
  \end{cases},
  \]
  where \node A, \node B, and \node C are all nodes of \node T, such
  that \node A is ancestral to \node B, which is ancestral to \node C.
\end{definition}

Since we give the nodes of a tree a total order, we can create a
bijection between the nodes of a tree and the first $n$ integers;
thus, the return values of the triplet location function and the 
last three arguments to a tree-projection decision function are all
elements of $\mathbb Z_{n - 1}$.

\begin{definition}
  Let $G=(\Sigma, \Nu, P, S)$ be an LR(1) grammar, and let $\beta B$
  be a viable prefix followed by $a$. Let \node T be a parse-proper
  tree with yield $\beta B a$, letting \node B and \node A be the
  nodes corresponding to $B$ and $a$, respectively.  Let $\rho$ be a
  tree-projection decision function, and let $\mu$ be a triplet
  location function.  We define an operator $\Pi_{\rho,\mu}$ which
  will transform \node T. If $\mu(\node T)=\emptyset$, then
  $\Pi_{\rho,\mu}$ has no effect. Assume instead that $\mu(\node
  T)=(\node A, \node B, \node C)$; let the parents of \node A, \node
  B, and \node C be $\node P_{\node A}$, $\node P_{\node B}$, and
  $\node P_{\node C}$, respectively, should they all exist---in
  particular, $\node P_{\node A}$.  The operation of $\Pi_{\rho,\mu}$
  is as follows.
  \begin{enumerate}
  \item If $\rho(\node T, \node A, \node B, \node C) = 0$, then do one
    of the following:
    \begin{enumerate}
    \item if $\node A$ is the root of \node T, then make $\node B$ the
      new root, but

    \item if \node A is not the root of \node T, then remove \node A
      as a child of $\node P_{\node A}$, and change the parent of
      \node B to $\node P_{\node A}$.
    \end{enumerate}
    
  \item Otherwise, if $\rho(\node T, \node A, \node B, \node C)=1$,
    then remove \node B as a child of $\node P_{\node B}$, and we
    change the parent of \node C to $\node P_{\node B}$.
  \end{enumerate}
  We call $\Pi_{\rho, \mu}$ the \defn{tree triplet surgery operator.}
\end{definition}

If $X$ is a totally ordered set, then we use the following total order
on $X \times X \times \dotsb \times X \equiv X^n$. Let $(a_1, a_2,
\dotsc, a_n), (b_1, b_2, \dotsc, b_n) \in X^n$; we say that $(a_1,
a_2, \dotsc, a_n) \le (b_1, b_2, \dotsc, b_n)$ in either of the
following cases:
\begin{enumerate}
\item $a_i = b_i$ for $1 \le i < j \le n$, and $a_j < b_j$; or
\item $a_i = b_i$ for $1 \le i \le n$.
\end{enumerate}

\begin{definition}
  Let $G=(\Sigma, \Nu, P, S)$ be an LR(1) grammar, and let
  $\beta B$ be a viable prefix followed by $a$. Let \node T be a
  parse-proper tree with yield $\beta B a$, letting \node B and \node
  A be the nodes corresponding to $B$ and $a$, respectively.  Let
  $\node W=(\node M_0,\node M_1,\dotsc,\node M_m)$ be the
  prefix-ancestral sequence for \node B and \node A, and let $( V_i
  )_{i=0}^n$ be the prefix-ancestral interstitial sequence for \node B
  and \node A.  We let $0\le j\le m$ be the least index such that
  there are nodes $\node N_1,\node N_2,\node N_3\in\node V_j$ such
  that
  \begin{gather*}
    \node N_1<\node N_2<\node N_3 \text{, and} \\
    \nodeLabel(\node N_1)=\nodeLabel(\node N_2)=\nodeLabel(\node N_3).
  \end{gather*}
  We say that the three nodes $(\node N_1, \node N_2, \node N_3)$ are
  a \defn{repetitive triple for $\node V_j$}. If no such $j$ exists,
  then let $\mu \sub B(\node T)=\emptyset$.  If such a $j$ does exist,
  let $(\node N_1, \node N_2, \node N_3)$ be a repetitive triple for
  $V_j$, such that there does not exist a repetitive triple $(\node
  M_1, \node M_2, \node M_3)$ satisfying
  \[
  (\node M_1, \node M_2, \node M_3) < (\node N_1, \node N_2, \node
  N_3).
  \] 
  Call the repetitive triple $(\node N_1, \node N_2, \node N_3)$ the
  \defn{active repetitive triple} for \node T, and let $\mu \sub
  B(\node T) = (\node N_1, \node N_2, \node N_3)$.
\end{definition}

\begin{definition}
  Let $G=(\Sigma, \Nu, P, S)$ be an LR(1) grammar, and let $\beta B$
  be a viable prefix followed by $a$. Let \node T be a parse-proper
  tree with yield $\beta B a$, letting \node B and \node A be the
  nodes corresponding to $B$ and $a$, respectively. Let \node X be the
  set of nodes ancestral to \node A, but not ancestral to \node B. Let
  $\node C_1, \node C_2, \node C_3 \in \node X$ be three nodes such
  that
  \begin{gather*}
    \node C_1 < \node C_2 < \node C_3 \text{, and} \\
    \nodeLabel(\node C_1)=\nodeLabel(\node C_2)=\nodeLabel(\node C_3),
  \end{gather*}
  Call $(\node C_1, \node C_2, \node C_3)$ a \defn{lookahead
    repetitive triple} (LA-repetitive triple) for \node T. If $(\node
  C_1, \node C_2, \node C_3)$ is a LA-repetitive triple for \node T,
  such that there does not exist a lookahead repetitive triple $(\node
  D_1, \node D_2, \node D_3)$ satisfying
  \[
  (\node D_1, \node D_2, \node D_3) < (\node C_1, \node C_2, \node
  C_3),
  \]
  then we refer to $(\node C_1, \node C_2, \node C_3)$ as the
  \defn{active} LA-repetitive triple for \node T. If there is no
  active LA-repetitive triple in \node T, then $\mu \sub A(\node
  T)=\emptyset$; if $(\node C_1, \node C_2, \node C_3)$ is the active
  LA-repetitive triple, then let $\mu \sub A(\node T)=(\node C_1,
  \node C_2, \node C_3)$.
\end{definition} 

We will use one of only two constructions for tree-projection decision
functions in the present work. Let $\rho_0$ be such that $\rho_0(\node
T, \node A, \node B, \node C)=0$ always. Our other construction for a
tree-projection decision function is more complex.  Let $G=(\Sigma,
\Nu, P, S)$ be an LR(1) grammar, and let $\beta B$ be a viable prefix
followed by $a$. Let \node T be a parse-proper tree with yield $\beta
B a$, letting \node B and \node A be the nodes corresponding to $B$
and $a$, respectively.  Let $\pi \in P$, where $\pi = A \to \alpha$,
and let $1 \le n \le |\alpha| + 1$.  Let $\node C_1$, $\node C_2,$ and
$\node C_3$ be three nodes such that $\node C_1$ is ancestral to
$\node C_2$, which is ancestral to $\node C_3$. We define $\rho_{\pi, n}(\node T, \node C_1, \node C_2, \node C_3)$ presently: if
there is a node \node N ancestral to $\node C_2$, such that
\begin{itemize}
\item $\node C_1 < \node N < \node C_2$,

\item $\nodeLabel(\node N)=\pi$, and

\item $H_{\node T}(\node N)=n$,
\end{itemize}
then let $\rho_{\pi, n}(\node T, \node C_1, \node C_2, \node
C_3)=0$; otherwise, let $\rho_{\pi, n}(\node T, \node C_1,
\node C_2, \node C_3)=1$.

\begin{definition}
  Let $G=(\Sigma, \Nu, P, S)$ be an LR(1) grammar, and let $\beta B$
  be a viable prefix followed by $a$. Let \node T be a parse-proper
  tree with yield $\beta B a$, letting \node B and \node A be the
  nodes corresponding to $B$ and $a$, respectively.  Let $\rho$ be a
  tree-projection decision function. We now define two special tree
  triplet surgery operators; let $\Phi_\rho=\Pi_{\rho, \mu \sub B}$
  and let $\Lambda_\rho=\Pi_{\rho, \mu \sub A}$.  Let $p$ and $q$ be
  such that $\Phi_\rho^{p + 1} \node T=\Phi^p \node T$ and
  $\Lambda_\rho^{q + 1} \node T=\Lambda^q \node T$, respectively;
  refer to $p$ and $q$ as the \defn{\PhiLimit} and \defn{\LambdaLimit}
  for \node T of $\Phi_\rho$ and $\Lambda_\rho$, respectively.  Define
  $\Psi_\rho \equiv \Lambda_\rho^q \Phi_\rho^p$, an operator we refer
  to as the \defn{proper projection operator}.
\end{definition}

We will establish the following conventions. Let $G=(\Sigma, \Nu, P,
S)$ be an LR(1) grammar, and let $\beta B$ be a viable prefix followed
by $a$. If \node T is a parse-proper tree and \node P is a node in
\node T, then we will use the symbol $\Psi_{\node T, \node P}$ to mean
the operator $\Psi_{\rho}$, with $\rho=\rho_{\pi, n}$, where
$\pi=\nodeLabel(\node P)$ and $n=H_{\node T}(\node P)$. We use the
symbol $\Psi_0$ to mean the operator $\Psi_{\rho_0}$.

\begin{lemma}
  \label{lemma:Phi T is B-proper}
  Let $G=(\Sigma, \Nu, P, S)$ be an LR(1) grammar, and let $\beta B$
  be a viable prefix followed by $a$. If \node T is a parse-proper
  tree with yield $\beta B a$, then $\Phi_\rho^p \node T$ is proper
  above \node B, where $p$ is the \PhiLimit for \node T.
\end{lemma}
\begin{proof}
  Let $G=(\Sigma, \Nu, P, S)$ be an LR(1) grammar, and let $\beta B$
  be a viable prefix, followed by $a$, for $B \in \Nu$. Let \node T be
  a parse-proper tree with yield $\beta B a$. Let $\rho$ be a
  tree-projection decision function, and let $p$ be the \PhiLimit of
  $\Phi_\rho$ for \node T. Let $(\node V_i)_{i=0}^n$ be the
  prefix-ancestral interstitial sequence for \node B and \node A.

  We proceed by induction on $p$. If $p = 0$, then there are no
  repetitive triples for \node T. There are thus no more than 2
  distinct nodes $\node B_1$ and $\node B_2$ such that $\nodeLabel(\node B_1)=\nodeLabel(\node B_2)$.  Therefore, \node T is proper
  above \node B. Since $\Phi_\rho \node T = \node T$, we conclude that
  $\Phi_\rho \node T$ is proper above \node B.

  Assume that $\Phi_\rho^{p \sub Z} \node Z$ is proper above \node B
  for every tree \node Z with \PhiLimit of $p \sub Z$, for some $p
  \sub Z \ge 0$. Assume also that $p = p \sub Z + 1$. Let $(\node B_1,
  \node B_2, \node B_3)$ be the active repetitive triple for \node
  T. We replace one of these three nodes with one of the remaining
  two; since $\nodeLabel(\node B_1)=\nodeLabel(\node B_2)=\nodeLabel(\node B_3)$, we conclude that $\Phi_\rho \node T$ is parse
  proper. As the \PhiLimit of $\Phi_\rho$ for $\Phi_\rho \node T$ is
  $p - 1$, the induction hypothesis implies that $\Phi_\rho^{p - 1}
  \Phi_\rho \node T$ is proper above \node B.
\end{proof}

\begin{lemma}
  \label{lemma:Lambda T is A-proper}
  Let $G=(\Sigma, \Nu, P, S)$ be an LR(1) grammar, and let $\beta B$
  be a viable prefix followed by $a$. If \node T is a parse-proper
  tree with yield $\beta B a$, then $\Lambda_\rho^q \node T$ is proper
  above \node A, where $q$ is the \LambdaLimit.
\end{lemma}
\begin{proof}
  Let $G=(\Sigma, \Nu, P, S)$ be an LR(1) grammar, and let $\beta B$
  be a viable prefix, followed by $a$, for $B \in \Nu$. Let \node T be
  a parse-proper tree with yield $\beta B a$. Let $\rho$ be a
  tree-projection decision function, and let $q$ be the \LambdaLimit
  of $\Lambda_\rho$ for \node T. Let \node X be the set of all nodes
  ancestral to \node A but not \node B.

  We proceed by induction on $q$. If $q = 0$, then there are no
  LA-repetitive triples for \node T. Thus, there are at most two nodes
  $\node D$ and $\node D'$ in \node X such that $\nodeLabel(\node
  D)=\nodeLabel(\node D')$. Therefore, \node T is proper above \node
  A. Since $\Lambda_\rho \node T = \node T$, we conclude that
  $\Lambda_\rho \node T $ is proper above \node A.

  Assume that, for any tree \node Y with a \LambdaLimit of
  $\Lambda_\rho$ that is $q \sub Y \ge 0$, that we know that
  $\Lambda_\rho^{q \sub Y} \node Y$ is proper above \node A.  Assume
  also that $q=q \sub Y + 1$. Let $(\node C_1, \node C_2, \node C_3)$
  be the active LA-repetitive triple for \node T. We know that
  $\nodeLabel(\node C_1)=\nodeLabel(\node C_2)=\nodeLabel(\node C_3)$,
  and since whichever node is replaced gets replaced by one with the
  same label, we have a parse-proper tree in $\Lambda_\rho \node
  T$. The \LambdaLimit of $\Lambda_\rho$ for $\Lambda_\rho \node T$ is
  $q - 1$. We therefore have, by the induction hypothesis, that
  $\Lambda_\rho^{q - 1} \Lambda_\rho \node T$ is proper above \node A.
\end{proof}

\begin{figure}[t!]
  \begin{center}
    \includegraphics[scale=.75]{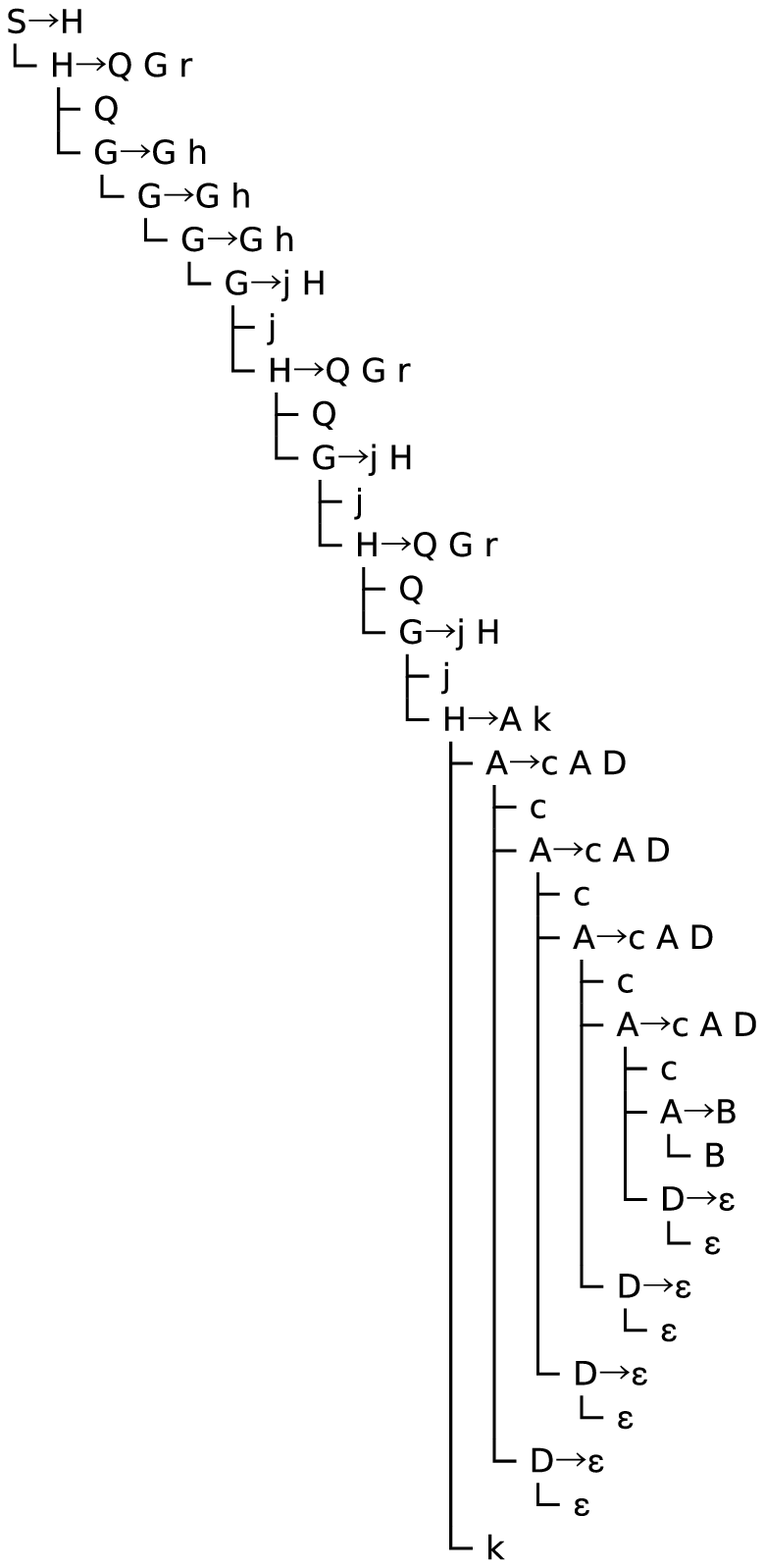}
    \includegraphics[scale=.75]{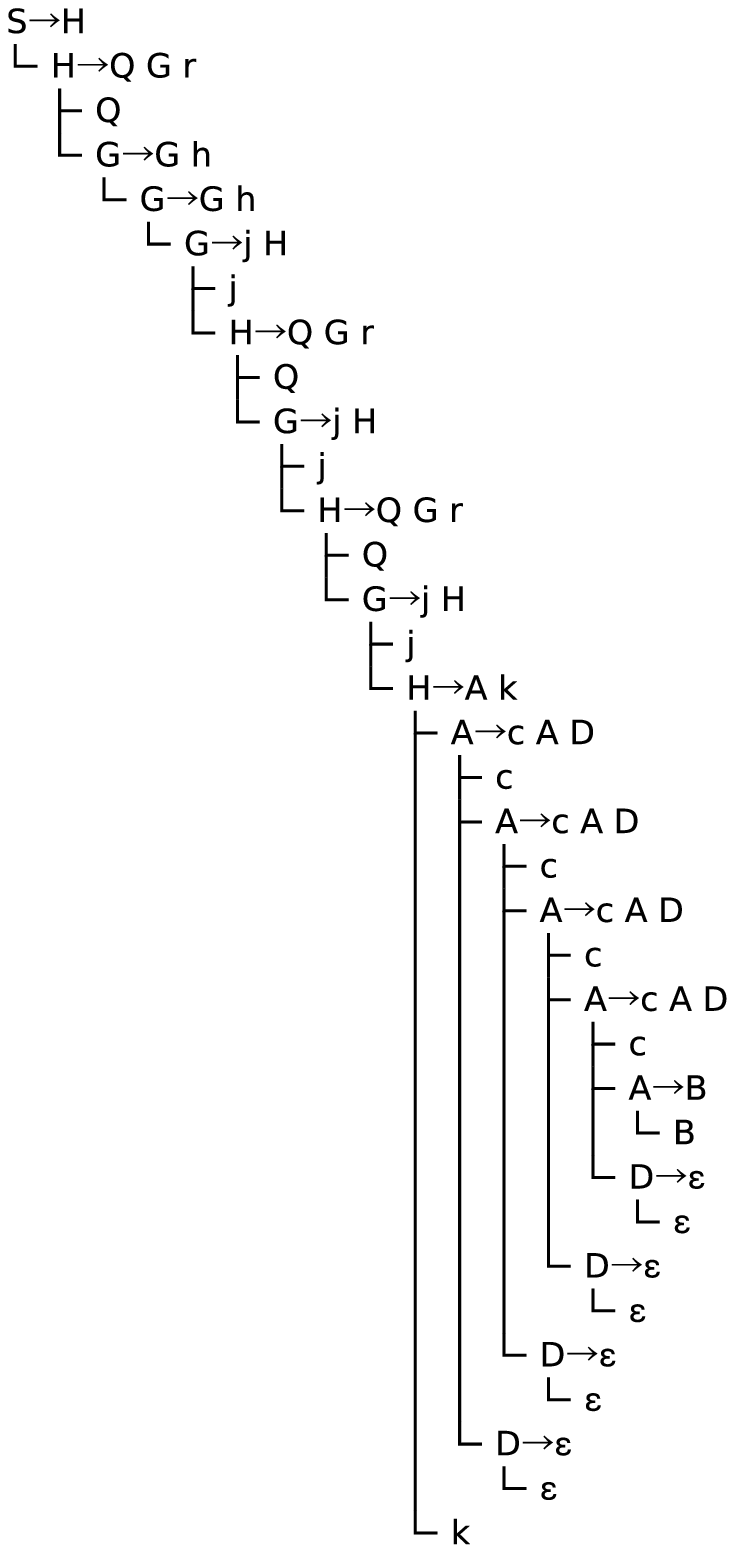}
  \end{center}
  \caption{\label{figure:projection example simplified tree} The
    simplified, and the proper simplified, trees for the viable prefix
    $qjqjqjccccB$, in former case, followed by the suffix $krrhhhr$,
    and in the later case, with lookahead $k$.}
\end{figure}

Let us consider an example. Let $G=(\Sigma, \Nu, P, S)$ be the LR(1)
grammar with 
\begin{align*}
  \Sigma &= \{q, r, h, j, c, b, k\},\\
  \Nu &= \{S, H, G, A, B, D, Q\}\text{, and}\\
  P &= \{ \begin{aligned}[t]
    &\production{S}{H},\\
    &\production{H}{Q G r \alt A k},\\
    &\production{G}{G h \alt G j H},\\
    &\production{A}{c A D \alt B},\\
    &\production{D}{\epsilon},\\
    &\production{B}{b},\\
    &\production{Q}{q}\}.
  \end{aligned}
\end{align*}
We now consider the string $x=qjqjqjccccbkrrhhhr$. It is easily
verified that $x \in \lang{G}$. Let \node T be the parse tree for
$x$. With the order that we have given trees in this work, we can
represent \node T graphically as in \figref{figure:projection example
  parse tree}.
\begin{figure}[t!]
  \begin{center}
    \includegraphics[scale=.75]{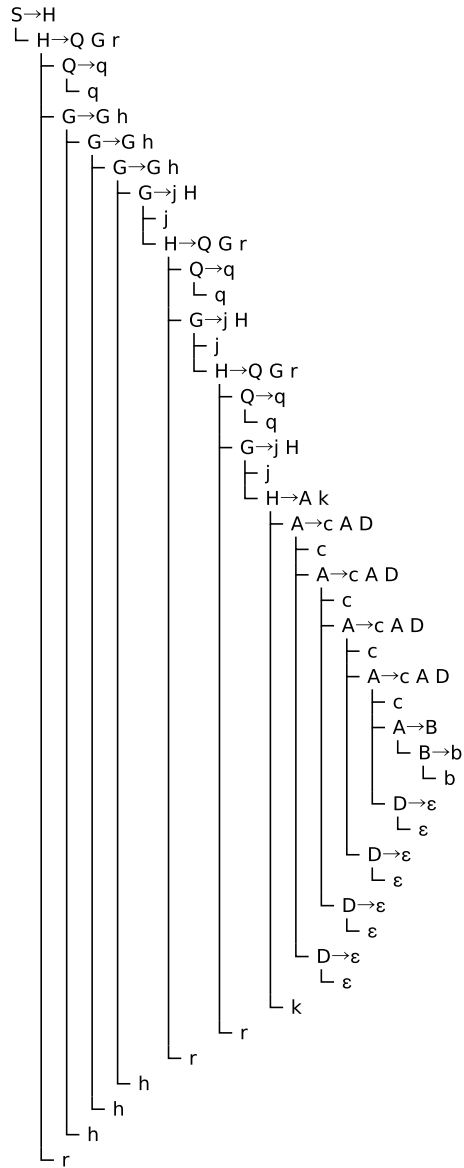}
  \end{center}
  \caption{\label{figure:projection example
      parse tree} The parse tree for the string $qjqjqjccccbkrrhhhr$.}
\end{figure}
We now consider a simplified tree and a proper simplified tree for the
viable prefix $qjqjqjccccB$, when followed by $k$; we have presented
these trees in \figref{figure:projection example simplified tree}.

We care about proper simplified trees because $Z_{\beta a}$ reproduces
the conservation function, yet $T_{\beta a}$ is finite.

\begin{theorem}
  Let $G=(\Sigma,\Nu,P,S,T,M)$. The proper simplified tree set is finite.
\end{theorem}
\begin{proof}
  Let $G=(\Sigma, \Nu, P, S)$ be an LR(1) grammar, and let $\beta B$
  be a viable prefix followed by $a$, for $B \in \Nu$. If will suffice
  to show that there is an upper bound to the height for elements of
  $T_{\beta B a}$.

  Let $\node T \in T_{\beta B a}$.  Let \node B and \node A be nodes
  in \node T corresponding to $B$ and $a$, respectively. Let $(\node
  V_i)_{i=0}^n$ be the prefix-ancestral interstitial sequence for
  \node B and \node A.

  Let $D \in \Nu$ such that
  \begin{equation}
    \label{eq:finite height:D derives epsilon in the longest derivation}
    D \derives^* \epsilon,
  \end{equation}
  where, for no $D' \in \Nu$ with $D' \ne D$ is it the case that $D'
  \derives^* \epsilon$ is a longer derivation than \eqref{eq:finite
    height:D derives epsilon in the longest derivation}. Let $N_D$ be
  the length of \eqref{eq:finite height:D derives epsilon in the
    longest derivation}.

  Let \node X be the greatest node ancestral to both \node B and \node
  A. Let $\node B \sub a$ be the child of \node X autoancestral to \node
  B and let \node Q be those children of \node X greater than $\node B
  \sub a$. The greatest child of \node X is in \node Q; let this child
  be \node {A_Q}.

  Every element in $\node Q \setminus \{ \node {A_Q} \}$ has a height
  not greater than $N_D$. Let $\node Y \sub A$ be the subtree rooted at
  \node {A_Q}. Let $\node X \sub A$ be the set of nodes in $\node Y \sub
  A$ ancestral to \node A. By the condition that \node T is proper above
  \node A, we have that $|\node X \sub A| \le 2 |P|$. It is possible
  that the height of $\node Y \sub A$ could exceed $|\node X \sub A| +
  1$, by not by $N_D$. That is, the height of $\node Y \sub A$ is less
  than $|\node X \sub A| + N_D$.

  Let $\node Z \sub B$ be the subtree with $\node B \sub a$ as its
  root. Let $\gamma$ be the yield of $\node Z \sub B$. Let $\Nu_\gamma
  \subset \Nu$ such that for all $C \in \Nu_\gamma$, we have that $C
  \derives^* \gamma$. Letting $\Nu_\gamma=\{C_1, C_2, \dotsc, C_m\}$,
  let $h_i$ be the height of the parse tree corresponding to the
  derivation $C_i \derives^* \gamma$. Finally, let $M \sub B = \max \{
  h_1, h_2, \dotsc, h_m \}$; clearly, the height of $\node Z \sub B$
  is less than $M \sub B$.

  Let $\delta \in (\Sigma \cup \Nu)^*$ such that $\delta \gamma =
  \beta B$. Within each $\node V_i$, for $0 \le i \le |\delta|$, there
  are not more than $2 |P|$ nodes, by the condition that \node T is
  proper above \node B. Therefore, if we let $\node T \sub S$ be the
  tree formed by removing from \node T all those nodes in \node Q, we
  can conclude that the height of $\node T \sub S$ is not more than
  $2|P| |\delta|$.

  Combining these results, we see that the height of \node T is not
  greater than
  \[
  2|P||\delta| + \max \{M \sub B, |\node X \sub A| + N_D|\}. \qedhere
  \]
\end{proof}

\begin{definition}
  Let $G=(\Sigma, \Nu, P, S, T, M)$.  Let $P_0$ be a production set
  over the terminal alphabet $\Sigma$ and the nonterminal alphabet
  $\Nu' \supset \Nu$. If $\pi=\production{A}{X_1 X_2 \dotso X_m}$ is
  a production in $P$ such that $Y_{\beta a}(\pi) \ne -1$, and either
  \begin{itemize}
  \item $\pi\in P_0$ if $Y_{\beta a}(\pi)=m+1$; or

  \item there is some $\phi\in P_0$ such that $\phi=\production{A}{X_1
      X_2 \dotso X_i S_1 S_2 \dotso S_p}$ where $i=Y_{\beta a}(\pi)$
    if $Y_{\beta a}(\pi)<m+1$;
  \end{itemize}
  then we say that \defn{$P_0$ simply conserves $\beta a$ for $P$}. If
  we replace $Y_{\beta a}$ with $Z_{\beta a}$, then we say that
  \defn{$P_0$ simply conserves $\beta a$ for $P$ properly}.
\end{definition}

\begin{lemma}
  Let the LR(1) grammar $G=(\Sigma,\Nu,P,S)$, the viable prefix
  $\beta$ ending in a nonterminal, and the terminal $a$, which follows
  $\beta$ all be given. Let $\node T\in F_{\beta a}$. For any node
  \node P in \node T, such that $H_{\node T}(\node P)\ne-1$, there
  exists a tree $\node T' \in T_{\beta a}$ such that, for some node
  $\node P'$ in $\node T'$, we have that
  \[
  H_{\node T}(\node P)=H_{\node T'}(\node P').
  \]
\end{lemma}
\begin{proof}
  Let $\beta=\alpha B$, where $B$ is a nonterminal. Let $\node T\in
  F_{\beta a}$, and let \node P in \node T, such that $H_{\node
    T}(\node P)\ne-1$. Let \node B and \node A correspond to $B$ and
  $a$, respectively.

  Let \node U be the set of nodes ancestral to \node B.  Let $\node
  W=(\node M_0,\node M_1,\dotsc,\node M_m)$ be the prefix-ancestral
  sequence for \node B and \node A, and let $( V_i )_{i=0}^n$ be the
  prefix-ancestral interstitial sequence for \node B and \node A.

  Let $\Omega \equiv \Psi_{\node T, \node P}$.  It can easily be
  verified by inspecting the definition of $\Psi_{\node T, \node P}$
  that there is a node $\node P'$ in $\Omega \node T$, corresponding
  to \node P, such that $H_{\Omega\node T}(\node P')=H_{\node T}(\node
  P)$.

  The simplified tree $\Omega \node T$ is proper above \node A and it
  is proper above \node B (Lemmas~\ref{lemma:Phi T is B-proper}
  and~\ref{lemma:Lambda T is A-proper}). Since there is some node
  $\node P \sub p$ in $\Omega \node T$ such that $H_{\Omega \node
    T}(\node P \sub p)=H_{\node T}(\node P)$, we have our conclusion.
\end{proof}

\begin{cor}
  Let the TLR grammar $G=(\Sigma,\Nu,P,S,T,M)$, the viable prefix
  $\beta$, the terminal $a$, which follows $\beta$, and a production set
  $P_0$ all be given. Then $P_0$ simply conserves $\beta a$ for $P$ if
  and only if it does so properly.
\end{cor}

\begin{theorem}
  Let the TLR grammar $G=(\Sigma,\Nu,P,S,T,M)$, the viable prefix
  $\beta$, the terminal $a$, which follows $\beta$, and the grammar
  transformation $\Delta\in\allDeltas G$ be given. Let $\Delta
  G=(\Sigma, \Nu_{\Delta G}, P_{\Delta G}, S, T, M)$. Then
  $\validDelta{\Delta}{G}{\beta a}$ if and only if $P_{\Delta G}$
  simply conserves $\beta a$ for $P$.
\end{theorem}
\begin{proof}
  Let $\validDelta{\Delta}{G}{\beta a}$. Let $x \in \Sigma^*$ be such
  that $\beta a x$ is a sentential form, and let $y \in \Sigma^*$ be
  such that $\beta a x \derives^* y$. Let \node T be the parse tree
  for $y$, and let \node B and \node A be the nodes in \node T
  corresponding to the symbols $B$ and $a$, respectively.

  Let $\pi$ be a free production; that is: $V_{\beta
    a}(\pi)=-1$. Thus, there are no nodes \node P in \node T meeting
  any of the first three criteria from \dref{def:conservation function
    for sentential form}. Therefore, there are no simplified trees
  with a node \node P such that $\nodeLabel(\node P)=\pi$.

  Let $\pi=A \to \alpha$ be a conserved production that is not
  entirely conserved; thus, $V_{\beta a}(\pi)=n$, for $-1 \ne n \le
  |\alpha|$. There is some node \node P ancestral to \node A such that
  $\nodeLabel(\node P)=\pi$, where the $\nth n$ child of \node P is
  autoancestral to \node A; this node does not get removed from \node
  T by $\mu$ (\dref{def:simplified tree}). Therefore, in the tree $
  \mu \node T$, there is a node $\node P_0$ such that $H_{\node T \sub
    s}(\node P_0)=n$. We consider the possibility that $Y_{\beta
    a}(\pi) > n$.

  Assume that $Y_{\beta a}(\pi) = n_0 > n$. That is, assume that there
  is some $\node T_1\in F_{\beta a}$---letting $\node B_1$ and $\node
  A_1$ be the nodes corresponding to the symbols $B$ and $a$,
  respectively---such that there is a node \node F in $\node T_1$ that
  whose $\nth{n_0}$ child is autoancestral to $\node A_1$.  As $\beta$
  is a viable prefix followed by $a$, there exists some
  $x_1\in\Sigma^*$ such that $\beta a x_1$ is a sentential form. Let
  $z\in\Sigma^*$ such that $\beta a x_1\derives^* z$; let $\node T_z$
  be the parse tree for $z$ . There is an obvious injective mapping of
  nodes $q\colon\node T_1\to\node T_z$ that preserves the structure of
  $\node T_1$. The node $q(\node F)$ in $\node T_z$ is such that
  $N_{\beta a x_1, \node T_z}(q(\node P))=n_0$, a contradiction. Thus, $Y_{\beta
    a}(\pi)=n$.

  Let $\phi=\production{C}{\gamma}$ be entirely conserved; since $\phi
  \in P_{\Delta G}$, we have that $P_{\Delta G}$ simply conserves
  $\beta a$ for $P$.  Therefore, the ``If'' direction is proved.

  Let $P_{\Delta G}$ simply conserve $\beta a$ for $P$. Let
  $\pi=\production{A}{\alpha}$ be such that
  \[
  Y_{\beta a}(\pi) \le |\alpha|.
  \]
  Is it possible that $V_{\beta a}(\pi) > Y_{\beta a}(\pi)$? Assume
  one such $x$ exists. Let $y\in\Sigma^*$ such that $\beta
  ax\derives^*y$, and let \node T be the parse tree for $y$. We have
  asumed that there exists some node P in \node T such that
  \[
  N_{\beta ax, \node T }(\node P)=Y_{\beta a}(\pi).
  \]
  But, since $N_{\beta ax, \node T}(\node P)>-1$, it must be the case that
  either:
  \begin{enumerate}
  \item \node P is an ancestor of \node B;

  \item \node P is an ancestor of \node A; or

  \item \node P shares an ancestor with \node A and \node B, but
    is an ancestor of neither, such that $\node B<\node P<\node
    A$.
  \end{enumerate}
  In any of these three cases, we would have that \node P would not be
  removed from \node T by $\mu$. Hence, there is some $\node T_0\in F_{\beta a}$
  such that there exists $\node P_0$ in $\node T_0$ such that
  \[
  H_{\node T_0}(\node P_0)>-1
  \]
  and $\nodeLabel(\node P)=\pi$, a contradiction.

  Let $\pi$ be such that $Y_{\beta a}(\pi)=n+1$. Since $\pi\in P_{\Delta
    G}$, we have that $P_{\Delta G}$ conserves $\beta a$ for $P$.
\end{proof}

\subsection{The Connection Between Parse Paths and Proper Simplified
  Trees}

We are now ready to justify the method of \algoref{algo:compute
  conservation func} as a means of determining if a grammar
transformation is valid. We have modeled the operation of the
Algorithm as an enumeration of parse paths, and we have examined a new
formulation for determining if a transformation is valid. We now show
that the method of \secref{sec:simplified tree test} is just another
way of looking at the operation of the Algorithm, in that each proper
simplified tree corresponds to a parse tree, and visa versa.

\subsubsection{A Mapping of Parse Paths to Proper Simplified Trees}

Let $G$ be a TLR grammar.  We let $\mathbb T_{G}$ be the set of all
trees whose nodes are labeled either with a terminal, nonterminal or
production from $G$.  Let the set of all parse paths for the viable
prefix $\beta$, followed by the terminal $a$ be
$\mathbb P_{\beta B,a,G}$. Additionally, let
\begin{align*}
  \mathbb U_{\beta B,a,G} &= \bigcup_{(U, J, s, D) \in \mathbb P_{\beta
  B,a,G}} U,\\
  \mathbb J_{\beta B,a,G} &= \bigcup_{(U, J, s, D) \in \mathbb P_{\beta
  B,a,G}} J,\\
  \mathbb S_{\beta B,a,G} &= \bigcup_{(U, J, s, D) \in \mathbb P_{\beta
  B,a,G}} s\text{, and}\\
  \mathbb D_{\beta B,a,G} &= \bigcup_{(U, J, s, D) \in \mathbb P_{\beta
  B,a,G}} D.
\end{align*}
If $G$, $\beta$ and $a$ are understood, we omit them. If we are in a
context where $G$, $\beta$ and $a$ are understood, then we use the
symbols $\mathbb U_0$ and $\mathbb U_1$ as synonyms for $\mathbb J$
and $\mathbb U$, respectively.

\begin{definition}
  Let $\node T$, a nonempty tree in $\mathbb T$, be given. Let $\node
  P$ be the greatest leaf in $\node T$; if $\node P$ is labeled with a
  grammar symbol, then let $\node P \sub A$ be the parent of $\node
  P$; if $\node P$ is labeled with a production, then let $\node P
  \sub A=\node P$. Define $\node P \sub A$ to be the \defn{attach
    point for \node T.}
\end{definition}

\begin{definition}
  \label{def:consumption function}
  Let the TLR grammar $G=(\Sigma, \Nu, P, S, T, M)$, and the viable
  prefix $\beta$, followed by the terminal $a$ be given.  Let $k \in
  \{0, 1\}$, and let $x \in \Sigma^*$ such that $|x|=k$.  Define $Q_k
  \colon \mathbb U_k \to \mathbb T$ as follows: the value of
  $Q_k(U)$---letting $U=(u_1, u_2, \dotsc, u_n)$, and letting
  $u_n=[\production{A}{\alpha \cdot \gamma}, x]$---is given by one of
  the following \ref{item:consumption function:last} cases.
  \begin{enumerate}
  \item 
    \label{item:consumption function:|U|=1}
    Assume that $|U|=1$. In this case, we must have, by
    \ruleref{item:k-parse precession:first item has dot at beginning}
    of \dref{def:k-parse precession}, that $\alpha=\epsilon$. Let $Q_k
    (U) = \node T \sub i$, where $\node T \sub i$ is the one-node tree
    whose node is labeled $\production {A}{\gamma}$.

  \item 
   % \label{item:consumption function:|U|>1 and alpha != epsilon}
    Assume that $|U|>1$, and that $\alpha \ne \epsilon$. Let
    $\alpha=\alpha' X$ for some $X \in (\Sigma \cup \Nu)$.  In this
    case, we let $\node T= Q_k(u_1, u_2, \dotsc, u_{n-1})$, and we let
    $\node P$ be the attach point for $\node T$.  We let $\node T'$ be
    that tree formed from $\node T$ by adding a child labeled $X$ to
    the children of $\node P$, such that this new node is greatest
    child of $\node P$ in $\node T$; let $Q_k(U) = \node T'$.

  \item
   % \label{item:consumption function:|U|>1 and alpha = epsilon}
    \label{item:consumption function:last}
    Assume that $|U|>1$, and that $\alpha = \epsilon$. In this case,
    we let $\node V= Q_k(u_1, u_2, \dotsc, u_{n-1})$, and we let
    $\node Q$ be the attach point for $\node V$.  We let $\node V'$ be
    that tree formed from $\node V$ by adding a child labeled
    $\production{A}{\gamma}$ to the children of $\node Q$, such that
    this new node is greatest child of $\node Q$ in $\node V'$; let
    $Q_k(U) = \node V'$.
  \end{enumerate}
  Call $Q_k$ the \defn{$k$\dash{}consumption function.}
\end{definition}

\begin{definition}
  Call the 1\dash{}consumption function  the \defn{upward link
    conversion function}; denote this function $Q \sub U$. Call the
  0\dash{}consumption function the \defn{ancestral link
    conversion function}; denote this function $Q \sub A$.
\end{definition}

\begin{definition}
  \label{def:downward link conversion function}
  Let the TLR grammar $G$, and the viable prefix $\beta$, which is
  followed by $a$, be given. Let $\mathbb T^*$ be the set of sequences
  of elements from $\mathbb T$. We will define $Q \sub D \colon
  \mathbb D \to \mathbb T^*$. Let $D \in \mathbb D$ be $(L_1, L_2,
  \dotsc, L_n)$. The value of $Q \sub D (D)$ is given by one of
  \ref{item:downward link conversion function:last} cases.
  \begin{enumerate}
  \item 
    \label{item:downward link conversion function:terminal link}
    Assume that $n=1$. In this case, let $L_1=(X_1 X_2 \dotso X_p,
    k)$. For $1 \le i < k$, let $\node V_i$ be the parse tree for the
    derivation $X_i \derives^* \epsilon$; let $\node V_k$ be the
    single-node tree whose node is labeled $X_k$. 

  \item
    \label{item:downward link conversion function:last}
    Assume that $n > 1$. In this case, let $L_1=(X_1 X_2 \dotso X_p,
    k)$. For $1 \le i < k$, let $\node V_i$ be the parse tree for the
    derivation $X_i \derives^* \epsilon$. Let $Q \sub D(L_2, L_3,
    \dotsc, L_n)=(\node Y_1, \node Y_2, \dotsc, \node Y_h)$. Let
    $\node V_k$ be the tree whose root:
    \begin{itemize}
    \item is labeled $P(L_1, L_2)$, and
    \item has $\node Y_1, \node Y_2, \dotsc, \node Y_h$ as its
      children.
    \end{itemize}
  \end{enumerate}
  Define $Q \sub D(D)=(\node V_1, \node V_2, \dotsc, \node V_k)$.
  Call $Q \sub D$ the \defn{downward link conversion function}.
\end{definition}

\begin{definition}
  If $G$ is a context-free grammar, and \node T is a parse-proper tree
  such that, for every interior node \node N, we have that $\childString(\node N)=\properChildString(\node N)$, then we say that \node T is
  \defn{parse-complete.}
\end{definition}

\begin{prop}
  A parse tree is parse-complete.
\end{prop}

\begin{definition}
  If $G=(\Sigma, \Nu, P, S)$ is an LR(1) grammar, and \node T is
  a parse-proper tree then we define a function $F \sub U\colon
  \mathbb T\to\mathbb T$, where $F \sub U(\node T)$ is given by tthe
  following: for every interior node $\node N$ such that $\properChildString(\node N) \ne \childString(\node N)$, we let $A_1, A_2, \dotsc, A_n
  \in \Nu^*$ be such that $\properChildString(\node N) = \childString(\node N)
  A_1 A_2 \dotso A_n$, and we make the trees $\node V_1,\node
  V_2,\dotsc, \node V_n$ be new children of \node N (in order), where
  for $1 \le i \le n$, the tree $\node V_i$ is the parse tree for the
  derivation $A_i \derives^* \epsilon$.
\end{definition}

\begin{prop}
  \label{prop:F makes a complete tree}
  If $U$ is an upward link, then $F \sub U(Q \sub U(U))$ is
  parse-complete.
\end{prop}

\begin{definition}
  \label{def:parse path conversion function}
  Let $(U, J, s, D)$ be a parse path for the viable prefix $\beta B$
  followed by $a$. We will define $R\colon \mathbb P \to \mathbb T$.
  The value of $R(U, J, s, D)$ is as follows. Let $\node T \sub U=F
  \sub U(Q \sub U(U))$, let $\node Y=Q \sub D(D)$, and let $\node T
  \sub J=Q \sub A(J)$.  There are then two cases to consider.
  \begin{enumerate}
  \item 
    \label{item:parse path conversion function:|I|=0}
    If we have that $|U|=0$, then let $\node T_0$ be the tree formed
    by:
    \begin{enumerate}
    \item 
      \label{item:add B for the sidelink}
      attaching a node labeled $B$ to the tree as the greatest child
      of the attach point for $\node T \sub J$; and then

    \item attaching each tree from $\node Y$, in order, following the
      node added in \ref{item:add B for the sidelink}, to the attach
      point for $\node T \sub J$.
    \end{enumerate}

  \item 
    \label{item:parse path conversion function:|I|>0}
    If we have that $|U|>0$, then let $\node T_0$ be the tree formed
    by:
    \begin{enumerate}
    \item 
      \label{item:add Z for the sidelink}
      attaching the root node of $\node T \sub U$ to the tree as the
      greatest child of the attach point for $\node T \sub J$; and
      then

    \item attaching the root node of each tree from $\node Y$, in
      order, following the node added in \ref{item:add Z for the
        sidelink}, to the attach point for $\node T \sub J$.
    \end{enumerate}
  \end{enumerate}
  Define $R(I,J,s,D)=\node T_0$; we call $R$ the \defn{parse-path
    conversion function}.
\end{definition}

\begin{lemma}
  If $(U, J, s, D)$ is a parse-path, then $R(U, J, s, D)$ is
  parse-proper.
\end{lemma}
\begin{proof}
  Let $k \in \{0, 1\}$, and let $x \in \Sigma^*$ such that $|x|=k$.
  Let $P_k \in \mathbb I_k$, where $P_k=(p_1, p_2, \dotsc, p_n)$. We
  proceed by induction on $n$. Since there are no interior nodes in
  $Q_k(P_k)$ when $|P_k| < 2$, we can use $|P_k|=2$ as our basis
  step. The first two items of $P_k$ are of the form
  \begin{align*}\ 
    &[\production {A_1} {\cdot A_2 \alpha_1}, x]\text{, and}\\
    &[\production {A_2} {\cdot \alpha_2},  x],
  \end{align*}  
  respectively. Let $\node T \sub Q=Q_k(P_k)$. Since the only child of
  the root, which is the only interior node, is labeled $\production
  {A_2} {\alpha_2}$, the child-string of the root is $A_2$, which is a
  prefix of $A_2 \alpha_1$.

  Let $P_k' \in \mathbb I_k$ be a parse-precession, such that
  $|P_k'|\ge 2$. Assume that we know that $Q_k(P_k)$ is a parse-proper
  tree; we now assume that $|P_x|=|P_k'|+1$. Let $\node T \sub P$ be
  the tree obtained after $|P_k'|$ applications of $Q_k$. There are
  two possibilities for $p_n$.
  \begin{enumerate}
  \item Assume that $p_n$ is of the form $[\production {A} {\alpha X
      \cdot \gamma}, x]$. In this case, the attach point $\node P \sub
    A$ will be labeled $\production {A} {\alpha X \gamma}$. The
    child-string of $\node P \sub A$ will be $\alpha$. The final
    application of $Q \sub k$ will attach a node labeled $X$ as the
    greatest child of $\node P \sub A$, forming the tree $\node
    T_0$. The parent of this new node---the counterpart of $\node P_A$
    in $\node T_0$---has the child-string $\alpha X$, which is a
    prefix of $\alpha X \gamma$.

  \item Assume that $p_n$ is of the form $[\production {C} {\cdot
      \delta}, x]$. In this case, the attach point of $\node T \sub P$
    will be labeled $[\production {D \sub A} {\zeta \sub A \cdot C
      \eta \sub A}, x]$. The child-string of $\node P \sub A$ is
    $\zeta \sub A$. After the final application of $Q \sub k$, the
    child-string will be $\zeta\sub{A} C$, which is a prefix of $\zeta
    \sub{A} C \eta \sub A$, as required.
  \end{enumerate}

  Thus, we have established that $Q \sub U(U)$ and $Q \sub A(J)$ are
  parse-proper trees.

  We turn now to the downward link. Let $D=(L_1, L_2, \dotsc,
  L_p)$. Assume that $p=1$. In this case, the value of $Q \sub D$ is a
  sequence of trees $\node Y=(\node T_1, \node T_2, \dotsc, \node
  T_m)$. The trees $\node T_1, \node T_2, \dotsc, \node T_{m-1}$ are
  all parse trees, so they are all parse-proper. In this case, the
  tree $\node T_m$ will be a single-node tree, hence, it is trivially
  parse-proper.

  Assume now that all elements of $Q \sub D(D')$ are parse-proper if
  $|D'| \ge 1$. Assume also that $p=|D'| + 1$. After $p - 1$
  applications of $Q \sub D$, we have a downward link
  $(\theta,k_\theta)$ and a sequence $\node Y=(\node V_1, \node V_2,
  \dotsc, \node V_q)$. By the induction hypothesis, each element of
  $\node Y$ is a parse-proper tree. The penultimate  application of $Q
  \sub D$ will result in a sequence of $k_\theta$ trees: the first
  $k_\theta - 1$ trees are parse-proper, as they are parse trees. We
  know that $P(L_1,L_2)$ is a production, where
  \[
  P(L_1, L_2)=\production {A \sub } {\nodeLabel(\treeRoot(\node V_1))
    \nodeLabel(\treeRoot(\node V_2)) \dotso \nodeLabel(\treeRoot(\node V_{k_\theta})) X_1 X_2 \dotso X_{n_\theta}}.
  \]
  The next application of $Q \sub D$ will result in a sequence of trees,
  all but the last of which are clearly parse trees; let the last tree
  in this sequence be $\node V'$. Since the elements of $\node Y$ are
  added in order, we have, letting the children of the root of $\node
  V'$ be labeled $\node C_1, \node C_2, \dotsc, \node C_{k_\theta}$,
  we can see that 
  \begin{align*}
    &\nodeLabel(\node C_1)=\nodeLabel(\treeRoot(\node V_1)),\\
    &\nodeLabel(\node C_2)=\nodeLabel(\treeRoot(\node V_2)),\\
    &\text{etc.}
  \end{align*}
  Since this is a prefix of the right side of $P(L_1, L_2)$, we have
  shown that every element of $Q \sub D(D)$ is parse-proper. 

  There are two ways that three trees $Q \sub A(J)$ and $Q \sub U(U)$
  are combined with the trees in $Q \sub D(D)$ to form the tree $R(U,
  J, s, D)$. Let $s=[\production {A_\sigma} {\alpha_\sigma X_\sigma
    \cdot \gamma_\sigma}]$.
  \begin{enumerate}
  \item Assume that $|U|=0$. In this case, we know that
    $X_\sigma=B$. By \ruleref{item:add B for the sidelink} of
    \dref{def:parse path conversion function}, we add a node labeled
    $B$ to the attach point.

  \item Assume instead that $|U| > 0$. In this case, we know that $u_1
    = [\production {X_\sigma} {\cdot \kappa_1}, a]$. Thus, $\treeRoot(Q \sub U(U)) = X_\sigma$. By \ruleref{item:add Z for the
      sidelink} of \dref{def:parse path conversion function}, we add a
    tree---namely, the tree $Q \sub U(U)$---whose root is labeled
    $\production {X_\sigma} {\kappa_1}$.
  \end{enumerate}

  The attach point $\node P_{0, \mathrm A}$ of $Q \sub A(J)$ is labeled $\production
  {A_\sigma} {\alpha_\sigma X_\sigma \gamma_\sigma}$; since $Q \sub A(J)$
  is parse-proper, the child-string of the attach point must be
  $\alpha_\sigma$. So after the application of either
  \ruleref{item:add B for the sidelink} or \ruleref{item:add Z for the
    sidelink}, whichever is appropriate, we have shown that the child
  string $\node P_{0, \mathrm A}$ now has the child-string $\alpha_\sigma
  X_\sigma$.

  Now, $L_1$ is a link, either partial-downward or terminal, for
  $\gamma_\sigma$. We can write
  \[
  \gamma_\sigma = X_1 X_2 \dotso X_{k_\theta} X_{k_\theta + 1} \dotso
  X_{k_\sigma}.
  \]
  After the application of \ruleref{item:add B for the sidelink} or
  \ruleref{item:add Z for the sidelink}, as appropriate, we add the
  elements of $Q \sub D(D)$ to the attach point. Clearly,
  \begin{gather*}
    X_1=\nodeLabel(\treeRoot(\node V_1)),\\
    X_2=\nodeLabel(\treeRoot(\node V_2)),\\
    \vdots\\
    X_{k_\theta - 1}=\nodeLabel(\treeRoot(\node V_{k_\theta - 1}));
  \end{gather*}
  but what of $X_{k_\theta}$ and $\node V_{k_\theta}$? If $L_1$ is a
  terminal link, then $\node V_{k_\theta}$ is a one-node tree whose
  node is labeled $X_{k_\theta}=a$. Otherwise, the root of $\node
  V_{k_\theta}$ is labeled $P(L_1, L_2)$, which is $\production {X}
  {\lambda}$. In either case, $\nodeLabel(\treeRoot(\node
  V_{k_\theta}))=X_{k_\theta}$.
\end{proof}

\begin{definition}
  Let the TLR grammar $G$, and the parse-proper tree $\node T$ be
  given. Let $\node L=(\node K_1, \node K_2, \dotsc, \node K_n)$ be
  the set of leaves, such that $\node K_1 < \node K_2 < \dotsb < \node
  K_n$. If $\nodeLabel(\node K_i) \in \Sigma \cup \Nu \cup \{ \epsilon
  \}$ for all $1 \le i \le n$, then we let
  \[
  \alpha=\nodeLabel(\node K_1) \nodeLabel(\node K_2) \dotso \nodeLabel
  (\node K_n).
  \]
  We call $\alpha$ the \defn{yield} of $\node T$.
\end{definition}

\begin{lemma}
  \label{lem:the yield of R is beta B a}
  The yield of $R(U, J, s, D)$ is $\beta B a$.
\end{lemma}
\begin{proof}
  Let $Q \sub D(D)=(\node V_1, \node V_2, \dotsc, \node V_{n \sub
    d})$. The yield of any of the trees $\node V_1, \node V_2, \dotsc,
  \node V_{n \sub d - 1}$ is clearly $\epsilon$. The yield of $\node
  V_{n \sub d}$ is clearly $a$.

  Let $k \in \{0, 1\}$, and let $x \in \Sigma^*$ such that $|x|=k$.

  Let $P_k=(p_1, p_2, \dotsc, p_n)$ be a \kparse{k} precession. We
  note that the only way a grammar symbol leaf is added to the tree
  $\node T=Q_k(P_k)$ is when we evaluate $Q_k$ on an item of the form
  \begin{equation}
    \label{eq:form of item that induces Q_k to add a symbol node}
    [\production {A} {\alpha X \cdot \beta}, x]
  \end{equation}
  If $p_{i_1}, p_{i_2}, \dotsc, p_{i_m}$ is the set of all items of
  the form \eqref{eq:form of item that induces Q_k to add a symbol
    node}, such that $i_1 < i_2 < \dotsb < i_m$, where
  \[
  p_{i_h} = [\production {A_h} {\alpha_h X_h \cdot \beta_h}, x],
  \]
  for $1 \le h \le m$, then the yield of $Q_k(P_k)$ is evidently the
  string $X_1 X_2 \dotso X_m$. But this is just the trace of
  $P_k$. Therefore, the concatenation of the yield of $Q \sub A(J)$
  and $Q \sub U(U)$ is $\beta B$, by \tref{thm:the trace of (U, J, s)
    is beta B}.

  The yield of $R(U, J, s, D)$ is clearly the yield of $Q \sub A(J)$,
  $Q \sub U(U)$, and $\node V_{n \sub d}$, concatenated, but this is
  just $\beta B a$.
\end{proof}

\begin{lemma}
  If $R(U, J, s, D)$ is a parse path, then $R(U, J, s, D)$ is a
  simplified tree.
\end{lemma}
\begin{proof}
  We wish to find a sentence $y \in \lang G$ such that, letting \node
  T be the parse tree for $y$, we have that $\Psi_0 \node T=R(U, J, s,
  D)$.

  Let $\mathscr N$ be the set of all leaves of $\Psi_0 \node T$ that
  are labeled with a nonterminal. For any nonterminal $A$, let $\node
  F_A$ be the parse tree for the derivation
  \[
  A \derives^* z_A,
  \]
  for some $z_A \in \Sigma^*$.  Let $\node T_1$ be that tree formed
  from $\node T$ by replacing every node $\node N \in \mathscr N$ with
  the tree $\node F_{\nodeLabel(\node N)}$.

  Let $\mathscr D$ be the set of all production nodes $\node J_0$ in
  $\node T_1$ such that the child-string of $\childString (\node J_0)
  \ne \properChildString(\node J_0)$. For any such node $\node J_0$, let
  $\properChildString(\node J_0)=\alpha_0 Z_1 Z_2 \dotso Z_m$, where
  $\alpha_0=\childString(\node J_0)$; let $\node Z_{\node J_0}$ be the
  sequence of trees $(\node F_{Z_1}, \node F_{Z_2}, \dotsc, \node
  F_{Z_m})$. Let $\node T_2$ be the tree formed by, for each $\node J
  \sub f \in \node D$, adding the elements of $\node Z_{\node J \sub
    f}$ as children of $\node J \sub f$, in order.

  Let $z$ be the yield of $\node T_2$. Clearly, the parse tree of $z$
  is $\node T_2$. Recalling the simple-tree projection operator $\mu$,
  consider the tree $\mu \node T_2$. We will examine the changes to
  $\node T_2$ made by $\mu$.

  For any node $\node N \sub z$ such that $\node N \sub z$ corresponds
  to one of the symbols in $\beta$, we note that $\node N \sub z$
  would be removed and replaced with a node labeled $\nodeLabel(\node
  N \sub z)$ by $\mu$. We would also remove any node $\node M \sub z$
  that is not an ancestor of any node representing one of the symbols
  in $\beta a$, provided that $\node M \sub z > \node A$, where $\node
  A$ is the node representing $a$.

  We consider the conditions under which we will add nodes to $\node
  T_1$ in the construction of $\node T_2$. Let $\node H \sub x$ be a
  node in $\node T_1$, such that we add children to it when
  constructing $\node T_2$.  There are two cases:
  \begin{enumerate}
  \item Assume that $\node H \sub z$ is some node in $Q \sub A(J)$. In
    this case, we note that in $Q \sub A(J)$---letting $\node H \sub
    J$ be the node corresponding to $H \sub z$ in $Q \sub A(J)$---the
    greatest child of $\node H \sub J$ is an ancestor of the greatest
    tree in $Q \sub D(D)$. Let this child be the $\nth{i}$ child of
    $\node H \sub J$ in. Clearly, the simplified tree must retain the
    first $i$ children.

  \item Assume that $\node H \sub z$ corresponds to some node in one
    of the trees in $Q \sub D(D)$. We must have that $\node H \sub z$
    corresponds to a node in the final element of $Q \sub D(D)$. Let
    this node be $\node H \sub D$. Let $\node A \sub D$ be the node in
    the final element of $Q \sub D(D)$ that is labeled $a$. We can
    easily show the following: ``if $\properChildString(\node H \sub D) \ne
    \childString(\node H \sub D)$, then the greatest child of $\node H
    \sub D$ is autoancestral to $\node A \sub D$.''  Thus, any nodes
    added as children of $\node H \sub D$ during the construction of
    $\node T_2$ will be removed when $\mu$ is applied to $\node T_2$.
  \end{enumerate}

  Can any node be removed from the part of $\node T_2$ corresponding
  to $Q \sub U(U)$? By \pref{prop:F makes a complete tree}, we have
  that there are no production nodes in $F \sub U(Q \sub U(U))$ that
  would have children removed.

  By \lref{lem:the yield of R is beta B a}, we can say that the only
  nodes which will be replaced during the construction of $\node T_1$
  from $\node T$ are the nodes corresponding to symbols in $\beta$
  which are labeled with nonterminals. Let $\node N_\beta$ be one of
  the nonterminal nodes that is replaced by a production nodes. The
  head of the production will be $\nodeLabel(\node N_\beta)$, and when
  we apply $\mu$ to $\node T_2$, we will this replacement node with a
  node labeled $\nodeLabel(\node N_\beta)$: that is, we will revert to
  $\node N_\beta$.

  So, we have examined all nodes and changes made applying $\mu$ to
  $\node T_2$, and we have arrived back at $\node T$. Therefore, $R(U,
  J, s, D)$ is the simplified tree for $\beta a$.
\end{proof}

\begin{theorem}
  If $(U, J, s, D)$ is a parse path, then $R(U, J, s, D)$ is a proper
  simplified tree. 
\end{theorem}
\begin{proof}
  Let $G=(\Sigma, \Nu, P, S)$ be an LR(1) grammar, and let $\beta
  B$ be a viable prefix, followed by $a$, where $B \in \Nu$. Let $(U,
  J, s, D) \in \mathbb P_{\beta B, a, G}$, and let $\node T=R(U, J, s,
  D)$.  Let $\node A$ and $\node B$ be the nodes in $\node T$
  corresponding to the symbols $B$ and $a$.

  There are two ways that $\node T$ may fail to be proper: it may be
  improper above either $\node A$ or $\node B$.

  Assume the former, for the sake of a contradiction. Let $D=(L_1,
  L_2, \dotsc, L_n)$.  This means that there are three distinct nodes
  $\node P \sub {A,1}$, $\node P \sub {A,2}$, and $\node P \sub {A,3}$
  in \node T, each of which are ancestral to $\node A$ but not $\node
  B$, such that $\nodeLabel(\node P\sub {A,1})=\nodeLabel(\node P\sub
  {A,2})=\nodeLabel(\node P\sub {A,3})$. Assume, without loss of
  generality, that $\node P \sub {A,1} < \node P \sub {A,2} < \node P
  \sub {A,3}$. There are thus three partial downward links $L_{i_1}$,
  $L_{i_2}$, and $L_{i_3}$, where $i_1 < i_2 < i_3$, corresponding to
  $\node P\sub {A,1}$, $\node P\sub {A,2}$, and $\node P \sub {A,3}$,
  respectively.  Now,
  \begin{align*}
    \nodeLabel(\node P \sub {A,1})&=P(L_{i_1}, L_{i_1 + 1}),\\
    \nodeLabel(\node P \sub {A,2})&=P(L_{i_2}, L_{i_2 + 1})\text{, and}\\
    \nodeLabel(\node P \sub {A,3})&=P(L_{i_3}, L_{i_3 + 1}).
  \end{align*}
  However, the fact that $P(L_{i_1}, L_{i_1 + 1})=P(L_{i_2}, L_{i_2 +
    1})=P(L_{i_3}, L_{i_3 + 1})$ is in contradiction with
  \dref{def:complete downward link to a}. Therefore, $\node T$ is
  proper above $\node A$.

  We now consider the latter of the two ways in which \node T may fail
  to be proper: that is, assume that \node T is not proper above \node
  B.

  We know, from the previous two Lemmas, that $\node T$ is a
  simplified tree for $\beta B a$.  Let \node W and $(\node
  V_i)_{i=0}^q$ be the prefix-ancestral set and prefix-ancestral
  interstitial sequence, respectively.

  We have assumed that there is some $i$ such that, within $\node
  V_i$, there are three nodes $\node C_1$, $\node C_2$, and $\node
  C_3$ such that
  \begin{gather*}
    \node C_1 < \node C_2 < \node C_3 \text{, and}\\
    \nodeLabel(\node C_1) = \nodeLabel(\node C_2) = \nodeLabel(\node
    C_3) .
  \end{gather*}
  Depending on whether or not the nodes in $\node V_i$ are ancestral
  to \node A or not, we define $k$, $x$, and $I$ as follows:
  \begin{itemize}
  \item if the nodes in $\node V_i$ are ancestral to \node A, then let
    $k=0$, $x=\epsilon$, and $I=J$;
    
  \item otherwise, the nodes in $\node V_i$ are not ancestral to \node
    A, in which case we let $k=1$, $x=a$, and $I=U$.
  \end{itemize}
  Either way, let $I=(r_1, r_2, \dotsc, r_m)$.

  Note that $\node C_1$ has no child labeled with a grammar symbol;
  thus it was created when $Q_k$ was applied to an item of the form
  $[\production {A_1} {\cdot \alpha_1}, x]$; likewise for $\node C_2$
  and $\node C_3$. The three nodes $\node C_1$, $\node C_2$, and
  $\node C_3$ thus correspond to three items $r_{t_1}$, $r_{t_2}$, and
  $r_{t_3}$, respectively. When $t_1 < q \sub t< t_3$, we have that
  $r_{q \sub t} \connectsBack{k} r_{q \sub t + 1}$, as required by
  \condref{item:k-parse precession:no repeats} of \dref{def:k-parse
    precession}, yet we have that $r_{t_1}=r_{q \sub
    t}=r_{t_2}$ when $q \sub t=t_2$, in violation of that Condition.

  Therefore, we have shown that $\node T$ is proper above $\node B$
  because the argument of the last paragraph applies to $Q \sub U(U)$
  and $Q \sub A(J)$. Since $\node T$ is proper above $\node A$, it is
  in fact a proper simplified tree.
\end{proof}

\subsubsection{The Interchangeability of Parse Trees and Parse Paths}

\begin{theorem}
  There is a bijection between the set of all parse paths and the proper
  simplified tree set for $\beta a$. 
\end{theorem}
\begin{proof}
  Let $G=(\Sigma, \Nu, P, S, T, M)$ be a TLR grammar, let $\beta \in
  (\Sigma \cup \Nu)^*$ and $B \in \Nu$ be such that $\beta B$ is a
  viable prefix followed by $a$.

  The bijection is $R$. We first begin by proving that $R$ is
  surjective.

  Let $\node T \sub s$ be a proper simplified tree for $\beta B$
  followed by $a$. We let $\node A$ and $\node B$ be the nodes
  corresponding to $a$ and $B$, respectively. Let $\node S$ be the
  greatest node in $\node T \sub s$ such that $\node S$ is an ancestor
  of both $\node B$ and $\node A$. Let $\node {R_B}$ be that child of
  $\node S$ that is autoancestral to $\node B$. Let $\node T \sub J$ be
  that tree formed by removing $\node {R_B}$ from $\node T \sub s$, along
  with the other children of $\node S$ that are greater that $\node
  {R_B}$. Let $\node T' \sub U$ be the subtree rooted at $\node
  {R_B}$. Finally, letting $\node V_1, \node V_2, \dotsc, \node
  V_{n \sub v}$ be those children of $\node S$ greater than $\node {R_B}$
  we let $\node Y=(\node V_1, \node V_2, \dotsc, \node V_{n \sub v})$.

  Let $\node T \sub U$ be that tree formed from $\node T \sub U'$ by
  removing all nodes $\node N_\epsilon$ if $\node N_\epsilon$ is the
  root of a subtree with $\epsilon$-yield.

  Let $\alpha \sub D$ be the right side of $\nodeLabel(\node S)$. We
  define a function $Q \sub D\inv\colon (\mathbb T^* \times (\Sigma
  \cup \Nu)^*) \to \mathbb D $ now. The value of $Q \sub D\inv( \node
  Y \sub D, \alpha \sub D)$ we give now, letting $\node Y \sub
  D=(\node V \sub {D,1}, \node V \sub {D,2}, \dotsc, \node V_{\mathrm
    D,n \sub D})$.
  \begin{enumerate}
  \item If $\node V_{\mathrm D,n \sub D}$ has but a single node, then
    let $L$ be a terminal link from $\alpha \sub D$ to $a$, with the
    value $(\alpha \sub D, n \sub D)$.  In this case, define $Q \sub
    D\inv(\node Y \sub D, \alpha \sub D) = (L)$.

  \item If $\node V_{\mathrm D,n \sub D}$ has multiple nodes, then we
    first let $L'$ be a partial downward link from $\alpha \sub D$ to
    $a$, with the value $(\alpha \sub D, n \sub D)$. Next, we label
    the children of the root of $\node V_{\mathrm D, n \sub D}$ as
    follows (in order): $\node W_1, \node W_2, \dotsc, \node W_{m \sub
      D}$, and we label the right side of $\nodeLabel(\treeRoot(\node
    V_{\mathrm D,n \sub D}))$ as $\alpha'$. Let $(L_1, L_2, \dotsc,
    L_{p \sub D})=Q \sub D\inv((\node W_1, \node W_2, \dotsc, \node
    W_{m \sub D}), \alpha')$. Define $Q \sub D\inv(\node Y \sub D,
    \alpha \sub D) = (L', L_1, L_2, \dotsc, L_{p \sub D})$.
  \end{enumerate}

  We will pause and establish an intermediate result. Let $\node Y \in
  \mathbb T^*$ be such that, should we let $\node Y=(\node V_1, \node
  V_2, \dotsc, \node V_{m_I})$, each of $\node V_1, \node V_2, \dotsc,
  \node V_{m \sub I-1}$ is a parse tree with $\epsilon$-yield and
  $\node V_{m \sub I}$ is a parse-proper tree whose yield is $a$;
  furthermore,
  \[
  \nodeLabel(\treeRoot(\node V_1)) \nodeLabel(\treeRoot(\node V_2))
  \dotso \nodeLabel(\treeRoot(\node V_{m \sub I}))
  \]
  is a prefix of $\alpha$. We will show, by induction on the height of
  $\node V_{m \sub I}$, that
  \begin{equation}
    \label{eq:claim:Q_D(Q_D\inv(Y, alpha))=Y}
    Q \sub D(Q \sub D\inv(\node Y, \alpha))=\node Y.
  \end{equation}

  Let the height of $\node V_{m \sub I} \equiv h \sub V$. Assume that
  $h \sub V = 1$. In this case, $\node V_{m \sub I}$ has only a single
  node, and that is labeled $a$. Thus, the value of $Q \sub
  D\inv(\node Y, \alpha)$ is the terminal link $(\alpha, m \sub I)$.
  We have, by \ruleref{item:downward link conversion function:terminal
    link} of \dref{def:downward link conversion function}, that $Q
  \sub D(Q \sub D\inv(\node Y, \alpha)=\node Y'$, where we are letting
  $\node Y'=(\node Q_1, \node Q_2, \dotsc, \node Q_{m \sub
    I'})$. However, we note that $\left \lvert \node Q_{m \sub I'}
  \right \rvert=m \sub I$, and that the first $m \sub I - 1$ elements
  of $\node Y'$ are parse trees for the derivations
  \begin{gather*}
    \nodeLabel(\treeRoot(\node V_1))  \derives^* \epsilon,\\
    \nodeLabel(\treeRoot(\node V_2)) \derives^* \epsilon,\\
    \vdots\\
    \nodeLabel(\treeRoot(\node V_{m \sub I - 1})) \derives^* \epsilon,
  \end{gather*}
  respectively. Finally, since $\node V_{m \sub I}$ is the one-node
  tree whose root is labeled $a$---again, by \ruleref{item:downward
    link conversion function:terminal link} of \dref{def:downward
    link conversion function}---we have \eqref{eq:claim:Q_D(Q_D\inv(Y,
    alpha))=Y} if $h \sub V=1$.

  Assume now that we have established \eqref{eq:claim:Q_D(Q_D\inv(Y,
    alpha))=Y} if $h \sub V =k \sub V$, where $k \sub V\ge 1$; assume
  also that $h \sub V=k \sub V+1$. When considering the evaluation of
  $Q \sub D\inv$ on $\node Y$, we recursively evaluate of $Q \sub
  D\inv$ on $(\node Y \sub h, \alpha \sub h)$, such that the final
  element of $\node Y \sub h$ is of height $k \sub V$.  We can
  therefore say, by the induction hypothesis, that $Q \sub D(Q \sub
  D\inv(\node Y \sub h, \alpha \sub h)=\node Y_h$. We return to the
  evaluation of $Q \sub D$. Note that $L_1$ is a partial downward link
  for $\alpha$ to $a$, the value of which we denote $(\alpha \sub I, m
  \sub I)$. We denote $\alpha=X_1 X_2 \dotso X_{|\alpha|}$. Let $\node
  V_1', \node V_2', \dotsc, \node V_{m \sub I - 1}'$ be parse trees
  for the derivations
  \begin{gather*}
    X_1 \derives^* \epsilon, \\
    X_2 \derives^* \epsilon, \\
    \vdots \\
    X_{m \sub I - 1} \derives^* \epsilon,
  \end{gather*}
  and let $\node V_{m \sub I}'$ be the tree formed by making each of
  the elements of $\node Y_h$ children of a node labeled $\production
  {X_{m \sub I}} \gamma$. However, this is just $\node Y$, so we have
  \eqref{eq:claim:Q_D(Q_D\inv(Y, alpha))=Y}.

  Let $k \in \{0, 1\}$, and let $y \in \Sigma^*$ such that $|y|=k$.
  Let $\mathbb T \sub P \subset \mathbb T$ be the set of all
  parse-proper trees whose leaves are either a grammar symbol or
  $\epsilon$. We define $Q_{k,y} \inv \colon \mathbb T \sub P \to
  \mathbb U_k$ for some $\node T \sub x \in \mathbb T \sub P$ as
  follows.
  \begin{enumerate}
  \item Assuming that $\node T \sub x$ has no nodes, we let
    \[
    Q_{k,y} \inv (\node T \sub x) = ().
    \]

  \item Assume that $\node T \sub x$ has multiple nodes, and assume
    also that the greatest leaf is labeled by the production
    $\production {A \sub x} {\alpha \sub x}$. Let $\node T \sub x '$
    be that tree formed from $\node T \sub x$ by removing the latter's
    greatest leaf, and let $(i_1, i_2, \dotsc, i_{n \sub r}) = Q_{k,y}
    \inv (\node T \sub x ')$. Finally, let
    \[
    Q_{k,y} \inv (\node T \sub x) \equiv (i_1, i_2, \dotsc, i_{n \sub
      r}, [\production {A \sub x} {\cdot \alpha \sub x}, y])
    \]

  \item Assume that $\node T \sub x$ has multiple nodes, and assume
    also that the greatest leaf $\node N \sub x$ is labeled by the
    grammar symbol $X \sub x$; let the parent of $\node N \sub x$ be
    labeled $E \to \gamma \sub x$. Let $m \sub x - 1$ be the number of
    siblings of $\node N \sub x$, and let $\node T \sub x ''$ be that
    tree formed from $\node T \sub x$ by removing $\node N \sub
    x$. Let
    \[
    (j_1, j_2, \dotsc, j_{p \sub x}) = Q_{k,y} \inv (\node T \sub x
    '').
    \]
    Let $\gamma \sub x = Z_1 Z_2 \dotso Z_{s \sub x},$ and let
    \[
    j' =[ \production {E} {Z_1 Z_2 \dotso Z_{m \sub x} \cdot Z_{m \sub
        x + 1} \dotso Z_{s \sub x}}, y].
    \]
    Finally, let
    \[
    Q_{k,y} \inv (\node T \sub x) \equiv (j_1, j_2, \dotsc, j_{s \sub
      x}, j').
    \]
  \end{enumerate}    

  We establish an intermediate result.  Let $k \in \{0, 1\}$, and let
  $z \in \Sigma^*$ such that $|z|=k$.  Let $\node T \sub y$ be a
  parse-proper tree, and whose root is a production node. We wish to
  show that
  \begin{equation}
    \label{eq:Q_k(Q_{k,z} \inv(T)) = T}
    Q_k(Q_{k,z} \inv (\node T \sub y))=\node T \sub y
  \end{equation}
  Letting the number of nodes in $\node T \sub y$ be $n \sub y$, we
  proceed by induction on $n \sub y$. Assume, for our basis step, that
  $n \sub y = 1$. In this case, the only node is labeled $\production
  {A \sub y} {\alpha \sub y}$. The application of $Q_{k,z} \inv$
  yields the sequence
  \[
  ([\production {A \sub y} {\cdot  \alpha\sub{y}}, z]);
  \]
  this is the input of $Q_k$. The application of $Q_k$, as given by
  \ruleref{item:consumption function:|U|=1} of
  \dref{def:consumption function}, will create a one-node
  tree whose node is labeled $\production {A \sub y} { \alpha \sub
    y}$. This is just $\node T \sub y$.

  Assume that we know that \eqref{eq:Q_k(Q_{k,z} \inv(T)) = T} holds
  for $k \sub y$\dash{}node trees, where $k \sub y \ge 1$. Assume also
  that $n \sub y = k \sub y + 1$. We consider the greatest leaf of
  $\node T \sub y$, a leaf that we label $\node F \sub y$. This leaf
  may be either a grammar symbol or a production node---we consider
  these cases separately.
  \begin{enumerate}
  \item Assume that $\nodeLabel(\node F \sub y)=\production {C \sub y}
    {\gamma \sub y}$. Let $\node T \sub y'$ be that tree formed from
    $\node T \sub y$ by removing $\node F \sub y$; in order to
    evaluate $Q_{k,z}\inv$ on $\node T \sub y$, we first evaluate
    $Q_{k,z}$ on $\node T \sub y'$. By the induction hypothesis,
    $Q_k(Q_{k,z} \inv (\node T \sub y ')) = \node T \sub y '$. The
    value of $Q_{k,z} \inv(\node T \sub y)$ will be the item
    $[\production{C \sub y} {\cdot \gamma \sub y}, z]$ appended to the
    sequence $Q_{k,z} \inv (\node T \sub y ')$.  Consider the
    evaluation of $Q_k$ when we evaluate the expression $Q_k(Q_{k,z}
    \inv (\node T \sub y ))$: we first recursively evaluate of $Q_k$
    with the sequence of items $Q_{k,z} \inv (\node T \sub y ')$,
    which will yield $\node T \sub y'$; the evaluation of $Q_k$ will
    be completed by adding a node, corresponding to the item
    $[\production{C \sub y} {\cdot \gamma \sub y}, z]$ to $\node T
    \sub y'$; this node will be labeled $\production{C \sub y} {\gamma
      \sub y}$, hence, $Q_k(Q_{k,z} \inv (\node T \sub y
    ))=\node T \sub y$.

  \item Assume that $\nodeLabel(\node F \sub y)=X$, for some $X \in
    \Sigma \cup \Nu$. This case is similar to the first. The label of
    the parent of $\node F \sub y$ we give the label $\production {D
      \sub y } {\delta \sub y X \zeta \sub y }$, such that $\node F
    \sub y$ has $\left \lvert \delta \sub y \right \rvert$ siblings.
    Let $\node T \sub y''$ be that tree formed from $\node T \sub y$
    be removing $\node F \sub y$. As before, when evaluating
    $Q_k(Q_{k,z} \inv(\node T \sub y))$, we form the final tree from
    $\node T \sub y''$ and the item $[\production {D \sub y } {\delta
      \sub y X \cdot \zeta \sub y }, z]$; this yields $\node T \sub
    y$.
  \end{enumerate}

  We can appeal to this result directly to conclude that there is a
  function $Q \sub A \inv$ such that, for any parse-proper tree $\node
  T \sub a$, we have that
  \begin{equation}
    \label{def:Q_A(Q_A\inv(T))=T}
    Q \sub A (Q \sub A \inv(\node T \sub a)) = \node T \sub a .
  \end{equation}
  Similarly, there is a function $Q \sub u \inv$ such that
  \[
  Q \sub U(Q \sub U \inv(\node T \sub u))=\node T \sub u.
  \]

  We now return to the trees we considered in the beginning of this
  proof.  Let $D = Q \sub D \inv (\node Y)$, and let $J = Q \sub A
  \inv (\node T \sub J)$. Let $\pi=\nodeLabel(\node S)$, recalling
  that $\node S$ is a node in the tree $\node T \sub s$. Let the index
  of the child of $\node S$ that is autoancestral to $\node B$ be $i
  \sub b$, we write $\pi=\production {F} {\eta \theta}$, such that
  $|\eta|=i \sub b$; thus we let $s=[\production {F} {\eta \cdot
    \theta}]$. Finally, we turn to $U$: if $\node T \sub U$ has a
  single node, then we let $U \equiv ()$, otherwise, $\node T \sub U$
  has multiple nodes, in which case we let $U \equiv Q \sub U
  \inv(\node T \sub U)$.

  Let $\node V \equiv R(U, J, s, D)$. If we can show that $\node V
  \sub s=\node T \sub s$, then we shall have shown that $R$ is
  surjective. We first note that $\node T \sub J = Q \sub A (J)$ and
  $\node Y = Q \sub D (D)$. Now there are two cases to consider.
  \begin{enumerate}
  \item Assume that $\node T \sub U$ has but a single node. In this
    case, we attach a single node labeled $B$ as the greatest child to
    $\node T \sub J$, as in \ruleref{item:parse path conversion
      function:|I|=0} of \dref{def:parse path conversion function};
    but since $\node T \sub U$ has but a single node, then this node
    must be labeled $B$.

  \item Assume that $\node T \sub U$ has multiple nodes. If this is
    the case, then $\node T \sub U = Q \sub U (U)$. We note that the
    tree $\node T \sub {U,0}$, as specified in \dref{def:parse path
      conversion function}, is in fact equal to $\node T \sub U '$, as
    specified in this proof. In that Definition, we construct $R(U, J,
    s, D)$ accoring to \ruleref{item:parse path conversion
      function:|I|>0} by attaching $\node T \sub U '$ to $\node T \sub
    J$ at the place that it originally resided.
  \end{enumerate}
  In either case, we finish by attaching the nodes from $\node Y$
  following the root of $\node T \sub U '$. Clearly, this yields
  $\node T \sub s$.

  Assume now, for the sake of a contradiction, that $R$ is not
  injective. That is: assume that there are two distinct parse-paths
  $(U_1, J_1, s_1, D_1)$ and $(U_2, J_2, s_2, D_2)$ such that
  \begin{equation}
    \label{eq:R is not injective}
    R(U_1, J_1, s_1, D_1)=R(U_2, J_2, s_2, D_2).
  \end{equation}

  There are several ways in which these parse paths could differ. We
  examine each of these ways, eliminating each in turn.

  Let $P_k$ and $R_k$ be two \kparse{k} paths, such that $Q_k(P_k) =
  Q_k(R_k)$; let $w \in \Sigma^*$ such that $|w|=k$. We will show that
  $P_k = R_k$. Assume, for the sake of a contradiction, that $P_k \ne
  R_k$. Since the number of nodes in $Q_k(P_k)$ is $|P_k|$, and since
  $Q_k(P_k)=Q_k(R_k)$, we have that $|P_k|=|R_k|$. Let $P_k=(p_1, p_2,
  \dotsc, p_{n \sub P})$ and let $R_k=(r_1, r_2, \dotsc, r_{n \sub
    R})$. Let $i \sub z$ be the least index such that $p_{i \sub z}
  \ne r_{i \sub z}$; let $p_h = [\production {A_{\mathrm P,h}}
  {\alpha_{\mathrm P,h} \cdot \gamma_{\mathrm P,h}}, w]$ and let $r_k
  = [\production {A_{\mathrm P,h}} {\alpha_{\mathrm P,h} \cdot
    \gamma_{\mathrm P,h}}, w]$, for $1 \le h \le n \sub R$.
  There are \ref{item:Q_x is injective:last} cases to consider.
  \begin{enumerate}
  \item Assume that $i \sub z = 1$; thus, $\alpha \sub P = \alpha \sub
    R = \epsilon$. Since $Q_k$ yields identical one-node trees when
    evaluated on the sequences $(p_1)$ and $(r_1)$, we must have that
    $A_{\mathrm P, 1} \to \gamma_{\mathrm P, 1} = A_{\mathrm R, 1} \to
    \gamma_{\mathrm R, 1}$.

  \item 
    \label{item:alpha P != epsilon}
    Assume that $i \sub z > 1$. As $p_{i \sub z - 1} = r_{i \sub z - 1}$, we have
    that
    \begin{align*}
      A_{\mathrm P, i \sub z - 1} &= A_{\mathrm R, i \sub z -
        1},\\
      \alpha_{\mathrm P, i \sub z - 1} &= \alpha_{\mathrm R, i \sub z -
        1}\text{, and}\\
      \gamma_{\mathrm P, i \sub z - 1} &= \gamma_{\mathrm R, i \sub z -
        1}.
    \end{align*}
    We have either that:
    \begin{enumerate}
    \item if $\alpha_{\mathrm P, i \sub z} = \epsilon$, then the node
      corresponding to $p_{i \sub z}$ is the first child of a node
      corresponding to $p_{i \sub z -1 }$; let $\node N_{\mathrm P}$
      in $Q_k(P_k)$ and $\node N_{\mathrm R}$ in $Q_k(R_k)$ be the
      nodes corresponding to $p_{i \sub z - 1}$ and $r_{i \sub z -
        1}$, respectively; the first children of $\node N_{\mathrm P}$
      and $\node N_{\mathrm R}$, respectively, is labeled
      \[    
      A_1 \to \gamma_1
      \text{ and }
      A_2 \to\gamma_2,
      \]
      such that these two productions are identical; thus, $p_{i \sub
        z} = r_{i \sub z}$;

    \item if $\alpha_{\mathrm P, i \sub z} \ne \epsilon$, then 
      \[
      A_{\mathrm P, i \sub z - 1} \to \alpha_{\mathrm P, i \sub z - 1}
      \gamma_{\mathrm P, i \sub z - 1} = A_{\mathrm P, i \sub z} \to
      \alpha_{\mathrm P, i \sub z} \gamma_{\mathrm P, i \sub z}
      \]
      such that 
      \[
      |\alpha_{\mathrm P, i \sub z - 1}| + 1 = |\alpha_{\mathrm P, i
        \sub z}|;
      \]
      note that the node corresponding to $p_{i \sub z - 1}$ is
      identical to the node corresponding to $r_{i \sub z - 1}$, and
      that the least sibling of $p_{i \sub z - 1}$ that is greater
      than $p_{i \sub z - 1}$ corresponds to $p_{i \sub z}$, and is
      labeled by the grammar symbol $Y$, such that
      \begin{equation}
        \label{eq:alpha has Y appended}
        \alpha_{\mathrm P,i \sub z - 1} Y = \alpha_{\mathrm P,i \sub z};
      \end{equation}
      the node corresponding to $r_{i \sub z - 1}$ must have a sibling
      identical to the one corresponding to $p_{i \sub z}$; this
      sibling corresponds to $r_{i \sub z}$, so we conclude from
      \eqref{eq:alpha has Y appended} that $p_{i \sub z} = r_{i \sub
        z}$.
    \end{enumerate}
    \label{item:Q_x is injective:last}
  \end{enumerate}
  In all of these cases, we see that we have $P_k=R_k$.

  We have established a result that allows us to claim that $U_1=U_2$
  and that $J_1=J_2$.

  Assume that $D_1 \ne D_2$; we clearly have that $Q \sub D(D_1) = Q
  \sub D(D_2)$---from this, we will derive a contradiction. Let $H$ be
  the height of the last element of $Q \sub D(D_1)$; thus
  \[
  |D_1|=H=|D_2|.
  \]
  Let $D_1=(L_1, L_2, \dotsc, L_H)$ and let $D_2=(L_1', L_2', \dotsc,
  L_H')$, and let $k \sub d$ be the least index such that $L_{k \sub
    d} \ne L_{k \sub d}'$. There are two cases.
  \begin{enumerate}
  \item Assume that $k \sub d = 1$.  Let $Q \sub D(D_1)=(Y \sub {L,1},
    Y \sub {L,2}, \dotsc, Y_{L, n \sub Y})$; we have that
    \[
    \gamma \sub L=\nodeLabel(\treeRoot(\node Y \sub {L,1})) \nodeLabel(\treeRoot(\node Y \sub {L,2})) \dotso \nodeLabel(\treeRoot(\node Y \sub {L,n \sub Y})).
    \]
    Since $L_1=(\gamma \sub L, |Q \sub D(D_1)|)$ and let $L_1'=(\gamma
    \sub L, |Q \sub D(D_2)|)$, and since $ |Q \sub D(D_1)|= |Q \sub
    D(D_2)|$, we conclude that $L_1=L_1'$.

  \item Assume that $k \sub d > 1$. Let $\node T \sub d$ and $\node T
    \sub d'$ be the final elements of $Q \sub D(D_1)$ and $Q \sub
    D(D_2)$, respectively. In this case, let $\node N \sub d$ be the
    node in $\node T \sub d$ such that $\node N \sub d$ is the
    greatest node with exactly $|D_1| - k \sub d$ ancestors; let
    $\node N \sub d'$ be similarly defined for $\node T \sub d'$. The
    parents of the nodes $\node N \sub d$ and $\node N \sub d'$,
    corresponding as they do to the identical links $L_{k \sub d - 1}$
    and $L_{k \sub d - 1}'$, are identical, and identically labeled
    with the production $\production {A \sub L} {\alpha \sub L}$. Both
    $L_{k \sub d}$ and $L_{k \sub d}'$ are links, either
    partial-downward or terminal, for $\alpha \sub L$. Let the number
    of siblings of $\node N \sub d$ be $n \sub d$, which is also the
    number of siblings of $\node N \sub d'$. Clearly,
    \begin{align*}
      L_{k \sub d} &= (\alpha \sub L, n \sub d)\text{, and}\\
      L_{k \sub d}' &= (\alpha \sub L, n \sub d).
    \end{align*}
    Thus, conclude that $L_{k \sub d} = L_{k \sub d}'$.
  \end{enumerate}
  As we have a contradiction either way, we have therefore that
  $D_1=D_2$.

  The only other possibility is that $s_1 \ne s_2$. Let $\node P \sub
  s$ be the attach-point of $Q \sub A(J_1)$, and let $\nodeLabel(\node
  P \sub s) = \production {A \sub s} {\alpha \sub s X \sub s \gamma
    \sub s}$ such that $\node P \sub s$ has $|\alpha \sub s|$
  children. The child-string of $\node P \sub s$ is clearly $\alpha
  \sub s$, and $L_1$ (the first element of $D_1$ or $D_2$) is a link
  for $\gamma \sub s$. Therefore,
  \[
  s_1 = [\production {A \sub s} {\alpha \sub s X \sub s \cdot \gamma
    \sub s}] = s_2.
  \]

  Since $(U_1, J_1, s_1, D_1)=(U_2, J_2, s_2, D_2)$, we have that $R$
  is injective, and therefore, a bijection.
\end{proof}

\begin{theorem}
  Let $G=(\Sigma, \Nu, P, S)$ be an LR(1) grammar, let $\beta B$ be a
  viable prefix followed by $a$, for $B \in \Nu$, and let $\mathbf P$
  be a parse-path for $\beta B$. If $\pi$ is a production, then
  \[
  L_{\mathrm V, \mathbf P}(\pi) = M(R(\mathbf P), \pi).
  \]
\end{theorem}
\begin{proof}
  Let $\mathbf P=(U, J, s, D)$.

  Let $U=(u_1, u_2, \dotsc, u_n)$, let $J=(j_1, j_2, \dotsc, j_m)$,
  let $s=[\production {E} {\eta \cdot \theta}]$, and let $D=(L_1, L_2,
  \dotsc, L_p)$.  For $1 \le i\sub{n} \le n$, let $u_{i \sub n} \equiv
  [ \production {A_{i \sub n}} {\alpha_{i \sub n} \cdot \gamma_{i \sub
      n}}, a]$; likewise, for $1 \le i\sub{m} \le m$, let $j_{i \sub
    m} \equiv [ \production {C_{i \sub m}} {\delta_{i \sub m} \cdot
    \zeta_{i \sub m}}]$; finally, for $1 \le i\sub{p} \le p$, we let
  $L_{i \sub p} \equiv (\theta_{i \sub p}, k_{i \sub p})$.  Let $\node
  T$ be the proper simplified tree for $\beta B$ followed by $a$. Let
  $\node T \sub U = Q \sub U(U)$, let $\node T \sub J = Q \sub A(J)$,
  and let $\node Y = Q \sub D(D)$.

  Let $\pi=\production A \alpha$ be a production, and let $L_{\mathrm
    V, \mathbf P}(\pi)=n \sub V$. We will show that
  \begin{equation}
    \label{eq:L <= M}
    L_{\mathrm V, \mathbf P}(\pi) \le M(\node T, \pi).
  \end{equation}

  If $n \sub V = -1$, then $M(\node T, \pi) \ge n \sub V$ very
  trivially. So assume that $n \sub V \ne -1$. There are six
  possibilities.
  \begin{enumerate}
  \item Assume that $V \sub \Pi (\pi) = n \sub V$. In this case, there
    is some $i \sub u$ such that
    \[
    \production {A_{i \sub u}} {\alpha_{i \sub u} \gamma_{i \sub u}}
    = \pi.
    \]
    Now 
    \[
    u_{i \sub u} = [\production {A_{i \sub u}} {\alpha_{i \sub u}
      \cdot \gamma_{i \sub u}}, a];
    \]
    let $l \sub u=i \sub u - |\alpha_{i \sub u}|$; evidently,
    \[
    u_{l \sub u} = [\production {A_{l \sub u}} {\cdot \gamma_{l \sub
        u}}, a];
    \]
    which is to say that $\alpha_{l \sub u} = \epsilon$.  When we
    evaluate $Q \sub U$ on the item $u_{l \sub u}$, a node labeled
    $\pi$ gets added to $Q \sub U((u_1, u_2, \dotsc, u_{l \sub u -
      1}))$; let this node be $\node N \sub u$. Since $\node N \sub u$
    is an ancestor of $\node B$, but not $\node A$, then by
    \dref{def:[proper] simplified conservation function}, we have that
    $M(\node T, \pi) = |\gamma_{l \sub u}| + 1$.

  \item Assume that $V_{\Pi,\epsilon}(\pi) = n \sub V$. This means
    that $\pi$ is used in the derivation 
    \[
    \gamma_{i_\epsilon} \derives^* \epsilon,
    \]
    for some $1 \le i_\epsilon \le n$. There must, according to
    \dref{def:parse path conversion function}, be some node labeled
    $\pi$ in on of the parse trees attached to $\node T \sub U$ to
    form $\node T \sub {U,0}$, as specified in the same
    Definition. Let $\node N \sub {U,0}$ be this node. Since $\node N
    \sub {U,0}$ shares an ancestor (corresponding to the attach-point
    of $\node T \sub J$) with both $\node B$ and $\node A$, we have
    that $M(\node T, \pi) = |\alpha| + 1$.

  \item Assume that $V_\Phi(\pi) = n \sub V$. This means that there
    is some $1 \le k \sub J \le m$ such that $\production {C_{k \sub
        J}} {\delta_{k \sub J} \zeta_{k \sub J}} = \pi$. We have in
    this case that
    \begin{equation}
      \label{eq:|delta| + 1 = n_V}
      |\delta_{k \sub J}| + 1 = n \sub V.
    \end{equation}
    As every production node of $\node T \sub J$ is the greatest of
    its siblings, we see that only the greatest child of a production
    node can be autoancestral to $\node B$. Let $k'=k \sub J -
    |\delta_{k \sub J}|$; When we evaluate $Q \sub A$ on $j_{k'}$, we
    add a node labeled $\pi$ to $\node T \sub J$---let this node be
    $\node N \sub J$. There will be at least $|\delta_{k \sub J}|$
    children of $\node N \sub J$. Thus, by \dref{def:[proper]
      simplified conservation function},
    \[
    H_{\node T \sub J} (\node N \sub J)=|\delta_{k \sub J}| + 1,
    \]
    and so $M(\node T, \pi) = |\delta_{k \sub J}| + 1$.
    
  \item Assume that $V_\Psi(\pi) = n \sub V$. This means that
    there is some $1 \le h \sub V \le p$ such that
    \[
    P(L_{h \sub v}, L_{h \sub v + 1}) = \pi,
    \]
    where 
    \begin{equation}
      \label{eq:value of downlink = n_V}
      k_{h \sub V + 1} = n \sub V.
    \end{equation}
    When evaluating $Q \sub D$ on $L_{h \sub V}$, we create a sequence
    of trees: the final tree $\node T \sub D$ in this evaluation will
    have a root labeled $\pi$, such that this root has $k_{h \sub V +
      1}$ children; by \dref{def:[proper] simplified conservation
      function}, the function
    \[
    H_{T \sub D}(\pi) = k_{h \sub V + 1}.
    \]
    We have, by \eqref{eq:value of downlink = n_V}, that
    \[
    M(\node T \sub D, \pi) = n \sub V.
    \]

  \item Assume that $V_{\Psi,\epsilon}(\pi) = n \sub V$. This
    means that there is some $1 \le h \sub V' \le p$, where,
    writing $\theta_{h \sub V'}= X_1 X_2 \dotso X_{N \sub D}$, we
    have that $\pi$ is used in the derivation
    \begin{equation}
      \label{eq:X_{l_D} => epsilon}
      X_{l \sub D} \derives^* \epsilon,
    \end{equation}
    for some $1 \le l \sub D \le N \sub D$. When, according to
    \dref{def:downward link conversion function}, we evaluate $Q \sub
    D$ on the downward link $L_{h \sub V'}$, we create a parse tree
    for derivation \eqref{eq:X_{l_D} => epsilon}; in this parse tree,
    there will be a node $\node P_{D,\epsilon}$ which is labeled
    $\pi$. This node $\node P_{D,\epsilon}$ shares an ancestor with
    both $\node B$ and $\node A$; moreover, $\node B < \node
    P_{D,\epsilon} < \node A$. Therefore, we have
    \[
    H_{\node T}(\pi) = |\alpha| + 1 = n \sub V.
    \]

  \item Assume that $V_\Omega(\pi) = n \sub V$. In this case, we first
    note that $\pi=\production {E} {\eta \theta}$. When we join the
    two trees $\node T \sub J$ and $\node T \sub U$ to the trees in
    $\node Y$, we find that the node that was the attach-point of
    $\node T \sub U$ now has $|\eta| - 1$ children inherited from
    $\node T \sub J$, one child corresponding to either the sidelink
    or the root of $\node T \sub U$, and one child for each of the
    elements of $\node Y$. There children are added in the order
    listed: note that the final element of $\node Y$ corresponds to a
    node that is autoancestral $\node A$. Let
    \[
    i \sub s = m - |\eta| - 1;
    \]
    since $|\eta| > 1$, we know that 
    \[
    j_{i \sub s}=[\production {E} {\cdot \eta \theta}].
    \]
    Let $\node P \sub s$ be the node corresponding to this item (added
    during the evaluation of $\node Q \sub A$); we have already
    established that $\node P \sub s$ is an ancestor of $\node A$ with
    $|\eta| + |\node Y|$ children, the last of which is autoancestral
    to $\node A$. Therefore, according to \caseref{item:parse path
      conservation function:V_Omega} of \dref{def:[proper] simplified
      conservation function}, we have that
    \[
    H_{\node T}(\node P \sub s) = n \sub V.
    \]
  \end{enumerate}

  Therefore, we have established \eqref{eq:L <= M}.

  Let $\phi=C \to \gamma$ be some production, and let $M(\node T,
  \phi)=n \sub T$. We dispense with the possibility that $n \sub T =
  -1$, as it is trivial; thus, assume that $n \sub T > -1$. There must
  be some node $\node P_\phi$ in $\node T$ such that $\nodeLabel(\node
  P_\phi)=\phi$, such that,
  \[
  H_{\node T}(\node P_\phi)=n \sub T.
  \]
  There are three cases to consider for the relationship of $\node
  P_\phi$ with the nodes $\node B$ and $\node A$ (with some subcases).
  \begin{enumerate}
  \item Assume that $\node P_\phi$ is an ancestor of $\node B$, while
    $\node P_\phi$ is not an ancestor of $\node A$. By the
    construction of $\node T \sub {U,0}$ in \dref{def:simplified
      tree}, we can see that $\node P_\phi$ has $|\gamma|$ children
    and so, the value of $H_{\node T}(\node P_\phi)$ is $|\gamma| +
    1$. There must be some item $u \sub B$ in $U$ corresponding to
    $\node P_\phi$; this item will be of the form
    \[
    u \sub B = [\production {C} {\cdot \gamma}, a].
    \]
    By \dref{def:parse path conservation function},
    \condref{item:parse path conservation function:V_Pi}, we can see
    that $V_\Pi(\phi) = |\gamma| + 1$.

  \item Assume that $\node P_\phi$ is an ancestor of $\node
    A$. In this case, we have that $\node P_\phi$ corresponds to
    some element of $(U, J, s, D)$, according to one of the
    following possibilities.
    \begin{enumerate}
    \item There could be some node in one of the elements of $\node Y$
      corresponding to $\node P_\phi$. This node, which we call $\node
      P \sub Y$, is in the last element of $\node Y$, a tree which we
      label $\node T \sub Y$. Let the number of ancestors of $\node P
      \sub Y$ in $\node T \sub Y$ be $h \sub a$. When evaluating $Q
      \sub D$ on $D$, we add the node $\node P \sub
      Y'$---corresponding to $\node P \sub Y$---during the evaluation
      of $Q \sub D$ on $L_{h \sub a + 1}$. This node $\node P \sub T'$
      will be the root of a subtree of $\node T \sub Y$; a subtree
      which was created when evaluating $Q \sub D$ on $L_{h \sub a +
        1}$.  Since $\node P \sub T'$ has $k_{k \sub a + 1}$ children,
      \[
      n \sub T=k_{h \sub a} ;
      \]
      thus, since $P(L_{h \sub a + 1}, L_{h \sub a + 2})=\phi$, we have
      therefore that
      \[
      V_\Psi(D)=k_{h \sub a + 1}=n \sub T.
      \]

    \item
      \label{item:P is in J, P's child that is ancestral to A not in
        J}
      Assume that the child of $\node P \sub \phi$ which is
      autoancestral of \node A does not correspond to any node in
      $\node T \sub J$. Let $\node P \sub A$ be the node in $T \sub J$
      corresponding to $\node P_\phi$; note that $\node P \sub A$ has
      $n \sub T - 1$ children in $T \sub J$. We added $\node P \sub A$
      to $\node T \sub J$ when evaluating $Q \sub A$ on the item $j_{m
        - n \sub T + 1}$. Let $j_{m - n \sub T + 1}$ be of the form
      \[
      [ D \sub J \to \cdot \zeta \sub J \eta \sub J];
      \]
      evidently, $j_m$ is of the form 
      \[
      [ D \sub J \to \zeta \sub J \cdot \eta \sub J].
      \]
      We know that $|\eta \sub J| > 1$, and so we write $\eta \sub J
      \equiv X \sub J \eta \sub s$; we must have that 
      \[
      s=[\production {D \sub J} {\zeta \sub J X \sub J \cdot \eta \sub
        s}] = [ \production {E} {\eta \cdot \theta}].
      \]
      Now, $\node P_\phi$ will have one child corresponding to each of
      the children of $\node P \sub A$; additionally, it will have one
      $\node P \sub c$, such that $\node P \sub c$ is either labeled
      $B$, or $\node P \sub c$ is the node corresponding to the root
      of $Q \sub U(U)$; finally, it will have one child for each of
      the elements of $\node Y$. Since $|\node Y|=k_1$, we have that
      \[
      n \sub T = |\eta| + k_1;
      \]
      thus, by \dref{def:parse path conservation function}, we conclude that 
      \[
      V_\Omega(\phi)=n \sub T.
      \]
      
    \item Assume that we do not have \caseref{item:P is in J, P's
        child that is ancestral to A not in J}, and that there is some
      item in $J$ corresponding to $\node P_\phi$. This item is of the
      form
      \[
      j_{i \sub A} = [\production {C} {\cdot \gamma}].
      \]
      As $\node P_\phi$ has $n \sub T - 1$ children, we must have that
      \[
      j_{i \sub A + n \sub T - 1} = [\production {C} {\delta \sub J
        \cdot \zeta \sub J}],
      \]
      where
      \[
      |\delta_J| = n \sub T - 1
      \]
      such that $\delta \sub J \zeta \sub J = \gamma$. Therefore,
      according to \dref{def:parse path conservation function},
      \[
      V_\Phi(\phi)=|\delta \sub J| + 1=n \sub T.
      \]
    \end{enumerate}
    In all three of these cases, we find that
    \[
    L \sub V(\phi) \ge M(\node T, \phi).
    \]

  \item Finally, assume that $\node P_\phi$ shares an ancestor with
    both $\node B$ and $\node A$, but is an ancestor of neither, yet
    $\node B < \node P < \node A$. There are two ways that this can
    happen.
    \begin{enumerate}
    \item Assume that $\node P_\phi$ shares an ancestor with $\node B$
      but not $\node A$, such that $\node B < \node P < \node A$.
      Among all such ancestors, there is one, which we label $\node Q
      \sub B$, such that $\node Q \sub B$ is the root of a subtree
      that shares no nodes with $Q \sub U(U)$, save Let $\node T \sub
      {U, 0}$ be an in \dref{def:parse path conversion function}.  Let
      the node in $\node T \sub {U, 0}$ corresponding to $\node
      P_\phi$ be $\node P \sub {U, 0}$. Let $\node R \sub B$ be that
      child of $\node Q \sub B$ that is autoancestral to $\node B$; by
      the construction of $\node T \sub {U, 0}$, we can see that the
      subtree rooted at $\node R \sub B$ corresponds to one of the
      derivations
      \[
      \gamma_{k_0} \derives^* \epsilon
      \]
      for some $1 \le k_0 \le n$. Therefore,
      \[
      V_{\Pi,\epsilon}(\phi) = |\gamma| + 1.
      \]

    \item Assume that $\node P_\phi$ shares no ancestor with $\node B$
      that is not also an ancestor of $\node A$. Let $\node Y=(\node
      V_1, \node V_2, \dotsc, \node V_{k_1})$.  Let $\node P_\phi$
      correspond to a node $\node P \sub {Y, \epsilon}$ in any of the
      trees in $\node Y$. If $\node P \sub {Y, \epsilon}$ is in any of
      $\node V_1, \node V_2, \dotsc, \node V_{k_1 - 1}$, say, $\node
      V_{h \sub Y}$, then this node $\node P \sub {Y, \epsilon}$ is
      the root of a subtree with yield $\epsilon$ in the tree $\node
      V_{h \sub Y}$. If $\node P_\phi$ corresponds to a node
      in $\node V_{k_1}$, since the yield of $\node V_{k_1}$ is $a$,
      and $\node P \sub {Y, \epsilon}$ is not an ancestor of $\node
      A$,  we have that $\node P \sub {Y,\epsilon}$ is again the root
      of a subtree with yield $\epsilon$. Therefore, 
      \[
      V_{\Psi,\epsilon}(\phi)=|\gamma| + 1.
      \]
    \end{enumerate}
  \end{enumerate}

  In all three of these cases, we have established that 
  \begin{equation}
    \label{eq:L >= M}
    L_{\mathrm V, \mathbf P}(\phi) \ge M(\node T, \phi).
  \end{equation}
  Therefore, by \eqref{eq:L <= M} and \eqref{eq:L >= M}, we have that
  \[
  L_{\mathrm V, \mathbf P}(\phi) = M(\node T, \phi). \qedhere
  \]
\end{proof}

Thus, we may say that $\mathbb P$ and $\mathbb T \sub P$ are
``isomorphic'' under $R$, with respect to conservation of productions.

% LocalWords:  sidelink TLR subtree vacuo lookahead autoancestral

%%% Local Variables: 
%%% mode: latex
%%% TeX-master: "tlr-plain"
%%% End: 

\section{Related Work}
\label{sec:related work}

Going back to the first days of high-level computer languages, the
general idea of a computer language whose parser could modify
itself---a construction called an ``extensible language''---was
considered and tried numerous times. However, these efforts were not
always met with success. Perhaps it was because compiler construction
as a discipline itself was not well understood, or that the
appropriate formal language theory had not been developed, or that the
extensible compilers were not powerful enough: for whatever reason,
extensible languages have largely fallen by the wayside.

One of the first serious extensible language projects was a variant of
Algol~60 called IMP \cite{361966}. Along with IMP, another well
regarded extensible language was ECL \cite{807976}. These languages
allowed programs to (in modern parlance) specify new productions for
the language, and supply a replacement template, much in the manner of
a macro definition. The parsers for such languages were apparently
complex, ad hoc, and arcane affairs; a programmer wishing to extend
such a beast needed to understand a fair bit of the internals of the
parser to extend it and understand the cause of problems.  

An often expressed goal for an extensible language would be to allow a
program to supply a new data type, along with associated infix
operations, so that programs dealing with matrices or complex numbers
could use the natural syntax.  The modern approach to this problem is
to use operator overloading in an object-oriented language. This begs
the question: why would a programmer want to engage in the difficult
endeavour of modifying the parser when a mechanism like operator
overloading suffices?

The high point of interest in extensible languages was likely the
International Symposium on Extensible Languages. The Proceedings of
this Symposium \cite{800006} contain several reports on real-world
extensible languages: there are many reports on languages like ECL
(e.g. \cite{807976}, \cite{807988}, and \cite{807990}); some more
general works on macro systems (e.g. \cite{807977}, and
\cite{807983}); and some survey works (e.g. \cite{807996}, and
\cite{807999}). The mood was upbeat, but a little over-optimistic;
indeed, as Cheatham put it, extensible languages had delivered:
``there exist languages and host systems which fulfill the goals of
extensibility'' \citeyear{807999}. But not everyone was upbeat: there
were reports of failures---not of implementation, but of extensibility
itself \cite{807998}.

The extensible language concept did not disappear, even after the
appearance of languages like C++. One example is \cite{141037}, a
language that turned out to be only partially successful, complex to
use, and crippled by performance problems.

Some of these systems use a self-modifying compiler, and operate in a
single pass, while others use a two-pass compiler. Almost all of them
do a textual substitution, at least on the conceptual level. We would
therefore consider the study of these systems to be a study of macro
systems: if a macro system admits patterns that are more complex than
a function call (e.g. the CPP macro system for C and C++ requires
macros to be of the form \verb-MACRO_NAME(PARAM1, PARAM2, ...)-), then
we may say that the macro system is a \defn{syntactic macro}
system---otherwise, we say that the macro system is a \defn{simple
  macro} system. The study of syntactic macros has continued in its
own right, and syntax macros are present in some modern languages,
including Scheme \cite{290234}.

As interesting summary of the issues related to advanced macro systems
is due to Brabrand and Schwartzbach \citeyear{503035}, who summarize
prominent macro systems, and present a new one of their own. The macro
system presented in \cite{503035} operates on partial parse-trees,
which illustrates the necessity of a macro-aware parser.

The aforementioned syntactic macro systems can use macros in very
powerful ways, achieving many of the goals of the early extensible
languages. It is certainly possible to implement a language with
syntax macros using a parser for what we have in the present work
termed a transformative parser, which would allow for an implicit
macro call. This has been done in \cite{digital}, where a syntax macro
programming language implemented using a transformative LL parser is
described. The latter system is powerful enough to extend a functional
language into an imperative language (like C), and it avoids the
problems associated with many macro systems.

Most likely due to their syntactic simplicity (even austerity),
syntax-macros are usually reserved for functional programming
languages. That is not to say that they cannot be used for a
syntactically-rich language like C. Exactly this was done for C by 
Weise and Crew \citeyear{155105}; one point of note is that, rather
than supply a static template with which to replace the macro
invocation, their system allows for the replacement to be generated by
running procedural code on the macro parameters; the replacement will
be an abstract syntax tree. Allowing the macro body to include code,
which will be run (most likely by an embedded interpreter) during
compilation, could be called \defn{compile-time computation}. Other
macro systems allow this---Scheme most notably. Another system which
allows for compile-time computation is C++: it has been discovered
that the C++ template system is Turing complete
\cite{Veldhuizen:PEPM99}. 

The choice to allow compile-time computation in the macro body has
significant advantages---see \cite{243447} for a survey of partial
evaluation---but there are many drawbacks. The biggest drawback is the
increased complexity of having two languages side by side: the
run-time language and the compile-time language. 

The present work is part of an effort to make a real-world programming
language using a transformative LR(1) parser. This programming
language would, it is hoped, prove useful to developers of
domain-specific embedded languages (DSEL). The subject of DSEL is of
much interest today: for example, see the recent survey piece by
Mernik, Heering, and Sloane \citeyear{1118892}. In order to develop
DSELs, it is usually necessary to make modifications to the compiler's
source code, a task hopefully made easier by using a programming
language with a transformative LR(1) parser at its core. 

Another task which requires the modification of a compiler's source
code is the extension of a programming language to add new features:
for example, adding aspect-oriented capabilities to Java
\cite{1052906}. This is done often enough (especially with Java in
recent years) that software systems dedicated to this task have
appeared \cite{polyglot}. Indeed, this was one of the original
motivations for early extensible languages \cite{807997}. However, for
the Java systems mentioned above, the compiler is extended prior to
compilation, so we may term this an \defn{offline grammar
  transformation}. Some other modern systems do allow for
transformation during parsing, but a special macro invocation must be
used to tell the parser to launch a subparser---i.e. the parse is not
self-modifying. This can be done with quoting, as in \cite{141510}.

The formal basis for the TLR parsing algorithm is the theory of
LR($k$) languages. Introduced by Knuth \cite{knuth:lr}, his original
paper is insightful; another reference for the theory of LR($k$)
parsing is \cite{aho-ullman:top}. Coming at LR($k$) parsing from the
more practical side is the classic ``Dragon Book'' \cite{dragon} (for
$k=1$); this last work is particularly recommended. We base our
transformative language on the LR($1$) languages because the parsing
algorithm for this class of languages is well-suited for a
transformative language parser: since a substring of a sentence can
define the syntax for some substring immediately to its right, we
evidently want to scan sentences from left to right; a backtracking
algorithm is undesirable because it is complex and expensive to
backtrack past a point at which a grammar transformation occurred;
finally, bounded lookahead is important because until a decision is
made as to whether or not a grammar transformation will take place at
a certain point, it is unknown which (context-free) grammar has the
lookahead under its purview. Also, in practice, LR(1) parsers are
designed to execute code fragments after reduction by certain
productions: this is a natural place to insert the grammar
transformation algorithm---as indeed, we have done in the present
work.

It is not surprising that a parser for a transformative language based
the LR(1) languages has been presented before. Burshteyn
\citeyear{boris} formalizes an idea of \defn{modifiable
  grammars}---roughly equivalent to a transformative grammar. Much of
the present work is concerned with allowable transformations: in
\secref{sec:naive approach}, we saw the negative consequences of
admitting completely arbitrary transformations. In \cite{boris}, the
language generated by a modifiable grammar is equivalent to the naive
language considered in \secref{sec:naive approach}, so it does not
avoid those pitfalls. 

We do note that, if we only add or remove a few productions to or from
a grammar, the LR(1) parsing tables for that grammar do not change
``too much.'' The algorithm we present in the present work requires
the parsing table to be completely regenerated upon acceptance of a
grammar transformation. This need not be the case: indeed, in
\cite{boris}, the parser is modified only inasmuch as the grammar
transformation (our terminology) requires it; the method of
incremental LR(1) parser generation is originally due to Heering,
Klint, and Rekers \citeyear{74834}.  

The canonical LR(1) parser, creating by way of the method of Knuth
\citeyear{knuth:lr}, is often eschewed for the LALR(1) parser, owing
to the fact that the LALR(1) parser (if it even exists) has much
smaller parsing tables---however, Spector \citeyear{57684} showed how
to construct a different LR(1) parser that is often similar in size to
the LALR(1) parser. It would probably be a profitable exercise to
investigate the use of these techniques in the TLR parsing algorithm
for a real-world system.

The conventional view is that programming languages are not
context-free: that is, a program which uses an undeclared identifier
is considered to be syntactically well-formed (and hence in the
context-free language generated by the grammar) but semantically
meaningless.\footnote{Although there is the view that a program
  containing an ``undeclared identifier'' error is semantically
  well-formed; we could think of the error message produced by
  compiling it as the semantic value of the program \cite{knuth:lr}.}
In order to catch this semantic gaffe, the parser performs a separate
(at least in principle) semantic analysis phase, which is usually
performed by procedural code; an alternative, declarative approach to
syntactic and semantic analysis is to instruct the parser to add a new
production for an identifier when it is declared, later to be removed
when that identifier goes out of scope. A parser which modifies its
grammar to remove the need to perform semantic analysis will be herein
referred to as an \defn{adaptable grammar}. Two notable attempts at
adaptable grammar systems are \cite{382639} and \cite{inria}; the
field is surveyed in \cite{101357}. There has been renewed interest in
adaptable grammars: see \cite{carmi} and \cite{quinn}.

We have not been terribly clear about the differences between syntax
and semantics: if we (rightly) assume that syntax is what the parser
does, then processing of identifiers and scopes is evidently syntax,
assuming a powerful enough parser. It is in fact very difficult to
delineate syntax and semantics \cite{987483}.

It should be possible to make many types of systems on top of a TLR
parser. In this section, we have discussed: extensible languages,
syntax macros, and adaptable grammars.  The techniques in the present
work should be general enough to be used to achieve any of these three
techniques. Systems with these as their goals have not fared well: it
is to be hoped that the problem in the past was a lack of
understanding of the fundamental parsing issues; which will hopefully
be obviated by TLR techniques.

% LocalWords:   substring lookahead

%%% Local Variables: 
%%% mode: latex
%%% TeX-master: "tlr-plain"
%%% End: 

\section*{Conclusion}

There has been much interest over the years in languages with
features---like syntax macros, extensibility, and adaptable
grammars---that are incompatible with a parser generated from a static
grammar. Numerous as hoc efforts have been made to make systems with
these features and a parser whose grammar is not fixed, with little
success. One problem with these earlier attempts is a lack of
understanding of the consequences of changing the grammar---a
deficiency that this work hopes to address.

Another explanation may be that the features are not audacious enough:
why would someone want to trouble themselves with the arcana of the
parser to achieve something adequately accomplished by operator
overloading, templates, or semantic analysis? If, however, the
features are compelling enough, then programmers might be willing to
write grammar transformations.  We attempted, in the present work, to
take some of the mystery out of parsing a transformative language.

Naive transformative languages, while straightforward and occasionally
the subject of study, do not allow us to make guarantees about
halting. Requiring valid transformations does allow us to make
guaranteeds about halting, although the test for validity is
complex. We observe that the utility of this test does not end with
ensuring halting of compilation: should \algoref{algo:gta} fail, then
we know that the transformation is invalid---a report on which
productions were not conserved could be a useful diagnostic; also, the
stack represents the different ways that parser is trying to match the
sentence, and requiring valid transformations means that, once a
parser starts trying to match a production, that it cannot go back and
reinterpret what it already saw.

Many questions remain to be answered.

The central result of the present work is the correctness of
\algoref{algo:TLR}---see \tref{thm:correctness of the TLR parsing
  algo}---this correctness relies on the transformations emitted by
the \deltaMachine all being in $\mathcal V$; is there a larger set of
transformations which could fill the role played by $\mathcal V$?

We allow a full Turing machine to form the basis of a
\deltaMachine. Theorem~3 of \cite{boris} states (in part) that: ``Each
automatic BUMG (bottom-up modifiable grammar) accepts a context-free
language;'' in that work, an automatic BUMG is the counterpart of a
TLR grammar whose \deltaMachine is essentially a finite automaton. We
therefore ask: what class of languages are generated by TLR grammars
whose \deltaMachine{}s are (essentially) finite automata? What class
of languages are generated by TLR grammars whose \deltaMachine{}s have
bounded tapes?

The most important question to answer is this: can a practical
transformative programming language be constructed?

%%% Local Variables: 
%%% mode: latex
%%% End: 

\bibliographystyle{plain}
\bibliography{tlr-plain}
\end{document}